\documentclass[conference]{IEEEtran}
\usepackage{amsmath}
\usepackage{amsfonts}
\usepackage{array}
\usepackage{graphicx, caption, subcaption}
\usepackage{xcolor}
\usepackage{tikz}
\usepackage{booktabs}
\usepackage{multicol,lipsum}
\usepackage{bm}
\usepackage{float}
\usepackage{cuted}
\usepackage{bbm}
\usepackage{multirow}
\usepackage[noend, ruled]{algorithm2e}
\usepackage{tcolorbox}
\usepackage{array}
\usepackage{breqn}
\usepackage{mathtools}
\SetKwProg{Init}{Initialize:}{}{}
\usepackage{hyperref}

\usepackage{filecontents}

\usepackage{enumitem}
\setlist{nosep}

\usepackage[hang,flushmargin]{footmisc}

\let\oldbibliography\thebibliography
\renewcommand{\thebibliography}[1]{%
  \oldbibliography{#1}%
  \setlength{\itemsep}{4pt}%
}

\usepackage{diagbox}
\usepackage{xspace}
\usepackage{graphicx}

\newcommand{\dpsgd}{DP-SGD}
\newcommand{\memguard}{MemGuard}
\newcommand{\mixupMMD}{MixupMMD}
\newcommand{\advreg}{AdvReg}

\newcommand{\attack}{Diff-Cali}

\DeclareMathOperator*{\argmax}{\arg\!\max}
\newcommand{\mypara}[1]{\noindent\textbf{#1.}$\;$}

\definecolor{revision}{RGB}{0,0,255}

\newcommand{\ignore}[1]{}
\newcommand{\RNum}[1]{\uppercase\expandafter{\romannumeral #1\relax}}

\usepackage{amsthm}

\allowdisplaybreaks

\usepackage{etoolbox}
\makeatletter
\patchcmd{\hyper@makecurrent}{%
    \ifx\Hy@param\Hy@chapterstring
        \let\Hy@param\Hy@chapapp
    \fi
}{%
    \iftoggle{inappendix}{
        \@checkappendixparam{chapter}%
        \@checkappendixparam{section}%
        \@checkappendixparam{subsection}%
        \@checkappendixparam{subsubsection}%
        \@checkappendixparam{paragraph}%
        \@checkappendixparam{subparagraph}%
    }{}%
}{}{\errmessage{failed to patch}}

\newcommand*{\@checkappendixparam}[1]{%
    \def\@checkappendixparamtmp{#1}%
    \ifx\Hy@param\@checkappendixparamtmp
        \let\Hy@param\Hy@appendixstring
    \fi
}
\makeatletter

\newtoggle{inappendix}
\togglefalse{inappendix}

\apptocmd{\appendix}{\toggletrue{inappendix}}{}{\errmessage{failed to patch}}

\newcommand\zlbox[1]{{\begin{tcolorbox}[ colback=white, left=0pt, right=0pt, top=0pt, bottom=0pt] #1  \end{tcolorbox}}}

\begin{document}


\title{ A Comprehensive Study of Privacy Risks in Curriculum Learning}

\author{
{\rm Joann Qiongna Chen}\\
University of California, Irvine
\and
{\rm Xinlei He, \ Zheng Li, \ Yang Zhang}\\
CISPA Helmholtz Center for Information Security
\and
{\rm Zhou Li}\\
University of California, Irvine
} 

\maketitle

\begin{abstract}
Training a machine learning model with data following a meaningful order, i.e., from easy to hard, has been proven to be effective in accelerating the training process and achieving better model performance. The key enabling technique is curriculum learning (CL), which has seen great success and has been deployed in areas like image and text classification. Yet, how CL affects the privacy of machine learning is unclear. Given that CL changes the way a model memorizes the training data, its influence on data privacy needs to be thoroughly evaluated. To fill this knowledge gap, we perform the first study and leverage membership inference attack (MIA) and attribute inference attack (AIA) as two vectors to quantify the privacy leakage caused by CL.

Our evaluation of 9 real-world datasets with attack methods (NN-based, metric-based, label-only MIA, and NN-based AIA) revealed new insights about CL. First, MIA becomes slightly more effective when CL is applied, but the impact is much more prominent to a subset of training samples ranked as difficult. 
Second, a model trained under CL is less vulnerable under AIA, compared to MIA.
Third, the existing defense techniques like \dpsgd{}, \memguard{} and \mixupMMD{} are still effective under CL, though \dpsgd{} has a significant impact on target model accuracy. 
Finally, based on our insights into CL, we propose a new MIA, termed \attack{}, which exploits the difficulty scores for result calibration and is demonstrated to be effective against all CL methods and the normal training method. With this study, we hope to draw the community's attention to the unintended privacy risks of emerging machine-learning techniques and develop new attack benchmarks and defense solutions.    
\end{abstract}
\section{Introduction}\label{sec: intro}
Key to the success of machine learning (ML), especially deep learning (DL), is the advancement of algorithms, software, and hardware in training models on large-scale datasets. 
The traditional way to train a neural network (NN) is by feeding the training pipeline with random mini-batches in a sequence sampled from the training dataset. In other words, NN is forced to ``remember'' samples repeatedly in random order. 
On the other hand, human always learns the easy concepts first and then the hard ones, as guided by curricula. Given that NN is inspired by the human brain~\cite{rumelhart1986learning}, curriculum learning (CL), which simulates human learning by ordering the training data with difficulty scores and repeating the order across training epochs, has been proposed ~\cite{bengio2009curriculum}. With a ``teacher'' network, the difficult scores can be generated ahead of the samples and guide the training process. Previous studies have shown that CL can achieve both fast learning speed and high test accuracy~\cite{soviany2021curriculum, wang2021survey}, and CL has been adopted in many application domains like computer vision~\cite{bengio2009curriculum,sakaridis2019guided,duan2020curriculum,soviany2020image}, natural language processing~\cite{bengio2009curriculum,spitkovsky2009baby,zhou2020uncertainty,guo2020fine,liu2020task}, and claiming prominent success~\cite{wang2021survey}. 

Despite the huge success of ML, the privacy issues of ML are becoming more and more concerning, given that the training data could contain a large amount of sensitive information. The two most notable privacy attacks are membership inference attack (MIA)~\cite{JSBZG19, SSSS17} and attribute inference attack (AIA)~\cite{SS20}, where MIA aims to infer whether a given data sample is used to train the target model and AIA aims to infer the sensitive attribute of a data sample. Numerous attacks have emerged and demonstrated that privacy threats are real (e.g., over 80\% MIA accuracy against CIFAR100~\cite{SZHBFB19}). Recent studies have also shown the data samples are not equally vulnerable under privacy attacks~\cite{YKT19}, and the attack effectiveness could differ across target classes~\cite{JSBZG19}, target individuals~\cite{LBWBWTGC18}, and subgroups~\cite{chang2021privacy}. Yet, all previous works assume standard, stochastic training is employed by the target model. Hence, one interesting and important research problem is {how  new training techniques impact privacy for the overall population and individual samples.}  
In this work, we specifically study the privacy risks of CL. We are particularly motivated because CL modifies the data order, which differs from many new techniques such as  contrastive learning~\cite{HZ21} and other self-supervised learning techniques~\cite{liu2021self}. In general, CL increases a model's overall performance and lets a model treat samples differently based on their difficulty levels\footnote{The terms ``difficulty level'' and ``difficulty score'' are interchangeable.} Furthermore, Shumailov et al. ~\cite{shumailov2021manipulating} studied the connection between data ordering and backdoor attacks, which indicates data ordering could have negative impacts. This further motivates us to investigate the privacy risks of CL.



\mypara{Our Study} 
We take a quantitative approach to measure the privacy risks of CL. We selected two popular CL methods, bootstrapping~\cite{hacohen2019power} and transfer learning~\cite{weinshall2018curriculum}, as the evaluation objects, and constructed two other curriculum, named baseline curriculum and anti-curriculum, to understand the impact of data ordering and repeating, respectively. We selected 9 real-world, large-scale datasets (6 are image datasets and 3 are tabular datasets), trained target models with those CL methods and a normal method, and attacked the models with representative MIA and AIA methods. 

Regarding MIA, our evaluation shows that the target models become slightly more vulnerable under CL, e.g., the average attack accuracy (trained on ResNet-18 with transfer leaning) on our selected image datasets ranges from 0.01\% to  2.46\%.
More importantly, we found CL has a much bigger impact on the samples within the difficult group compared to the easy group, with the biggest gap of 4.23\% in terms of attack accuracy for CIFAR100 (ResNet-18 is the architecture). 
This observation sustains both image and non-image datasets.
We found the reason is that the data order reinforces the learning process hence making the model memorize difficult samples better, which is supported by measuring the memorization scores.
Regarding AIA, we found CL does not increase the attack accuracy, which can be explained by the fact that the sensitive attribute to be inferred is not influenced by data ordering and repeating. 



In addition to understanding the attacks, we also study existing defenses under the CL settings, including \dpsgd~\cite{abadi2016deep},  \memguard~\cite{JSBZG19}, \mixupMMD{}~\cite{li2021membership} and \advreg{}~\cite{NSH18}. 
 The result shows that none of them can mitigate the threats from MIA without dampening the model accuracy. 
 In particular, \dpsgd{} is effective in curbing MIA, and the drop of attack accuracy for the CL setting is even more than the normal setting. However, the privacy provided by  \dpsgd{} is at the cost of dropping the classification accuracy of the target model. 
 
Inspired by CL and a recent MIA that calibrates membership scores to achieve better attack accuracy~\cite{watson2021importance}, we consider the difficulty score as input for calibration and proposed a new MIA method, named \attack{} (difficulty calibrated MIA) . Our attack cannot only bring the difficult samples to a more vulnerable stage but also achieves a higher true-positive rate at low false-positive rate regions.



\mypara{Contributions} 
The contributions of this work are summarized below.
\begin{itemize}
    \item We take the first step to understanding the privacy risks introduced by CL.
    \item We conduct a comprehensive analysis to quantify the privacy risks  and our results show CL introduces disparate impacts to samples separated by difficulty levels.
    \item We propose a new MIA that exploits the difficulty scores for better attack performance.
\end{itemize}

\section{Preliminary}\label{sec: prelim}

\subsection{Curriculum Learning}\label{subsec: prelim_cl}

Curriculum learning (CL)~\cite{bengio2009curriculum} is designed to emulate the concept of the human learning process. The general idea is to have a \textit{curriculum} that imposes a structure on the training data so the ``student'' ML models can learn from the easier samples before the harder ones. As a result, training ML models under CL observes a shorter duration of convergence and higher testing accuracy\cite{bengio2009curriculum,weinshall2018curriculum,graves2017automated,hacohen2019power}. 
For example, Weinshall et al. proposed to use transfer learning to build the curriculum and achieved 0.5\% to 3.5\% higher accuracy than a model trained in the normal setting~\cite{weinshall2018curriculum}. 
CL has gained significant interest from the ML community, powering real-world applications in many domains. ~\autoref{sec: related_work} provides a more detailed survey.

Below, we formalize CL following the definition of Hacohen et al.~\cite{hacohen2019power}.  Let $\mathcal{X}=\{X_i\}^N_{i=1}=\{(x_i, y_i)\}^N_{i=1}$ be the training dataset, where $N$ is the number of samples, $x_i$ is a data point, and $y_i$ is the label of $x_i$. $T$ is the ML model to be trained. Assuming Stochastic Gradient Descent (SGD) is used for optimization, and each training iteration takes a mini-batch of $\mathcal{X}$, and a sequence of $M$ mini-batches $\mathcal{B}_1, \dots,  \mathcal{B}_M$ will be used for each epoch. 
The standard training procedure will sample $\mathcal{X}$ uniformly to generate the mini-batches. Instead, CL uses a \textit{difficulty measurer} $f(\mathcal{X}, C)$ to generate difficult scores for $\mathcal{X}$, and a \textit{training scheduler} sorts $\mathcal{X}$ by the difficult scores in an ascending order ahead of training. 
$C$ is the curriculum, and we will elaborate on its common options in \autoref{subsec:cl_design}. A sequence of subsets $\mathcal{X}'_1, \dots, \mathcal{X}'_M \subseteq \mathcal{X}$ are extracted from $\mathcal{X}$ after sorting, and the size of  $\mathcal{X'}_i$ is determined by a \textit{pacing function} $g(i)$. A mini-batch $\mathcal{B}_i$ is sampled uniformly from $\mathcal{X}'_i$. 
\autoref{alg:cl} summarizes the process. Noticeably, slight changes can be applied (e.g., skip the step of mini-batch sampling), but they should not affect the conclusions drawn from this study.

\begin{algorithm}[h]
\LinesNumbered
\caption{Curriculum learning framework.}\label{alg:cl}
\KwIn{Training dataset $\mathbf{\mathcal{X}=\{X_i\}^N_{i=1}}$, difficulty measurer $\mathbf{f(\mathcal{X}, C)}$, pacing function $\mathbf{g(i)}$, number of iterations $\mathbf{M}$, number of epochs $\mathbf{E}$, target model $\mathbf{T}$}
$\mathcal{X} \gets f\left(\mathcal{X}, C\right)$;\\
\For{$e \in 1, \dots, E$}{
    \For{$i \in 1, \dots, M$}{
    $\mathcal{X}'_i \gets \mathcal{X}[1, \dots, g(i)]$;\\
    $\mathcal{B}_i \gets sample(\mathcal{X}'_i)$;\\
    $T \gets train\left(T, \mathcal{B}_i\right)$
    }
}

\end{algorithm}

\subsection{Privacy Risks in Machine Learning}\label{subsec: prelim_mia}

Prior works have shown that the ML models could memorize sensitive information from the training data, which can be inferred by an adversary who keeps querying the model. Two major types of attacks are MIA~\cite{SSSS17,NSH18,NSH19,SZHBFB19} and AIA~\cite{MSCS19,SS20}, which have been extensively studied. 
The detailed literature survey of privacy attacks and other attacks is left to~\autoref{sec: related_work}. 
\mypara{Membership Inference Attack (MIA)}
Given a target model $T$ and any adversary's external knowledge $K$, the goal of MIA is to determine whether a data sample $x$ was used to train the model.
Formally, we have: 
\begin{equation}
    \mathcal{A}_{MI}: x, T, K \mapsto {1} \mbox{ or } {0}
\end{equation}
where $T$ is the target model and $K$ is the adversary's external knowledge, e.g., the distribution of the training data for $T$. $1$ ($0$) denotes the sample is a member (non-member). 

MIA can lead to serious privacy threats.
For example, given a model trained on clinical records of cancer patients to determine the medicine dosage~\cite{JSBZG19}, the attacker can learn whether a person has cancer by applying MIA to the model.
We follow previous work~\cite{SSSS17,SZHBFB19,SM21,LZ21,CTCP21}
and assume that the adversary only has black-box access to $T$, which means that the adversary can only query $T$ with the data sample and obtain its corresponding output. Then, $\mathcal{A}_{MI}$ predicts membership with the output of $T$. \autoref{subsec:basic_mia} elaborates the details.

\mypara{Attribute Inference Attack (AIA)}
Different from MIA, the goal of AIA is to infer attributes of a data sample that are not related to the target model's original classification task. A specific attack scenario is when AIA is used to infer some hidden sensitive attributes.
For instance, a target model is trained to conduct gender classification, while AIA aims to infer the political view of a data sample.
Such attribute is often hidden when training the target model. However, due to the intrinsic \textit{overlearning} property of ML~\cite{SS20}, a target model may try to capture attributes not directly relevant to its task. Note that AIA is different from property inference attack (PIA)~\cite{GWYGB18} which infers a property about the entire dataset rather than a sample: e.g., PIA can tell whether a training dataset is gender-balanced.

Instead of having direct access to the sample, we follow previous work~\cite{MSCS19,SS20} and consider the adversary only has its \textit{representation} (e.g., embedding) generated by a target model $T$.
Formally, AIA can be defined as:
\begin{equation}
    \mathcal{A}_{AI}: h \mapsto s
\end{equation}
where $h$ is a sample's representation provided by $T$ and $s$ is the sample's sensitive attribute predicted by $\mathcal{A}_{AI}$. ~\autoref{subsec:basic_aia} elaborates the details.
\section{Datasets and Target Models}
\label{sec: evaluation_setup}

In this work, we aim to quantify the privacy risks introduced by CL through the lens of MIA and AIA. To this end, we select popular datasets and models that are used for ML classification tasks. In our study, a total of 9 unique datasets are used, with 8 datasets used for MIAs and 3 datasets used for AIA. Among these datasets, 6 of them are image datasets, while the remaining 3 datasets consisted of non-image data.

\mypara{MIA Datasets} We use the following 8 datasets, which are also adopted by previous work~\cite{HZ21,LJQG21,mireshghallah2020privacy,SSSS17} to study MIA. They are CIFAR100~\cite{krizhevsky2009learning}, Tiny ImageNet~\cite{le2015tiny}, Place100, Place 60\cite{zhou2017places}, SVHN~\cite{netzer2011reading}, Purchase\cite{SSSS17}, Texas hospital stays\cite{SSSS17} and Locations~\cite{yang2016participatory}. We focus on image datasets mainly (the first 5 datasets), but tabular datasets are also evaluated. 
\begin{itemize}
    \item \mypara{CIFAR100~\cite{krizhevsky2009learning}} This dataset consists of $60,000$ colored images in $100$ classes, with $600$ images per class. The size of each image is $32\times 32$.
    \item \mypara{Tiny ImageNet~\cite{le2015tiny}} This is a subset of the ImageNet dataset\cite{deng2009imagenet}. It contains $100,000$ colored images of $200$ classes ($500$ for each class). The size of each image is $64\times 64$. 
    \item \mypara{Place100} This dataset is a subset of Places365\cite{zhou2017places} dataset, which is composed of more than 1.8 million images with 365 scene categories. Place100 is generated by randomly selecting $100$ scene categories with $600$ random images per category. 
    \item \mypara{Place60} This dataset is similar to Place100, except that it has $60$ classes containing $1,000$ images each.
    \item \mypara{SVHN~\cite{netzer2011reading}} The Street View House Numbers (SVHN) dataset is a real-world image dataset containing over $600,000$ digit images. This dataset includes images of house numbers taken from Google Street View images. The size of each image is $32\times 32$.
    \item \mypara{Purchase}
    This is a tabular dataset about purchase styles. Following Shokri et al.~\cite{SSSS17}, we leverage the Purchase-100 dataset (abbreviated as Purchase) and use $10,000$ records for training. The dataset itself contains $197,324$ records from 100 classes, where each record has 600 binary features.
    \item \mypara{Texas hospital stays}
    This dataset contains information about inpatient stays in several health facilities. Following Shokri et al.~\cite{SSSS17}, our task is to predict a patient’s main procedure. After pre-processing, the resulting dataset has $67,330$ records and $6,170$ binary features.  
    \item \mypara{Locations~\cite{yang2016participatory} }
    The original dataset was released by Foursquare about its mobile users’ location ``check-ins'', which has 11,592 users and 1,136,481 check-in records. Our task is to predict the user’s geo-social type (128 in total). Here we use the version pre-processed by Shokri et al.~\cite{SSSS17}, which has 446 binary features.
\end{itemize}

\mypara{AIA Datasets} 
Datasets with multiple attributes are required for AIA. To this end, we adapt Place100 and Place60 used as MIA datasets to AIA setting as they both contain multiple attribute labels. More specifically, the model for Place100 outputs whether a sample is an indoor scene, while the sensitive attribute is the category of the scene, which contains $100$ labels. Place60 has the total number of categories as $60$. In addition to Place100 and Place60, we introduce UTKFace~\cite{zhang2017age} specifically for AIA study.

\begin{itemize}
    \item \mypara{UTKFace~\cite{zhang2017age}} This is a large-scale facial dataset, which consists of over $20,000$ face images with annotations of age, gender, and ethnicity. In our evaluation, we set gender classification as the the task for target model, and the sensitive attribute to be inferred is ethnicity, which includes $5$ classes. 
\end{itemize}

\mypara{Target Models} We adopt three popular neural network architectures as the target model's architecture for the image datasets. They are ResNet-18~\cite{HZRS16}, ResNet-34~\cite{HZRS16} and MobileNet~\cite{SHZZC18}. 
We choose these models because variants of ResNet are still achieving SOTA (State of The Art) or near SOTA performance in image classification, and MobileNet is widely used on mobile devices.
We adopt cross entropy as the loss function and SGD as the optimizer. We train all models for $200$ epochs with a batch size of $128$. The learning rate is set to $0.1$\footnote{This learning rate is empirically chosen and has a very limited effect on attack accuracy. For example, when using a learning rate of  $0.001$, the MIA accuracy is affected by less than $0.2\%$ when attacking a ResNet-18 model trained on CIFAR100.}.
For the non-image dataset Purchase and Location, we choose a 3-layer MLP with the same number of epochs and batch size. The number of neurons in the hidden layer is 256. For the Texas dataset, we use a 5-layer MLP with 512 neurons in the hidden layer because this dataset contains more features. To avoid fortuitous outcomes, all experiments are repeated 5 times with the standard deviation presented.
\section{Methodology}\label{sec:methodology}

In this section, we describe the curriculum designs experimented with by our study, the implementation of the basic MIA and AIA, our proposed MIA, and the defense techniques to be tested.

\subsection{Curriculum Designs}
\label{subsec:cl_design}

We choose two popular curriculum learning (CL) methods, which also have open-sourced implementations~\cite{CL1, CL2}, to train the target model. We expect our major observations (described in~\autoref{sec: evaluation}) are also applicable to other CL methods, like self-paced curriculum~\cite{kumar2010self,jiang2015self}, and automated curriculum~\cite{graves2017automated}, because they share similar high-level ideas (e.g., self-paced curriculum differs from bootstrapping only in that self-paced curriculum does not let the curriculum completely guide its learning process). Below we explain the two CL methods.

\begin{itemize}
    \item \mypara{Bootstrapping~\cite{hacohen2019power}} The target model $T$ is first trained without CL, then it serves as a difficulty measurer ($f$ in~\autoref{alg:cl}) to order the training samples by their loss. 
    \item \mypara{Transfer learning~\cite{weinshall2018curriculum}} Different from bootstrapping, a pre-trained model is used for the difficulty measurer. We use inception-v3~\cite{szegedy2016rethinking}\footnote{It is a widely-used image recognition model that achieves over 78.1\% accuracy on the ImageNet dataset~\cite{deng2009imagenet}.} as the pre-trained model to evaluate the image datasets.   
    The evaluation on tabular datasets with transfer learning is skipped, as we did not find a widely used pre-trained model in such a setting.
\end{itemize}  

To better assess the improvement brought by the above two CL methods and their vulnerabilities under attacks, we establish two other methods for comparison. 

\begin{itemize}
    \item \mypara{Baseline curriculum} It uses a random curriculum that is irrelevant to the data samples' difficulty. This curriculum is then used across all training epochs. The normal training process is different in that a random order is drawn for every training epoch.
    \item \mypara{Anti-curriculum} It shares the same difficulty measurer with bootstrapping but arranges the samples from difficult to easy, reversing the outcome of bootstrapping. 
\end{itemize}

For the pacing function $g$, we choose varied exponential pacing~\cite{hacohen2019power}, exponentially increasing the fraction of data by steps (a step denotes the iterations with the same output of $g$). According to ~\cite{hacohen2019power}, different pacing functions perform similarly.

In summary, the four CL methods differ in the difficulty measurer and each CL method feeds training data using the same curriculum (or ordering) across all epochs. The baseline and anti-curriculum methods help us understand the contribution of data ranking and order fixing separately (e.g., anti-curriculum can be considered as using a wrong curriculum but still repeating the order across epochs as advised by CL).

As described in~\autoref{subsec: prelim_cl}, CL can accelerate the training process to reach higher accuracy. 
We first validate this claim by evaluating the training performance and the testing accuracy and comparing them to the normal training method, which does not use any curriculum as guidance.

\begin{table*}[]
\centering
\small
\begin{tabular}{l|ccccc}
\toprule
{\backslashbox{Dataset}{Method}} & Normal & Bootstrapping & Anti-curriculum & Baseline & Transfer Learning     \\
\midrule
Tiny ImageNet &   0.3842  $\pm$ 0.0027  &  \textbf{0.4002}  $\pm$ 0.0043  &   0.3776   $\pm$ 0.0036 &  0.3798 $\pm$ 0.0035  &   0.3803 $\pm$ 0.0043           \\
CIFAR100      &    0.6081 $\pm$ 0.0053&	\textbf{0.6232} $\pm$ 0.0078&	0.5991 $\pm$ 0.0098&	0.6099 $\pm$ 0.0045&	0.6127 $\pm$ 0.0221 \\
Place100      &  0.2992 $\pm$ 0.0054&	\textbf{0.3159} $\pm$ 0.0059&	0.2967 $\pm$ 0.0037&	0.3088 $\pm$ 0.0060&	0.3007 $\pm$  0.0053                     \\
Place60       &    0.4756 $\pm$ 0.0041&	\textbf{0.4903} $\pm$ 0.0040&	0.4815 $\pm$ 0.0025&	0.4847 $\pm$ 0.0071&	0.4707 $\pm$ 0.0154                      \\
SVHN          &  0.9592 $\pm$ 0.0004&	0.9598 $\pm$ 0.0006&	0.9566 $\pm$ 0.0005&	0.9593 $\pm$ 0.0006&	\textbf{0.9599} $\pm$ 0.0006                   \\
Purchase      &   0.4931 $\pm$ 0.0055&	\textbf{0.5324} $\pm$ 0.0037&	0.4760 $\pm$ 0.0055&	0.5289 $\pm$ 0.0043       & \multicolumn{1}{c}{-} \\
Texas         &       0.4809	$\pm$ 0.0072&\textbf{0.4975} $\pm$ 0.0066&	0.4606 $\pm$ 0.0101&	0.4877    $\pm$ 0.0095   & \multicolumn{1}{c}{-} \\
Location      &     0.5861 $\pm$ 0.0107&	\textbf{0.5914} $\pm$ 0.0027&	0.5563 $\pm$ 0.0156&	0.5838 $\pm$ 0.0077     & \multicolumn{1}{c}{-}\\
\bottomrule
\end{tabular}
\caption{Target model's average test accuracy on different datasets. ResNet-18 is used for all image datasets, and MLP for non-image datasets Purchase, Texas, and Location. Transfer learning CL does not apply to non-image datasets. The target model accuracy is relatively low except for SVHN because we use a subset of the original training data.}
\label{fig: target_acc}
\end{table*}

\begin{figure*}[!h]
    \centering
    \includegraphics[width=0.98\textwidth]{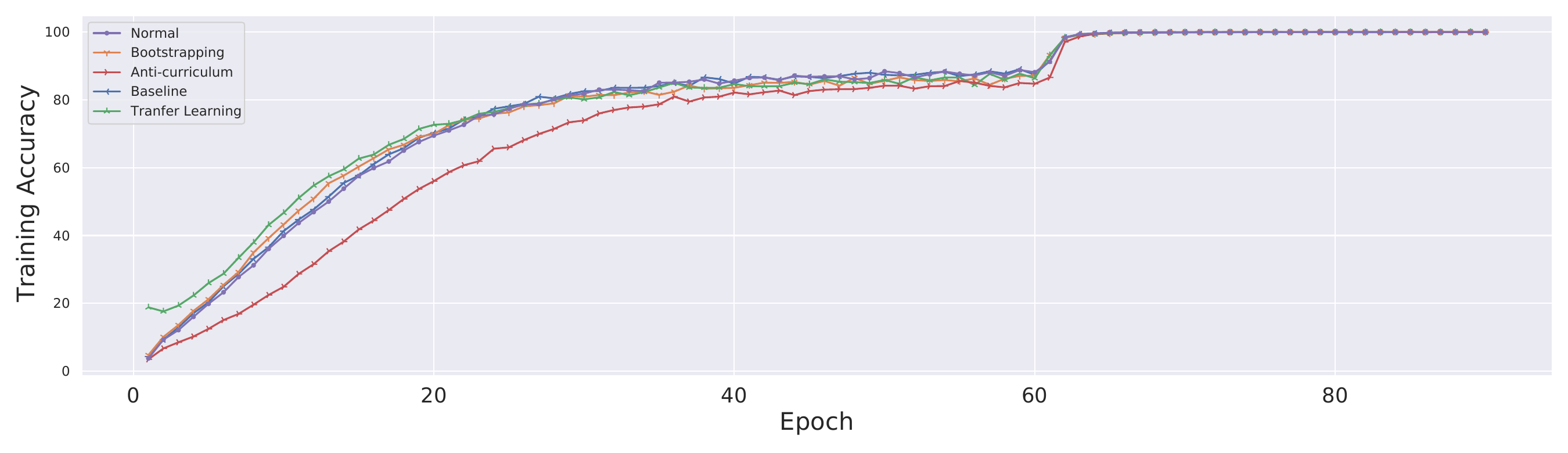}
    \caption{The training accuracy of different training methods with ResNet-18 on CIFAR100 along the increase of epochs (total of 90 epochs). Bootstrapping, transfer learning, and baseline reach higher accuracy faster and converge to a better result.
}
    \label{fig: speed_convergence}
\end{figure*}
\autoref{fig: target_acc} validates the effectiveness of CL.  At least one of the four CL methods can outperform the normal training by 0.06\% to 4.42\%, and the corresponding average training accuracy is given in ~\autoref{app:training_acc} (\autoref{tb: trainning_acc}). The maximum standard deviation in \autoref{fig: target_acc} is 0.0221 while 32 out of 37 results have a standard deviation less than 0.01. This indicates the difference among various CL methods is statistically significant.
It is worth noticing that bootstrapping and transfer learning always outperform normal training, and anti-curriculum performs the worst consistently. 
Interestingly, we observe that the baseline performs as well as the transfer learning curriculum for  Place100 and Place60, which means the transfer learning curriculum does not suit these two datasets well.
\autoref{fig: speed_convergence} validates the major motivation of adopting CL, i.e., reaching higher accuracy while converging faster. 
Throughout most of the training, bootstrapping and transfer learning reach higher accuracy faster than all the other methods. At the same time, it takes the longest for the anti-curriculum to reach the same training accuracy compared to all other methods. This indicates that repeating a meaningful data order improves training. 
This observation aligns with the discovery from previous work~\cite{wu2021when, hacohen2019power}.
Finally, CL is expected to have a disparate impact on classification accuracy across samples. Besides the analysis in ~\autoref{sec: evaluation}, we also use t-distributed stochastic neighbor embedding (t-SNE) to visualize this impact. More details including the visualization are in ~\autoref{app:t-SNE}.

\subsection{Basic MIA}
\label{subsec:basic_mia}

After providing a high-level overview of MIA in~\autoref{subsec: prelim_mia}, we now delve into the details, focusing on the three well-known attacks: NN-based (Neural Network-based)~\cite{SSSS17,SBBFZ20}, metric-based~\cite{SM21}, and label-only attacks~\cite{LZ21,CTCP21}. 

NN-based attack assumes a vector of \textit{prediction posteriors} (e.g., confidence scores or loss) of all class labels can be returned by the target model $T$ when querying $T$ with a data sample $x$. It is also assumed that the adversary has a \textit{shadow dataset} ($\mathcal{D}$) that has the same distribution and format as $T$'s private training dataset. $\mathcal{D}$ is used to train a set of \textit{shadow models} $\mathcal{S}$ \ignore{$=S_1, \dots, S_k$ ($k$ also equals to the number of disjoint subsets $\mathcal{D}$)} that behave similarly as $T$ (e.g., having the same architecture as $T$ like previous work~\cite{SSSS17,SZHBFB19,SM21}). 
The attacker trains an \textit{attack model} $\mathcal{A}_{MI}$ using $\mathcal{S}$. In particular, the attacker queries every shadow model $\mathcal{S}$ with the samples from its own training dataset and a disjoint testing dataset. The prediction posteriors of all samples and whether they are in training (denoted member) or testing (denoted non-member) are used as input to train $\mathcal{A}_{MI}$. 
Finally, the attacker queries $T$ with a sample of interest $x$ and uses the prediction posteriors as the input to $\mathcal{A}_{MI}$ to predict the membership status.

Compared to the NN-based attack, the model $\mathcal{A}_{MI}$ of metric-based attacks does not need to be trained. Instead, $\mathcal{A}_{MI}$ generates a privacy risk score from the output of $T$ and compares it to class-specific thresholds.

For the label-only attack, it assumes only the prediction label instead of the prediction posteriors are returned from $T$. Still, the adversary can continuously add adversarial perturbations to the input sample $x$ until its prediction label has been changed.
The key insight is that the magnitude of the adversarial perturbation is larger for the member sample as $T$ gives a more confident prediction. $\mathcal{D}$ and $\mathcal{S}$ can be used to select a threshold to separate the perturbation magnitudes of members and non-members.

\mypara{MIA Models}
Following the original setting of the NN-based attacks~\cite{SSSS17}, we adopt a 3-layer MLP with 64 and 32 hidden neurons, and 2 output neurons, as our attack model $\mathcal{A}_{MI}$. We use cross-entropy as the loss function and Adam as the optimizer with a learning rate of $0.01$. $\mathcal{A}_{MI}$ is trained for 100 epochs. For metric-based attacks, we follow the implementation of Song et al.~\cite{SM21} and consider 4 metrics, including correctness, confidence, entropy, and modified entropy. 
The associated attack methods are named metric-corr, metric-conf, metric-ent, and metric-ment. For label-only attacks, we leverage the implementation from ART~\cite{ART}.

Related research has shown that NN-based attacks often, though not universally, achieve better attack accuracy compared to metric-based and label-only attacks~\cite{SSSS17,SZHBFB19,HZ21}. Thus we use NN-based attack (specifically black-box-top3) for most of our evaluation in~\autoref{sec: evaluation}.

\subsection{Our Proposed MIA}
\label{subsec: our_attack}
Given that CL orders training samples by difficulty, impacting the model, we investigate the potential enhancement of MIA when the target model is trained under CL. For this purpose, we propose a novel MIA method called \textit{\attack} specifically tailored for CL. We first introduce calibrated MIA, which serves as inspiration for designing \attack, followed by the details of \attack.

\mypara{Calibrated MIA} 
Recently, Watson et al.~\cite{watson2021importance} proposed to use a calibrated membership score instead of the standard membership score (e.g., loss) to determine whether a sample is a member. 
Assume $s(T, x)$ is the original membership score, where $T$ is the target model, and $x$ is a sample. 
The calibrated membership score $s_{cal}(T ,x)$ is defined as follows:
\begin{equation}\label{eq:calibrated}
    s_{cal}(T, x) =  s(T,x) - \mathbb{E}_{\mathcal{S}\leftarrow \mathcal{A}(\mathcal{D})}{[s(\mathcal{S},x)]}
\end{equation}
where $\mathcal{S}$ are shadow models\footnote{$\mathcal{S}$ are named as reference models in ~\cite{watson2021importance}, which resemble shadow models~\cite{SSSS17} as they are also trained on the same data distribution of $T$.} that behave similarly as $T$, $\mathcal{D}$ is the shadow dataset, function $s(T,x)$ and $s(\mathcal{S},x)$ output the membership scores from target and shadow models, $\mathcal{A}$ randomly samples subsets of $\mathcal{D}$ to train $\mathcal{S}$, and $\mathbb{E}$ computes the expectation of $s(\mathcal{S},x)$. 
Finally, $s_{cal}(T, x)$ is compared to a fixed threshold $\theta$, and a sample is considered a member if $s_{cal}(T, x) \geq \theta$.

Previous MIA methods could have a high false positive rate (FPR) on non-members, often over-represented in the samples to be tested by the attacker. Equation~\ref{eq:calibrated} addresses this issue by using the \textit{difference} between the target model and shadow models to derive the membership signal: if $x$ is non-member to $\mathcal{S}$, it is also more likely non-member to $T$, therefore $s_{cal}(T, x)$ should be small. The evaluation results in ~\cite{watson2021importance} show the area under ROC curve (AUC) can be improved ``by up to 0.10'' (e.g., after calibrating the loss-based membership score with Equation~\ref{eq:calibrated}).

\mypara{Difficulty Calibrated MIA (\attack)} 
Calibrated MIA compares $s_{cal}(T, x)$ of all samples to a fixed threshold  $\theta$, and we argue that $\theta$ can be \textit{calibrated as well}.
We observe that a CL curriculum re-orders the samples by their difficulty before the target model is trained, and such strategy changes how a sample is memorized and vulnerable under MIA (see Section~\ref{subsec: evaluation_mia} and Section~\ref{subsec:mem}).  More specifically, we observe that CL makes the target model more vulnerable to MIA, especially for difficult samples (Finding 1 in Section~\ref{subsec: evaluation_mia}). Therefore, we can update $\theta$ according to the curriculum and make the attack model more accurate.
We assume the attacker can generate a curriculum similar as the one used by the target model. For example, the attacker can use the publicly released pre-trained model to generate the curriculum. Alternatively, the attacker can train shadow models similar to the target model and then build a curriculum according to loss from them. 

We implement this idea for NN-based MIA. When the attack model $\mathcal{A}_{MI}$ outputs the prediction posteriors for an input $x$, the posterior of the label ``member'' is compared against $\theta$, and $x$ is predicted as a member when the posterior is larger. 
When training $\mathcal{A}_{MI}$, we adjust $\theta$ based on samples' difficulty level to improve the training accuracy, and the pseudo-code is shown in \autoref{alg: our_attack}. Specifically, in each epoch, the calibrated membership scores $s_{cal}(T,\mathcal{D})$ are generated for $\forall x \in \mathcal{D}$, and we use the loss to compute $s$ (Line 2). Next, we try to find the threshold $\theta_0$ (ranging from 0 to 0.1 based on our empirical study) that achieves the best accuracy in separating members and non-members from $\mathcal{D}$ (Line 3).
After that, $\mathcal{A}_{MI}$ is updated by minimizing the training loss on $\mathcal{D}$ (Line 4) through adjusting the threshold with the following function:
\begin{equation}
    g(x, C, \theta_0) = \frac{\left(|\mathcal{D}|-C(x)\right)\left({\theta_0 - 0.0001}\right)}{|\mathcal{D}|-1}+0.0001
    \label{eq:func_g}
\end{equation}
where $C(x)$ indicates the rank of sample $x$ given by curriculum $C$. The rank for the easiest sample is 1, while the most difficult is $|\mathcal{D}|$. $g(x, C, \theta_0)$ is to assign a threshold $\theta$ from $[0.0001, \theta_0]$ ($0.0001$ is the initial threshold suggested by \cite{watson2021importance}) to each $x$ based on its difficulty level (determined by a curriculum $C$), that is, calibrating threshold of each $x$ based their difficulty level. 
The most difficult sample compares to $0.0001$,  the easiest one compares to $\theta_0$, and others compare to $\theta$ that is ranged in $[0.0001, \theta_0]$. The more difficult $x$ has a smaller threshold, meaning that we are lowering the bar for them to be predicted as members compared to the easy samples.
During the testing phase, the threshold for a sample $x$ is also adjusted with $g(x, C, \theta_0)$.

\begin{algorithm}[h]
\LinesNumbered
\caption{Training the attack model and adjusting threshold under \attack. ``pred'' is ``prediction''. 
}
\label{alg: our_attack}
\KwIn{Target model $\pmb{T}$ , reference model $\pmb{\mathcal{S}}$, shadow dataset $\pmb{\mathcal{D}}$, labels of shadow dataset $\pmb{L}$, attack model $\pmb{\mathcal{A}_{MI}}$, curriculum $\pmb{C}$, number of epochs $\pmb{E}$}

\For{$e \in 1, \dots, E$}{
$s_{cal}\left(T,\mathcal{D}\right) = s\left(T,\mathcal{D}\right)-s\left(S,\mathcal{D}\right)$;\\
$\theta_0= \underset{\theta}{\argmax}$  pred$\left(\mathcal{A}_{MI},L,s_{cal}\left(T,\mathcal{D}\right)\right)$;\\
$\mathcal{A}_{MI} \gets train\left(\mathcal{A}_{MI}, s_{cal}\left(T,\mathcal{D}\right), g\left(x, C, \theta_0\right)\right)$;\\
}
\end{algorithm}

\attack\ follows the direction of addressing the issue caused by over-represented non-members~\cite{watson2021importance, carlini2021membership}. On top of those works, \attack\ is customized under CL to amplify the effects of MIA. 
To demonstrate the benefit of \attack{}, we compare it with the score-based membership attack after difficulty calibration with default threshold in ~\cite{watson2021importance} (Cal). 

Overall, \attack{} outperforms Cal by 4.0\% to 9.9\% of attack accuracy while maintaining the same AUC. Besides, \attack{} improves MIA's TPR at extremely low FPR, making the difficult sample more vulnerable. The detailed evaluation of \attack{} across all metrics such as  attack accuracy, confidence score, and TPR at low FPR are presented in Section ~ref{subsec: diff_cali}.

Some recent works suggest to use class-specific thresholds~\cite{SM21}. We did not adjust the threshold by classes because our threshold has been fine-tuned with difficulty levels. 

\subsection{Basic AIA}
\label{subsec:basic_aia}

Song et al. proposed an inference-time attack and model-repurposing attack~\cite{SS20} for AIA, and here we focus on the first attack and follow the same setting as this work.
We consider the model evaluation to be partitioned~\cite{SS20} or the model is trained under federated learning~\cite{MSCS19}. The target model $T$ is split into two parts, i.e., an encoder and a classifier, and the adversary has black-box access to the encoder $E$.
The attacker has an auxiliary dataset $D$ containing pairs of $(x,s)$ where $s$ is the sensitive attribute. 
The embeddings $h$ can be generated by querying $E$, i.e., $h = E(x), \forall x \in D$. All pairs of $(h,s)$ will be used to train the attack model $\mathcal{A}_{AI}$ and later used to predict the values of $s$ in the target model $T$.

\mypara{AIA Model} Our $\mathcal{A}_{AI}$  is a 3-layer MLP with 128 hidden neurons in each hidden layer. We use cross-entropy as the loss function and SGD as the optimizer with a learning rate of $0.01$. The attack model is trained for $100$ epochs. The dimension of each sample’s embedding (i.e., the second to the last layer's output) is 512 for ResNet-18, 512 for ResNet-34, and 1024 for MobileNet. 
To train the target model $T$, we use the label for the original classification task (e.g., gender). To train $\mathcal{A}_{AI}$, we use the label from another field (e.g., race).

\subsection{Defense Methods}\label{subsec: defense_methodology}

Some defense methods have been proposed to reduce the success rate of privacy attacks, in particular, MIA. We are interested in how they perform under curriculum learning and our proposed attack. To this end, we select \dpsgd~\cite{abadi2016deep},  \memguard~\cite{JSBZG19}, {\mixupMMD~\cite{li2021membership} and \advreg~\cite{NSH18}}. \dpsgd{} and \memguard{} represent two directions in privacy protection, while \mixupMMD{} and \advreg{} are two more recent defense methods.
Below we explain the four defense methods.

\section{Evaluation Results}\label{sec: evaluation}

\begin{table*}[]
\centering
\small
\begin{tabular}{l|ccccc}
\toprule
{\backslashbox{Dataset}{Method}} & Normal & Bootstrapping & Anti-curriculum & Baseline & Transfer Learning     \\
\midrule
Tiny ImageNet &   0.9193  $\pm$ 0.0000  &  0.9385  $\pm$ 0.0000  &   0.9116   $\pm$ 0.0001 &  0.9207 $\pm$ 0.0000  &   \textbf{0.9439} $\pm$ 0.0000           \\
CIFAR100      &    0.8577 $\pm$ 0.0011&	\textbf{0.8751} $\pm$ 0.0001&	0.8376 $\pm$ 0.0001&	0.8582 $\pm$ 0.0001&	0.8718 $\pm$ 0.0001 \\
Place100      &  0.9425 $\pm$ 0.0000&	0.9549 $\pm$ 0.0001&	0.9335 $\pm$ 0.0001&	0.9416 $\pm$ 0.0001&	\textbf{0.9617} $\pm$  0.0001                     \\
Place60       &    0.8773 $\pm$ 0.0022&	\textbf{0.8987} $\pm$ 0.0001&	0.8625 $\pm$ 0.0001&	0.8827 $\pm$ 0.0001&	0.8902 $\pm$ 0.0001                      \\
SVHN          &  0.5570 $\pm$ 0.0000&	\textbf{0.5605} $\pm$ 0.0002&	0.5514 $\pm$ 0.0001&	0.5599 $\pm$ 0.0003&	0.5580 $\pm$ 0.0003                  \\
Purchase      &  \textbf{0.9524} $\pm$ 0.0016&	0.9453 $\pm$ 0.0024&	0.9118 $\pm$ 0.0122 &	0.9458 $\pm$ 0.0015       & \multicolumn{1}{c}{-} \\
Texas         &       0.6749	$\pm$ 0.0092&\textbf{0.7068} $\pm$ 0.0139&	0.5950 $\pm$ 0.0161&	0.7039    $\pm$ 0.0122   & \multicolumn{1}{c}{-} \\
Location      &     0.9153 $\pm$ 0.0066&	\textbf{0.9194} $\pm$ 0.0048&	0.8980 $\pm$ 0.0038&	0.9169 $\pm$ 0.0038     & \multicolumn{1}{c}{-}\\
\bottomrule
\end{tabular}
\caption{Accuracy of NN-based MIA on models trained on 8 datasets. Transfer learning CL does not apply to non-image dataset Purchase, Texas and Location.}
\label{fig: overall attack acc}
\end{table*}

In this section, we present the evaluation results of MIA and AIA when CL is applied to train the target model. We also attempt to explain the observations from the angle of data memorization and show the impact of CL on the existing defenses.  We highlight our insights with text boxes.

\mypara{Evaluation setup} 
To evaluate MIA and AIA, we split each dataset described in \autoref{sec: evaluation_setup} into three disjoint parts: one for training the target model, one for training a shadow model, and one for testing both the target and shadow model. 

To evaluate the defense methods, we split each dataset into five parts as some advanced methods need reference datasets for training. More details about the defenses can be found in \autoref{subsec: defense}. All experiments were \textit{repeated 5 times} to minimize the fortuitous outcomes, and the mean value and standard deviation were reported.

\mypara{Evaluation metrics} 
First, we compute the attack accuracy, measured by the correct predictions (member/non-member) versus all predictions, to assess the effectiveness of MIA/AIA, and the classification accuracy of the target model to assess the impact of curriculum learning and defenses. 
Second, to better understand the attack results, we retrieve the confidence scores of members and non-members, respectively. Note that the confidence score indicates the likelihood of a sample being classified as a member or non-member.
Third, we compute the true-positive rate (TPR) at the false-positive rate (FPR) of the attacks. As noted by Carlini et al.~\cite{carlini2021membership}, attacks should emphasize the member guesses over non-member guesses, so they should be evaluated by considering TPR at low FPR. This cannot be precisely modeled by the overall accuracy, precision, or recall. 

\subsection{Evaluation of Basic MIA}\label{subsec: evaluation_mia}

We start with the experiments on the 5 image datasets (CIFAR100, Tiny ImageNet, Place100, Place60, and SVHN), using ResNet-18 as the target model architecture and later ResNet-34 and MobileNet for comparison. 
The evaluation of the tabular datasets (Purchase, Texas hospital stays, and Locations) is presented at the end. The attack models are described in~\autoref{subsec:basic_mia}.

\mypara{MIA Accuracy} We found that models trained using meaningful CL methods (i.e., bootstrapping and transfer learning) are slightly \textbf{more vulnerable} to MIA. 
\autoref{fig: overall attack acc} shows the accuracy of NN-based black-box-top3 MIA~\cite{SSSS17} by datasets and CL methods. 
The biggest attack accuracy improvement observed for image datasets is 2.46\%  (Tiny ImageNet with transfer learning) while the biggest improvement for non-image datasets is 3.20\%  (Texas with bootstrapping).
Among different CL methods, bootstrapping and transfer learning are the most vulnerable, with an average of 1.29\% and 1.44\% improvement in the attack accuracy against the normal training, respectively.
For baseline CL, the attack accuracy decreases for Place100, whereas a slight increase is observed for the attack accuracy on other datasets.
For anti-curriculum CL, the attack accuracy decreases for all  datasets.
This result indicates both the data repeating (reflected by the results of baseline) and ordering (reflected by the results of bootstrapping and anti-curriculum) of CL (explained in~\autoref{subsec:cl_design}) contribute to the vulnerability under MIA. The consistent performance of bootstrapping and anti-curriculum indicates that \textbf{data ordering plays a bigger role}. 

Regarding the impact of datasets, we found more complex datasets (e.g., with more classes of labels) tend to have higher attack accuracy in general. For example, the average MIA accuracy is  94.39\% for Tiny ImageNet (200 classes),  87.18\% for CIFAR100 (100 classes),  96.17\% for Place100 (100 classes),  89.02\% for Place60 (60 classes), and 55.80\% for SVHN (10 classes), all under transfer learning. The same effects have also been observed in other works like~\cite{SSSS17}.

Regarding the metric-based and label-only attacks, the result is similar to the NN-based attack, as suggested by the evaluation on CIFAR100, shown in \autoref{fig: various attack acc}.  
The only exception is metric-corr, which performs worse than other attacks with bootstrapping. This result can be explained by the assumption of metric-corr that the target model is trained to predict correctly on its training data, which may not generalize well on the test data.
In the rest of the evaluation, we fix the attack model to black-box-top3, and the NN-based attack in the rest of the paper primarily refers to black-box-top3, unless indicated otherwise. 

\begin{figure}[h]
    \centering
    \begin{subfigure}[t]{.49\linewidth}
    \centering\includegraphics[width=\linewidth]{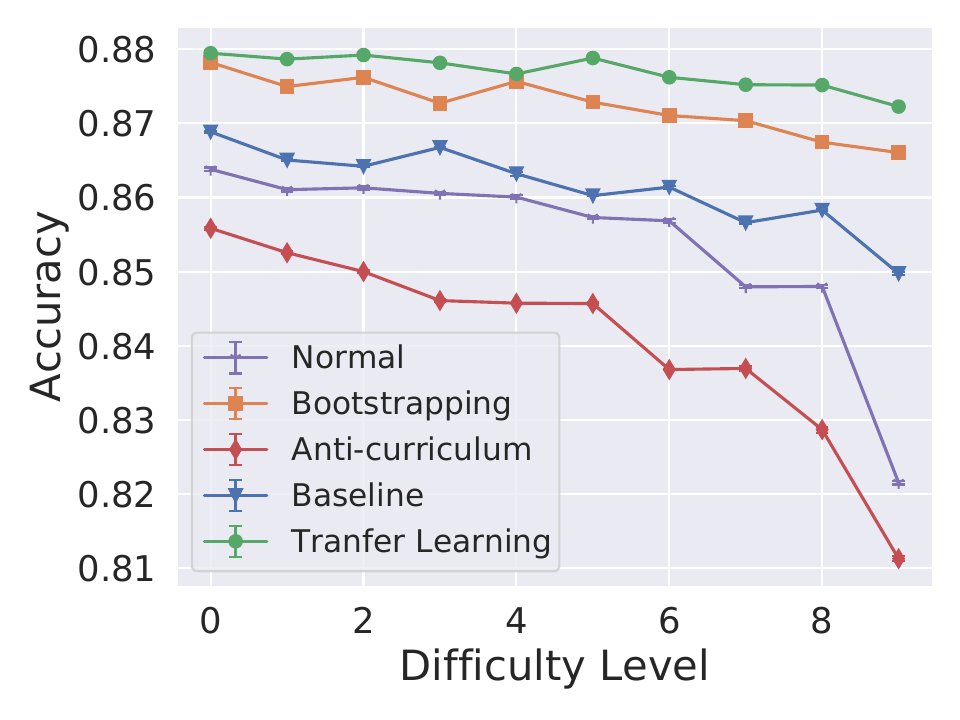}
    \caption{CIFAR100}\label{subfig: resnet18_cifar100_attack_acc}
  \end{subfigure}
  \begin{subfigure}[t]{.49\linewidth}
    \centering\includegraphics[width=\linewidth]{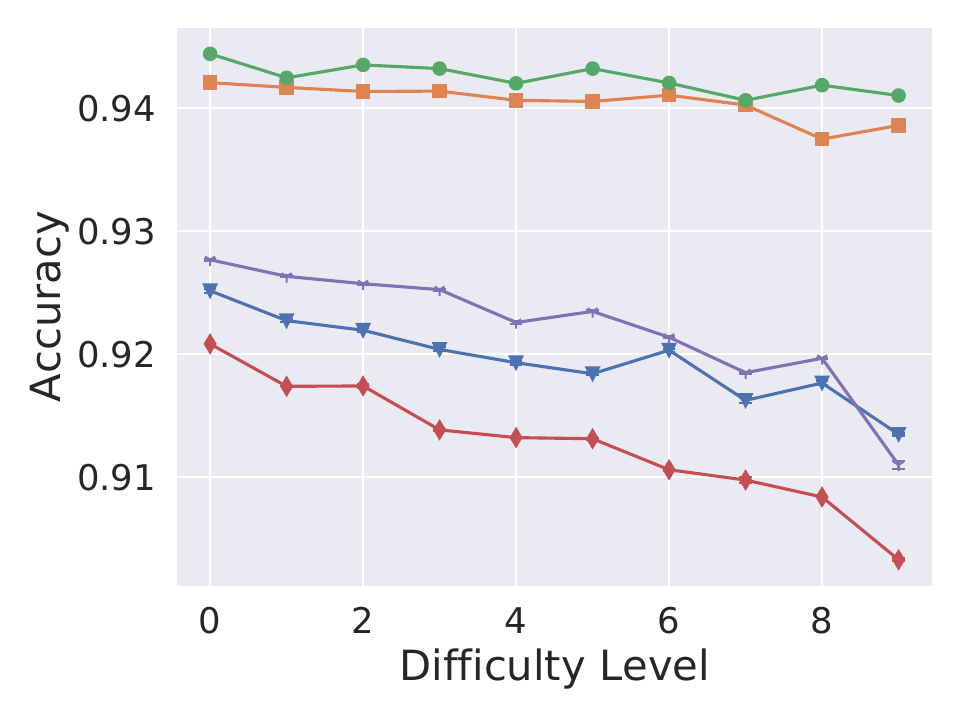}
    \caption{Tiny ImageNet}\label{subfig: resnet18_tiny_attack_acc}
  \end{subfigure}
    \caption{MIA accuracy on CIFAR-100, Tiny ImageNet. ResNet-18 is used for target model training.
    }
    \label{fig: bucket_attack_acc}
\end{figure}

\begin{table*}[h]
\centering
\small
\begin{tabular}{l|ccccc}
\toprule
{\backslashbox{Attack}{Method}} & Normal & Bootstrapping & Anti-curriculum & Baseline & Transfer Learning     \\
\midrule
NN-based~\cite{SSSS17} &   0.8572  $\pm$ 0.0011  &  \textbf{0.8751}  $\pm$ 0.0001  &   0.8376   $\pm$ 0.0002 &  0.8582 $\pm$ 0.0001  &   0.8718 $\pm$ 0.0001           \\
Metric-corr~\cite{SM21}     &    0.6920 $\pm$ 0.0000&	0.6820 $\pm$ 0.0000&	0.6905 $\pm$ 0.0000&	\textbf{0.6930} $\pm$ 0.0000&	0.6855 $\pm$ 0.0000 \\
Metric-conf~\cite{SM21}     &  0.8600 $\pm$ 0.0000&	\textbf{0.8810} $\pm$ 0.0000&	0.8458 $\pm$ 0.0000&	0.8553 $\pm$ 0.0000&	0.8740 $\pm$  0.0000                     \\
Metric-ent~\cite{SM21}      &    0.8490 $\pm$ 0.0000&	\textbf{0.8750} $\pm$ 0.0000&	0.8320 $\pm$ 0.0000&	0.8435 $\pm$ 0.0000&	0.8685 $\pm$ 0.0000                      \\
Metric-ment~\cite{SM21}         &  0.8620 $\pm$ 0.0000&	\textbf{0.8820} $\pm$ 0.0000&	0.8463 $\pm$ 0.0000&	0.8568 $\pm$ 0.0000&	0.8760 $\pm$ 0.0000                  \\
Label-only~\cite{ART}    &   0.8200 $\pm$ 0.0082&	\textbf{0.8263} $\pm$ 0.0082&	0.7963 $\pm$ 0.0117&	0.8050 $\pm$ 0.0045       & 0.8088 $\pm$ 0.0074\\
Cali~\cite{watson2021importance}        &       0.7889	$\pm$ 0.0012&\textbf{0.8272} $\pm$ 0.0009&	0.7532 $\pm$ 0.0004&	0.7781    $\pm$ 0.0025   & 0.8148    $\pm$ 0.0013  \\
Diff-Cali      &     0.8519 $\pm$ 0.0003&	\textbf{0.8670} $\pm$ 0.0006&	0.8382 $\pm$ 0.0006&	0.8438 $\pm$ 0.0008     & 0.8614    $\pm$ 0.0006 \\
\bottomrule
\end{tabular}
\caption{Average accuracy of NN-based, metric-based, label-only and our \attack{} attacks on models trained on CIFAR100 with ResNet-18. }
\label{fig: various attack acc}
\end{table*}

\autoref{fig: bucket_attack_acc} shows the attack accuracy of samples from different difficulty levels. More specifically, we construct the test dataset as half member samples and half non-member samples. Member samples are divided into different difficulty levels while non-member samples across each difficulty level are fixed.  \autoref{fig: bucket_attack_acc}  demonstrates that using a meaningful curriculum (i.e., bootstrapping and transfer learning) makes the model more vulnerable, especially for the difficult samples. 

\begin{figure*}[t]
    \centering
    \begin{subfigure}[t]{.245\linewidth}
      \centering\includegraphics[width=\linewidth]{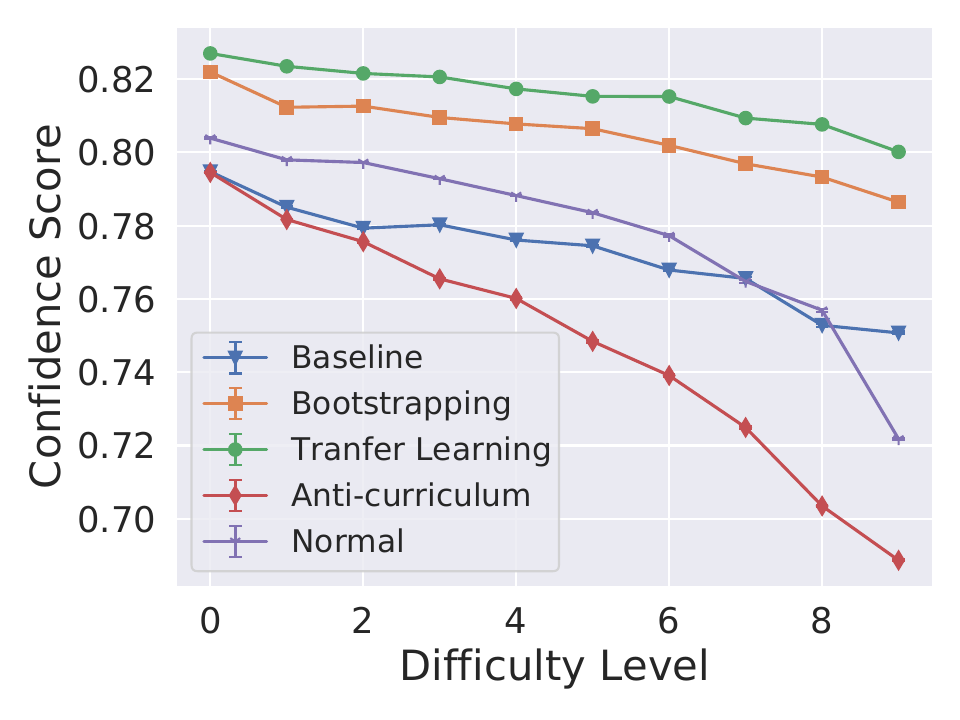}
    \caption{Members of CIFAR100}\label{subfig: cifar100_mem_confidence_score}
  \end{subfigure}
  \begin{subfigure}[t]{.245\linewidth}
    \centering\includegraphics[width=\linewidth]{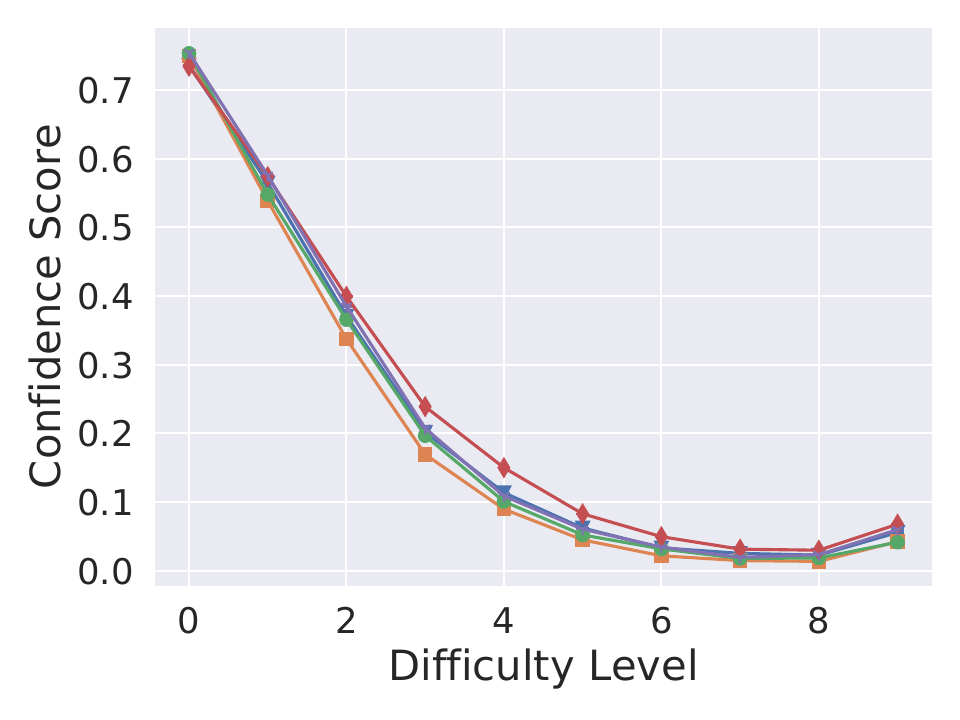}
    \caption{Non-members of CIFAR100}
  \label{subfig: cifar100_nonmem_confidence_score}
  \end{subfigure}
      \begin{subfigure}[t]{.245\linewidth}
    \centering\includegraphics[width=\linewidth]{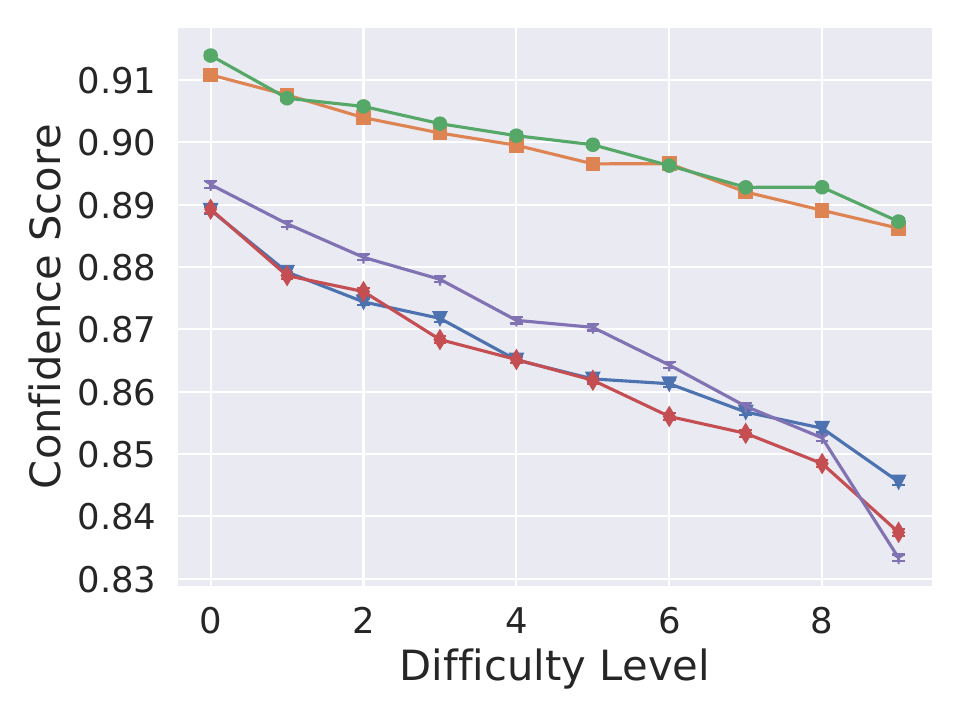}
    \caption{Members of Tiny ImageNet}\label{subfig: tiny_mem_confidence_score}
  \end{subfigure}
  \begin{subfigure}[t]{.245\linewidth}
    \centering\includegraphics[width=\linewidth]{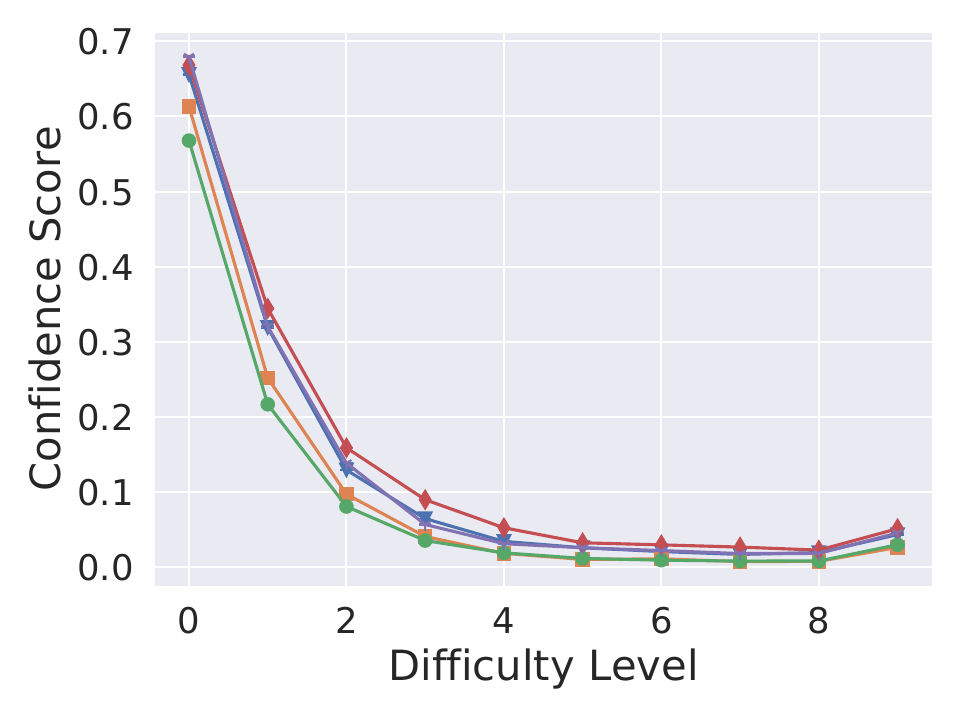}
    \caption{Non-members of Tiny ImageNet}
  \end{subfigure}
    \caption{Attack model's confidence score for both member and non-member samples on CIFAR-100 and Tiny ImageNet. ResNet-18 is used for target model training, and data samples are arranged according to their difficulty scores from bootstrapping.}
    \label{fig: confidence_score}
\end{figure*}

\mypara{Confidence Score} Since the key contribution of CL is to factor in the samples' difficulty levels during the training procedure, here we evaluate how difficulty levels impact the samples' vulnerability individually. Intuitively, the difficult samples should be harder to attack. 
However, since CL forces the model to learn the samples in a repetitive manner, we want to find out whether samples will be remembered by the model differently.
To assess and quantify the possible privacy risk discrepancy caused by CL, we first arrange samples according to their difficulty level. Then, we use the confidence score and attack accuracy to analyze individual samples. Note that we train a separate model and use the sample loss given by this model as a guide to determine how difficult a sample is. This model is used solely for getting the difficulty levels of all samples and is different from the target model in our following evaluation.  

\autoref{fig: confidence_score} depicts the attack model's confidence score by samples' difficulty levels. Though the difficult samples are not more vulnerable than the easy samples, \textbf{the gap in confidence scores is much narrower} (especially for the confidence score of members). 
Take the target model in CIFAR100 as an example, our attack model can recognize the most difficult member samples (scored as difficulty level $9$) from this model with over $7.83\%$ (absolute growth from 72.19\% to 80.02\%) more confidence, thanks to transfer learning (\autoref{subfig: cifar100_mem_confidence_score}). Interestingly, for the most difficult member samples, it is even possible for  anti-curriculum to have a higher confidence score compared to the normal training ((\autoref{subfig: tiny_mem_confidence_score}) ). This observation indicates that enforcing difficult samples to the training process first does not necessarily make the model more likely to forget them. If we perceive feeding difficult samples first to a model as negative, the repetition of a curriculum can possibly compensate for such a negative effect, i.e., making the target model memorize the difficult samples  better than a normal ML where these samples are presented at random times throughout training.

\begin{figure}[h]
    \begin{subfigure}[t]{.49\linewidth}
    \centering\includegraphics[width=\linewidth]{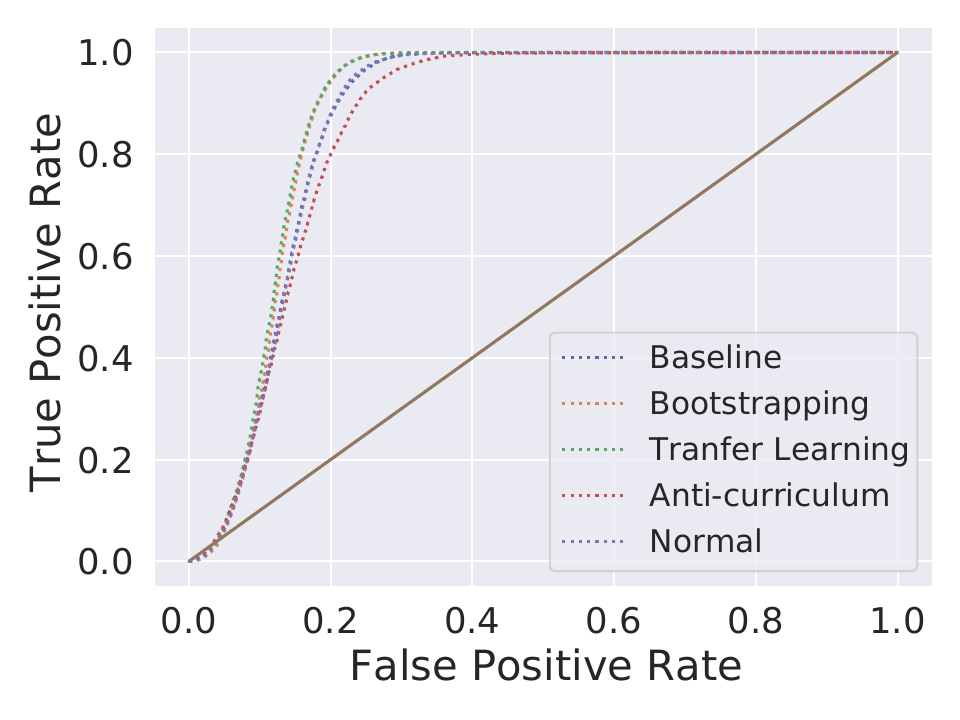} \caption{Linear scale NN-based MIA}\label{fig: tpr_tiny_nnbased}
  \end{subfigure}
  \begin{subfigure}[t]{.49\linewidth}
    \centering\includegraphics[width=\linewidth]{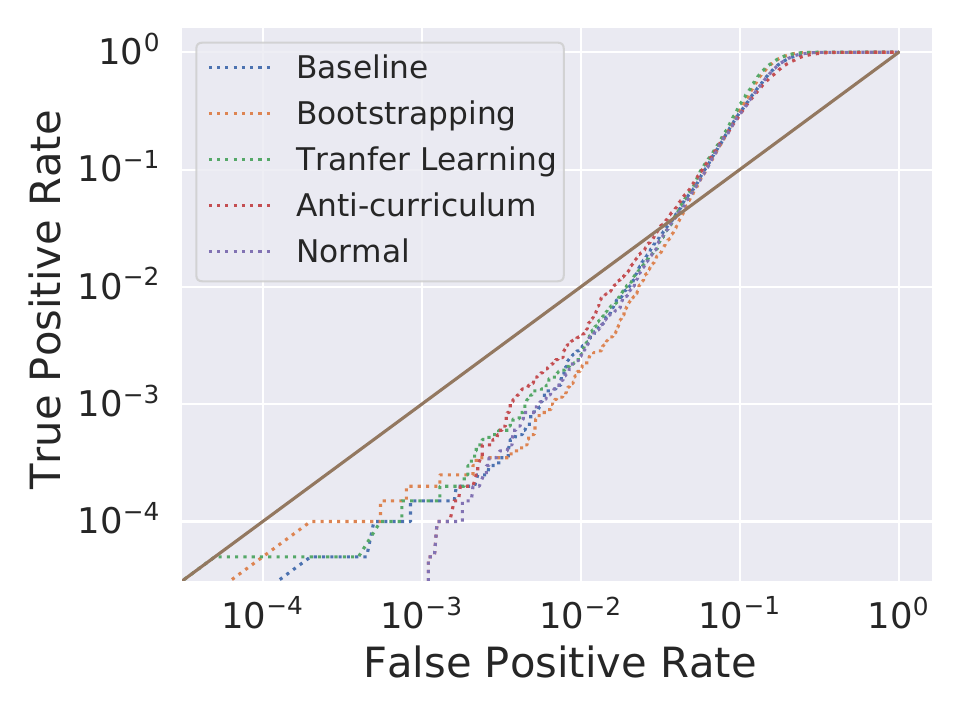} \caption{Log scale NN-based MIA}\label{fig: log_tpr_tiny_nnbased}
  \end{subfigure}
  
  \begin{subfigure}[t]{.49\linewidth}
    \centering\includegraphics[width=\linewidth]{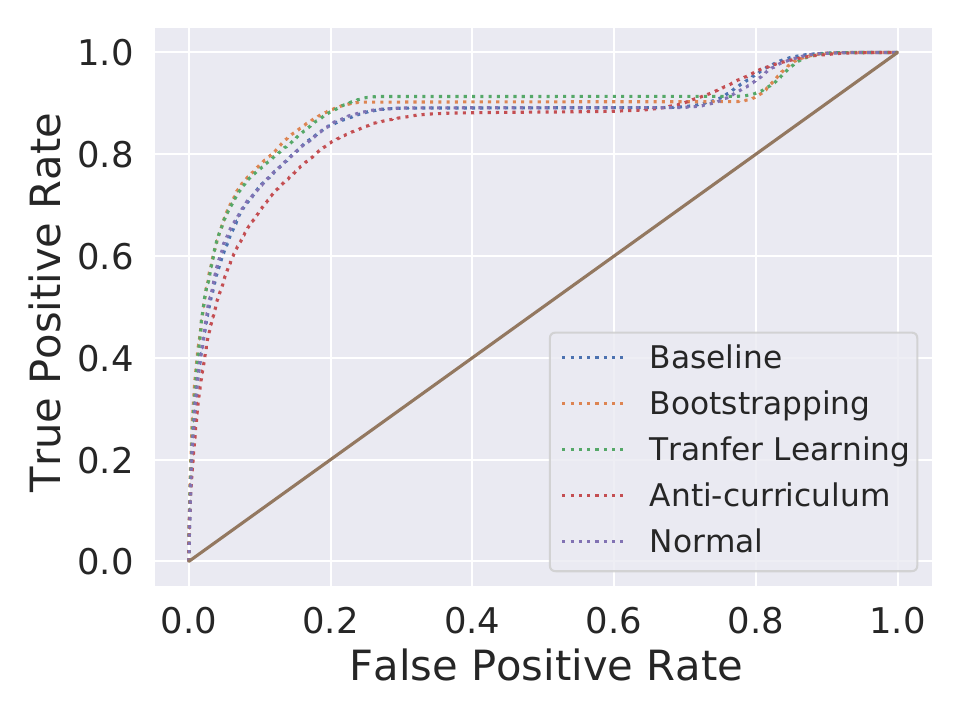}
    \caption{Linear scale \attack{}}\label{fig: tpr_tiny_diffcalibased}
  \end{subfigure}
  \begin{subfigure}[t]{.49\linewidth}
    \centering\includegraphics[width=\linewidth]{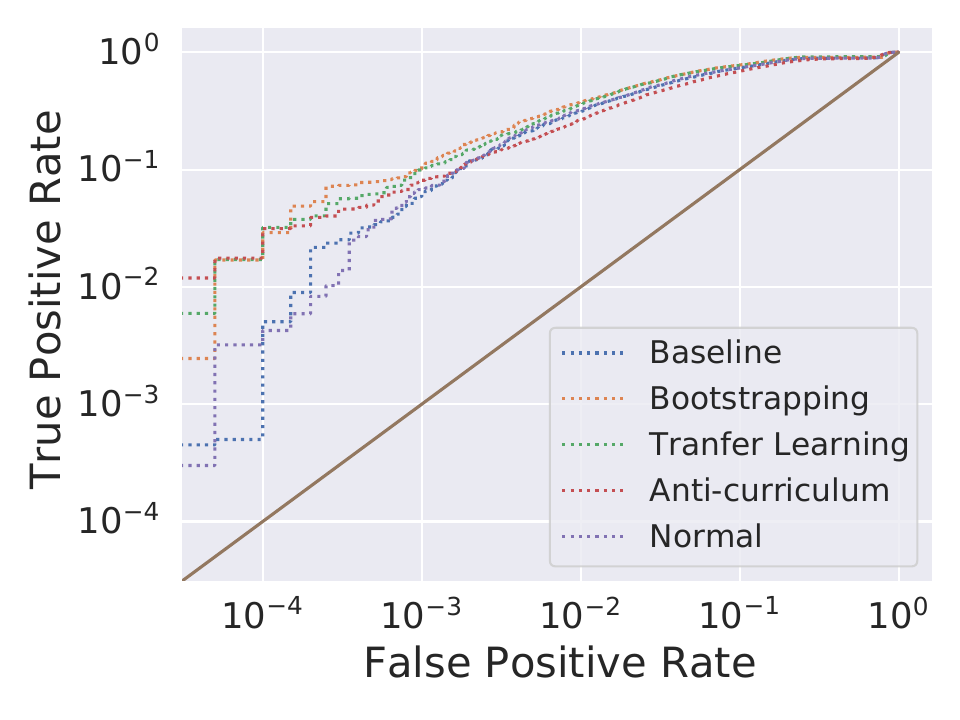}
    \caption{Log scale \attack{}}\label{fig: log_tpr_tiny_diffcalibased}
  \end{subfigure}
    \caption{TPR/FPR of NN_based MIA and \attack{} under different training method trained with ResNet-18 on CIFAR100. 
    }
    \label{fig: tpr_fpr_ours}
\end{figure}
\mypara{TPR at Low FPR}
In addition to the attack accuracy, we measured the relation between TPR at low FPR, as explained in ``evaluation metrics'' (\autoref{sec: evaluation}). Following Carlini et al.~\cite{carlini2021membership}, we present the ROC curve for the attacks with both linear scaling and log scaling to emphasize the low-FPR regime. \autoref{fig: tpr_tiny_nnbased} and~\autoref{fig: log_tpr_tiny_nnbased} demonstrate the ROC curve for  NN-based attack. The results show that using curriculum increases ROC. The TPR of transfer learning and bootstrapping are generally higher than the others except at extremely low FPR ($<10^{-4}$).  This indicates CL introduces disparate impact to members and non-members for most samples. Moreover, the NN-based attack fails to achieve a TPR better than random chance at any FPR below $0.045$, indicating potential for further improvement.

\begin{figure}[h]
    \centering
    \begin{subfigure}[t]{.49\linewidth}
    \centering\includegraphics[width=\linewidth]{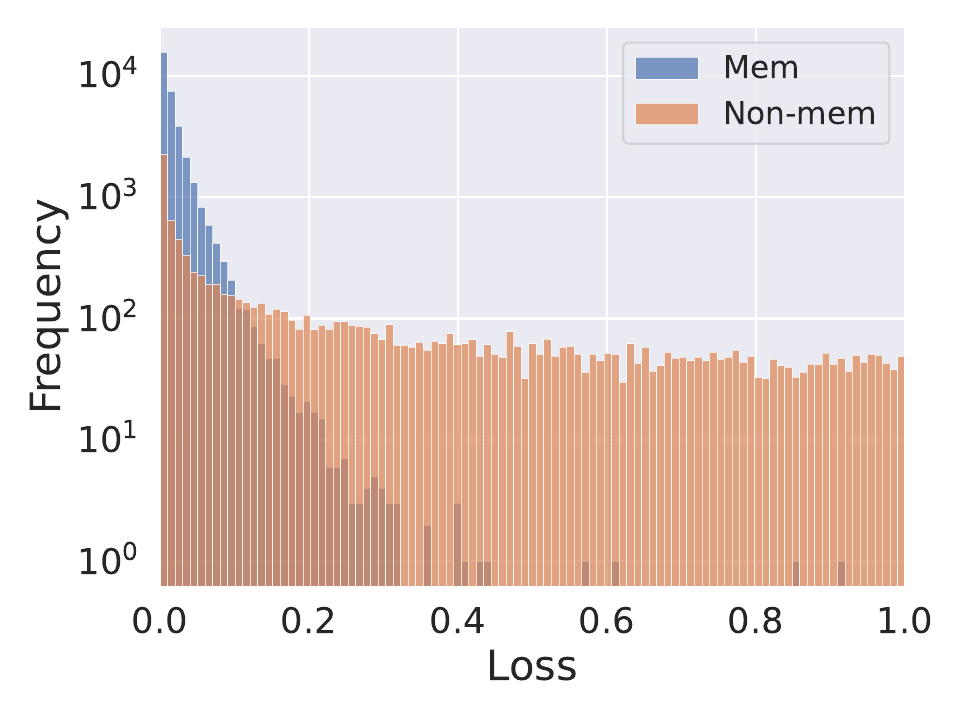
    }
    \caption{Normal training}\label{subfig: loss_resnet18_tiny_normal}
  \end{subfigure}
  \begin{subfigure}[t]{.49\linewidth}
    \centering\includegraphics[width=\linewidth]{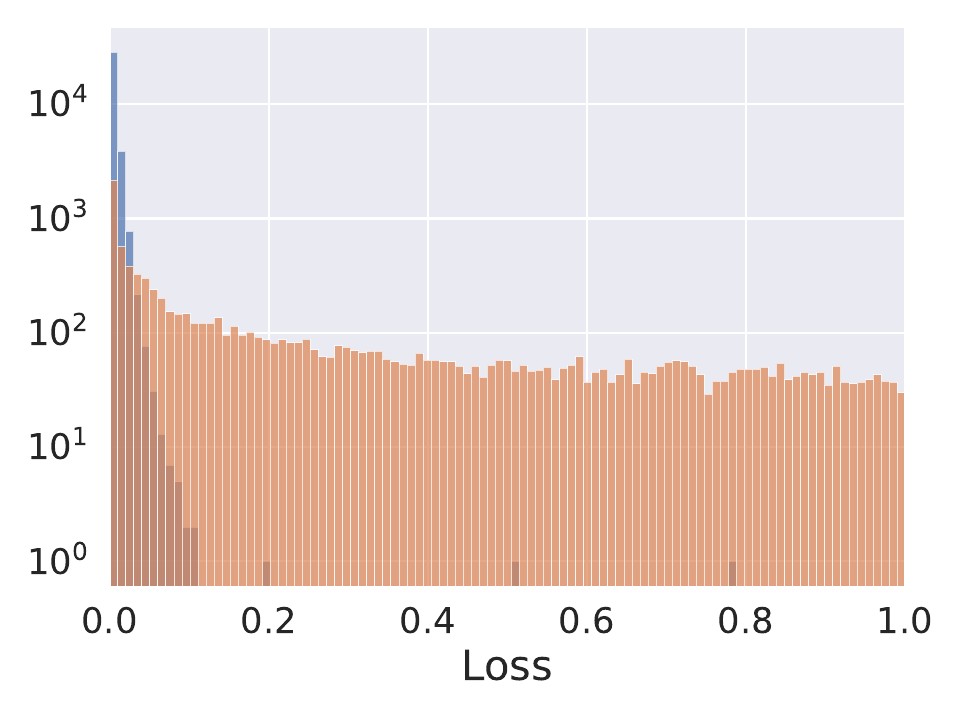
    }
    \caption{Bootstrapping}\label{subfig: loss_resnet18_tiny_cl}
  \end{subfigure}
    \caption{Loss distribution for models trained on Tiny ImageNet with ResNet-18.
    }
    \label{fig: loss_distribution}
\end{figure}

\begin{table*}[t]
\centering
\small
\begin{tabular}{l|ccccc}
\toprule
{\backslashbox{Architecture}{Method}} & Normal & Bootstrapping & Anti-curriculum & Baseline & Transfer Learning     \\
\midrule
ResNet-18 &   0.8572  $\pm$ 0.0011  &  \textbf{0.8751}  $\pm$ 0.0001  &   0.8376   $\pm$ 0.0002 &  0.8582 $\pm$ 0.0001  &   0.8718 $\pm$ 0.0001           \\
ResNet-34     &    0.8564 $\pm$ 0.0001&	\textbf{0.8746} $\pm$ 0.0003&	0.8481 $\pm$ 0.0002&	0.8559 $\pm$ 0.0002&	0.8715 $\pm$ 0.0002 \\
MobileNet     &  0.7979 $\pm$ 0.0001&	0.8308 $\pm$ 0.0000&	0.7763 $\pm$ 0.0002&	0.8318 $\pm$ 0.0000&	\textbf{0.8430} $\pm$  0.0001                    \\
\bottomrule
\end{tabular}
\caption{The average accuracy of NN-based attacks on models trained on different network architectures with CIFAR100.}
\label{fig: various_nets_acc}
\end{table*}
\mypara{Loss Distribution} 
The previous evaluation presents a macro-level understanding of CL's impact on MIA. Here we present a micro-level analysis by examining the loss distribution between members and non-members in models trained with normal and CL methods.
Due to the space limitation, here we only show the results of ResNet-18 trained on Tiny ImageNet in \autoref{fig: loss_distribution} which shows a clearer discrepancy in terms of the loss distributions comparing to other datasets. Note that the loss scores are normalized.
As one can see, there is a clear difference between their loss distributions, e.g., bootstrapping makes the overall members' loss much lower and the members' loss distribution less overlapped with non-members', especially for those members with higher difficulty levels.  
In \autoref{subsec:mem}, we also reason this observation from the perspective of data memorization.

\begin{figure}[h]
    \centering
    \begin{subfigure}[t]{.49\linewidth}
    \centering\includegraphics[width=\linewidth]{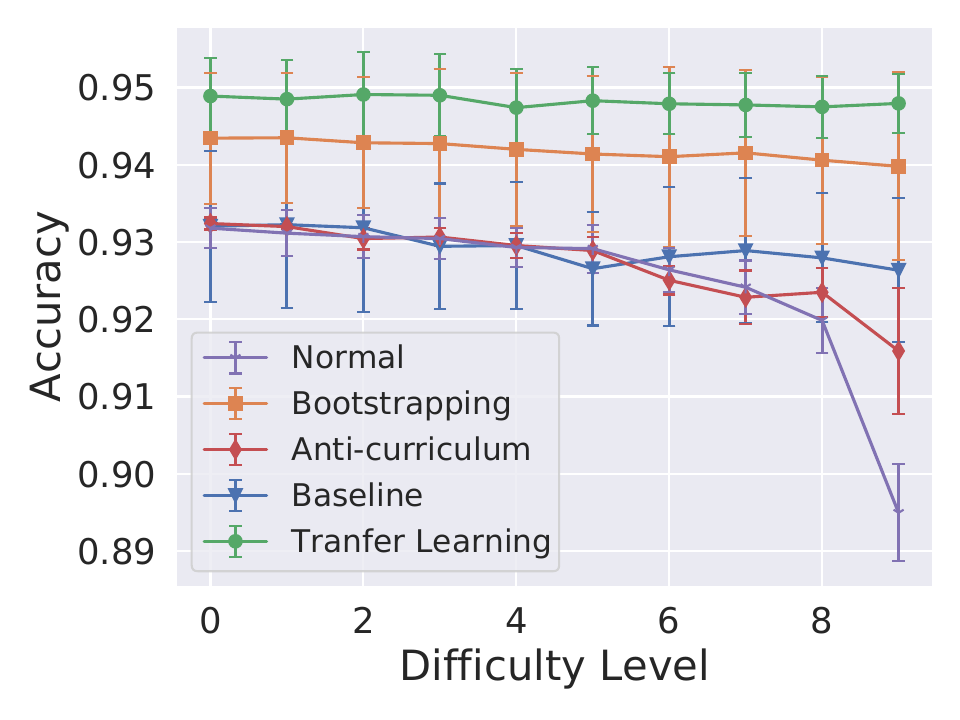} \caption{ResNet-34}
  \end{subfigure}
  \begin{subfigure}[t]{.49\linewidth}
    \centering\includegraphics[width=\linewidth]{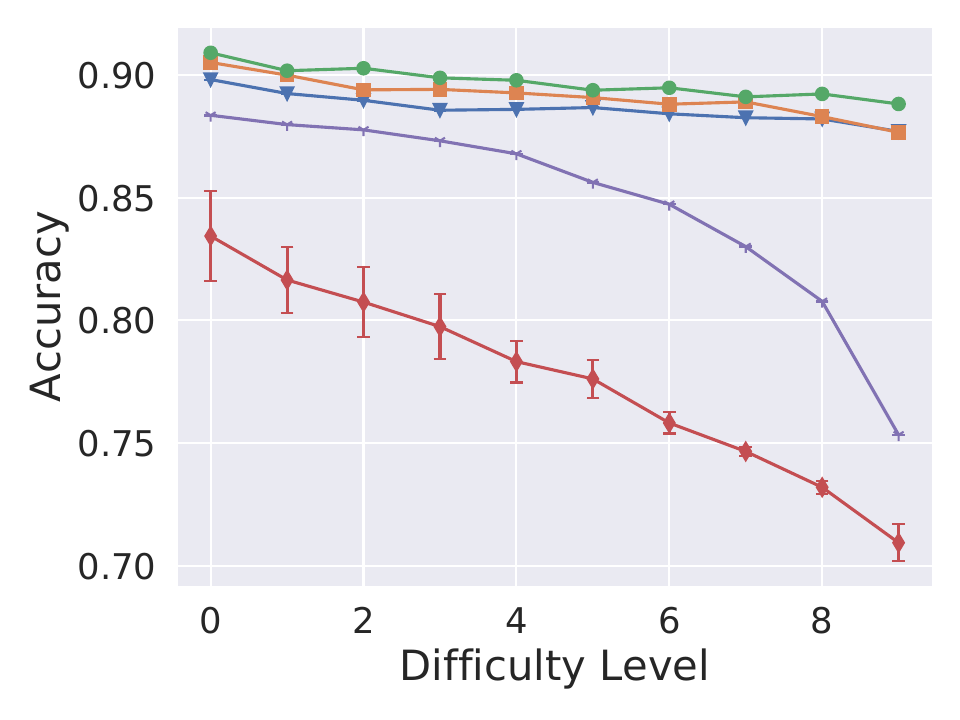}
    \caption{MobileNet}
  \end{subfigure}
 
    \caption{MIA accuracy for target model trained on Tiny ImageNet with ResNet-34 and MobileNet, respectively.}
    \label{fig: nets_attack_acc}
\end{figure}

\mypara{Target Model Architectures}
To study the impact of the architecture of the target model, we launched MIA against ResNet-34 and MobileNet and compare the results against ResNet-18. 
\autoref{fig: various_nets_acc} demonstrates the average attack accuracy of MIA when target models are trained with ResNet-18, ResNet-34, and MobileNet, respectively. It shows that they all share a similar trend of how CL affects MIA.  Though MobileNet turns out to be less vulnerable ($5.85\%$ and 5.93\%  less attack accuracy compared to ResNet-34 and ResNet-18, respectively), bootstrapping, transfer learning, and baseline all increase the overall attack accuracy.
\autoref{fig: nets_attack_acc} demonstrates the results by difficulty levels on ResNet-34 and MobileNet when training with Tiny ImageNet, which can be viewed together with  \autoref{subfig: resnet18_tiny_attack_acc} about ResNet-18. Though MobileNet turns out to be less vulnerable ($4\%$ less attack accuracy compared to ResNet-34 and ResNet-18), 
bootstrapping, transfer learning, and baseline all increase the overall attack accuracy and narrow down the gap between difficult and easy samples. As such, 
the privacy concerns in CL cannot be addressed by changing the target models' architectures.
This observation is consistent with other works~\cite{LZ21,HZ21} about MIA vs. architectures.

\mypara{Non-image Datasets} 
As shown in \autoref{fig: overall attack acc}, most experiments remain to have the same trend they are showing in image datasets. For Purchase, however, attack accuracy on normal training is 0.71\% higher than bootstrapping for example. This shows that CL does not always empower MIA more. In ~\autoref{app: purchase_fig} of \autoref{app:appendix}, we show the confidence score of members and non-members on Purchase, and the result is similar to the image datasets, where difficult samples are more vulnerable.

In the meantime, we found the changes caused by different CL methods are more drastic on the non-image datasets, compared to the image datasets. For example, Texas has a more prominent attack accuracy drop (8.0\%) on anti-curriculum. The non-image datasets are relatively simple, containing only binary features after pre-processing, hence they are more likely to be impacted by CL. ~\autoref{fig: target_acc} also shows the target model accuracy varies more for the non-image datasets under CL.

\zlbox{Finding 1: CL makes the target model more vulnerable to MIA, especially for difficult samples~\label{zlbox:finding1}.
}

\zlbox{Finding 2: Both data ordering and data repeating  make a model more vulnerable under MIA, while data ordering plays a bigger role in influencing the vulnerability of a model under MIA.
}

\subsection{Analysis with Data Memorization}
\label{subsec:mem}

The previous experiments show CL makes the difficult samples more vulnerable. Here, we attempt to explain this observation with a more principled analysis. Recent works~\cite{feldman2020does,feldman2020neural} suggest the effectiveness of MIA could be tied to how well the target model \textit{memorizes} individual data sample. The notion of memorization is formally defined as~\cite{feldman2020does}:
\begin{align}
    \textbf{mem}(\mathcal{A}, \mathcal{D}, i):= \underset{T \sim \mathcal{A} \left(\mathcal{D}\right)}{\mathsf{Pr}}\left[T\left(x_{i}\right)=y_{i}\right] - \underset{T \sim \mathcal{A}\left(\mathcal{D}^{\backslash i}\right)}{\mathsf{Pr}}\left[T\left( x_{i}\right)=y_{i}\right]
    \label{eq:mem}
\end{align}
where $\mathcal{A}$ denotes the training algorithm, $\mathcal{D}$ denotes the training dataset, $T$ is the trained model, $(x_i, y_i)$ denotes one sample with its ground-truth label, and $\mathcal{D}^{\backslash i}$ denotes $\mathcal{D}$ with $i$-th sample removed. The model is likely to memorize the data sample if adding $(x_i, y_i)$ to training significantly changes the model's prediction on $y_i$. Though~\autoref{eq:mem} models the memorization of a single data sample, we can easily extend it to quantify the memorization of multiple samples at once. 

\begin{figure}[!h]
    \centering
    \centering\includegraphics[width=\linewidth]{
    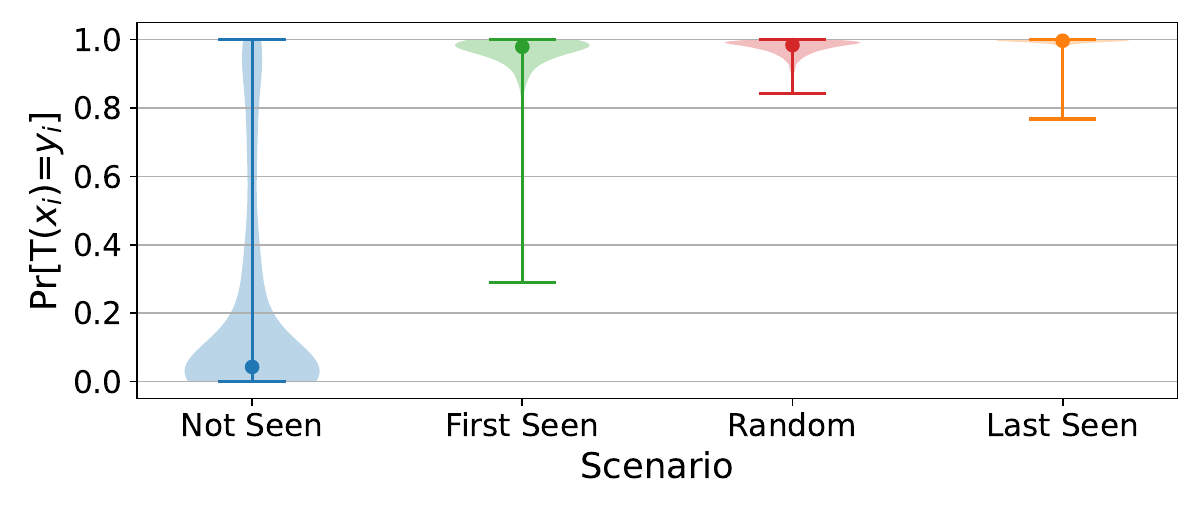
    }
    \caption{Memorization: violin plots of prediction probability of 800 most difficult samples, according to bootstrapping CL. The horizontal bars of each violin represent the minimum and maximum of the prediction probability. 
    }
    \label{fig: violin}
\end{figure}


Specifically, we evaluate ResNet-18 trained with CIFAR100.
We first leave out 800 most difficult data samples (4\% of all samples) and train a model without these data via bootstrapping (``not seen''). Then, we train the model under CL according to data memorization: the curriculum makes the 800 data samples either be seen at the beginning (``first seen''), end (``last seen''), or random places (``random'') of each training epoch.  
~\autoref{fig: violin} depicts the prediction probability of the true labels of the 4 scenarios. 
Data memorization under CL can be assessed by comparing ``first seen'', ``last seen'', and ``random'' to ``not seen'', following the idea of ~\autoref{eq:mem}.
We observe that other than ``not seen'', the other three scenarios memorize the difficult samples fairly well (higher prediction probability of the true class).
It turns out that data ordering has a strong impact on data memorization, e.g., ``last seen'' provides the strongest memorization compared to ``first seen'' and ``random''. 
The difficult samples are more vulnerable under CL because they are memorized better after data ordering.

Another concept often considered to be connected to memorization is data valuation. Here, we elaborate on the topic of data Shapley and study if our observation in this section can be explained from the angle of data valuation.
Specifically, we choose Shapley value~\cite{ghorbani2019data} as the metric, as it has the ``strongest theoretical foundation'' in data valuation research~\cite{hammoudeh2022training}. 
In essence, the data with high Shapley values are ones that on average contribute significantly to a model's prediction performance. 
We follow most of the experiment steps in this section and only change how the samples are selected for ``not seen'' (i.e., selected based on their Shapley values rather than difficulty levels).

\begin{figure}[!h]
    \centering
    \centering\includegraphics[width=\linewidth]{
    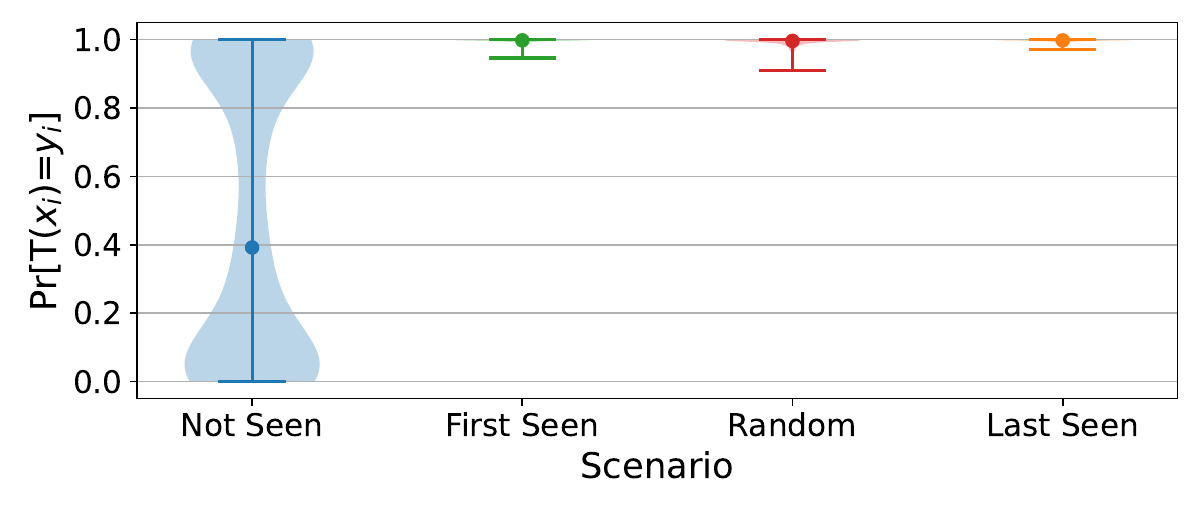
    }
    \caption{Shapley: violin plots of prediction probability of 800 most valuable samples according to KNN-Shapley. 
    }
    \label{fig: shap_violin}
\end{figure}

\begin{figure}[!h]
    \centering
    \centering\includegraphics[width=\linewidth]{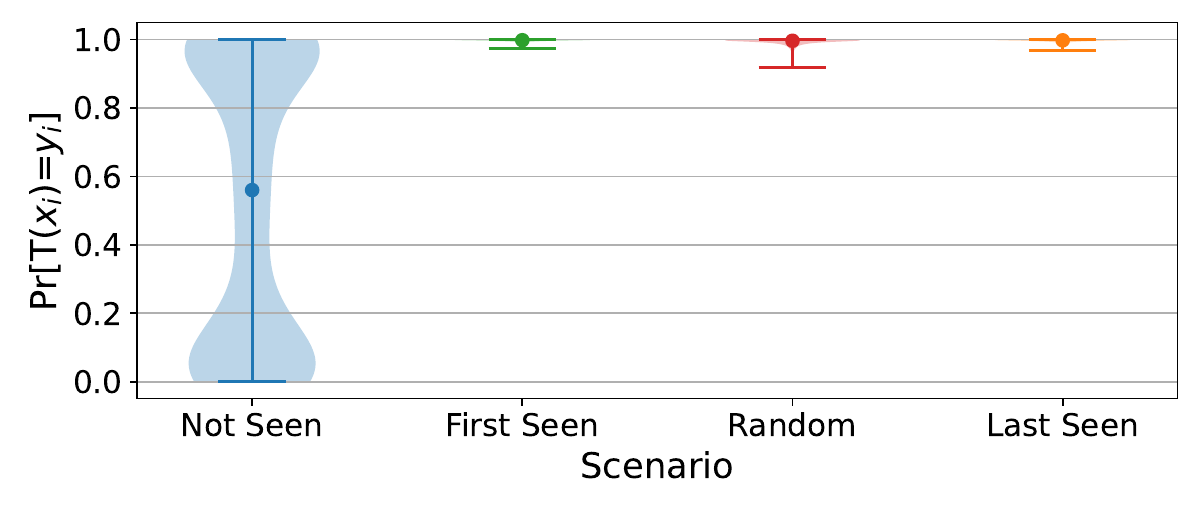
    }
    \caption{Reverse Shapley: violin plots of prediction probability of 800 least valuable samples according to KNN-Shapley. 
    }
    \label{fig: reverse_shap_violin}
\end{figure}

\mypara{KNN-Shapley}
Calculating Shapley values is intractable for a DNN model that is trained on a large dataset, as it requires a model to be retrained for $2^n$ times, where $n$ is the number of data points, to assess the contribution of one data point versus all possible subsets 
of the training set~\cite{hammoudeh2022training}. 
To address this scalability issue, Jia et al.~\cite{jia2021scalability} proposed KNN-Shapley, which uses a lightweight KNN \textit{surrogate} model to reduce the overhead of model retraining. The time complexity is reduced to $O(nlogn)$ and still, a good approximation of Shapley values can be obtained. As such, we use KNN-Shapley to calculate the Data Shapley values. 

\autoref{fig: shap_violin} and \autoref{fig: reverse_shap_violin} show the prediction probability of true label with 800 most and least valuable data samples according to KNN-Shapley. From the results of "not seen", we observe that the least valuable data have higher prediction accuracy on average (51\%), meaning that their absence in training has less impact compared to the more valuable data as presented in \autoref{fig: shap_violin}. Similarly, feeding the least valuable data first or at last to the training does not affect the prediction much.

Then, we compare the impact of difficulty level and Shapley value on data memorization, from \autoref{fig: violin} and \autoref{fig: shap_violin}.
Though both show that the absence of the most difficult or valuable data leads to poor prediction and seeing these data lastly benefits more than seeing them first during training, these changes are much more drastic for difficult samples (\autoref{fig: violin}) than the valuable samples (\autoref{fig: shap_violin}). For example, the median prediction probability of the ``not seen'' difficult samples and valuable samples are  39.19\% and 56.01\%. 
As such, the data reordering of CL makes the difficult samples more vulnerable, but not so for the valuable samples.

\zlbox{Finding 3: CL forces the model to memorize the difficult samples harder, which makes them more vulnerable.
}

\subsection{Evaluation of \attack}\label{subsec: diff_cali}
In order to fully utilize the information of difficulty levels exposed by CL, we propose \attack{} as described in \autoref{subsec: our_attack}.  Overall, the NN-based attack still has a slightly better attack accuracy compared to \attack{}, but \attack{} has higher confidence scores for difficult samples and has better TPR at the low FPR regime.

\begin{figure}[!h]
    \centering
    \begin{subfigure}[t]{.49\linewidth}
    \centering\includegraphics[width=\linewidth]{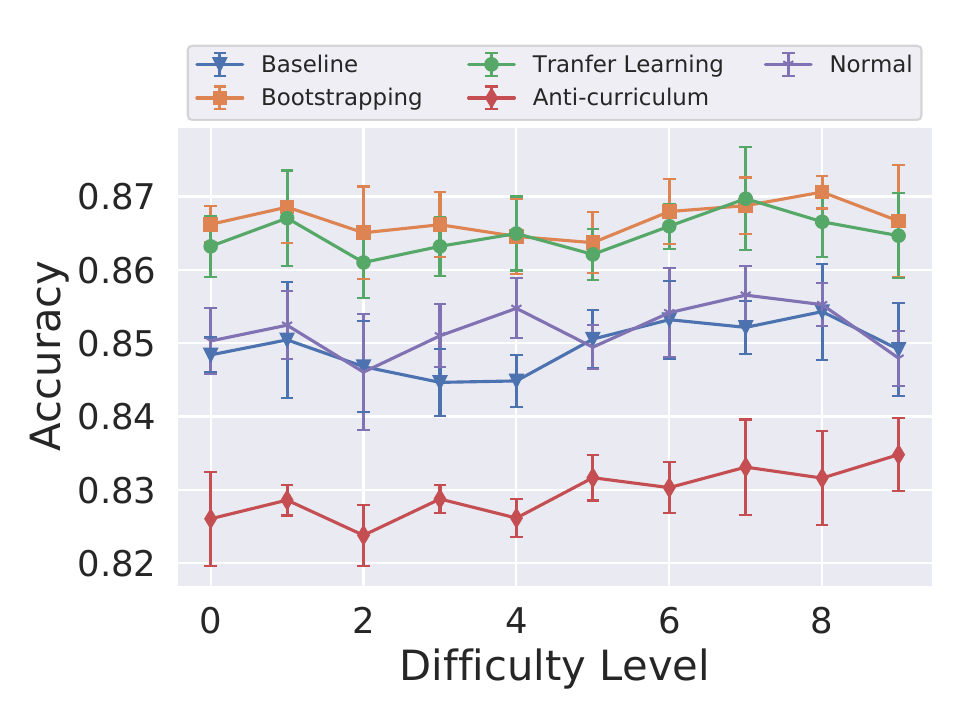}
    \caption{ CIFAR100}\label{subfig: ours_mia_cifar100}
  \end{subfigure}
  \begin{subfigure}[t]{.49\linewidth}
    \centering\includegraphics[width=\linewidth]{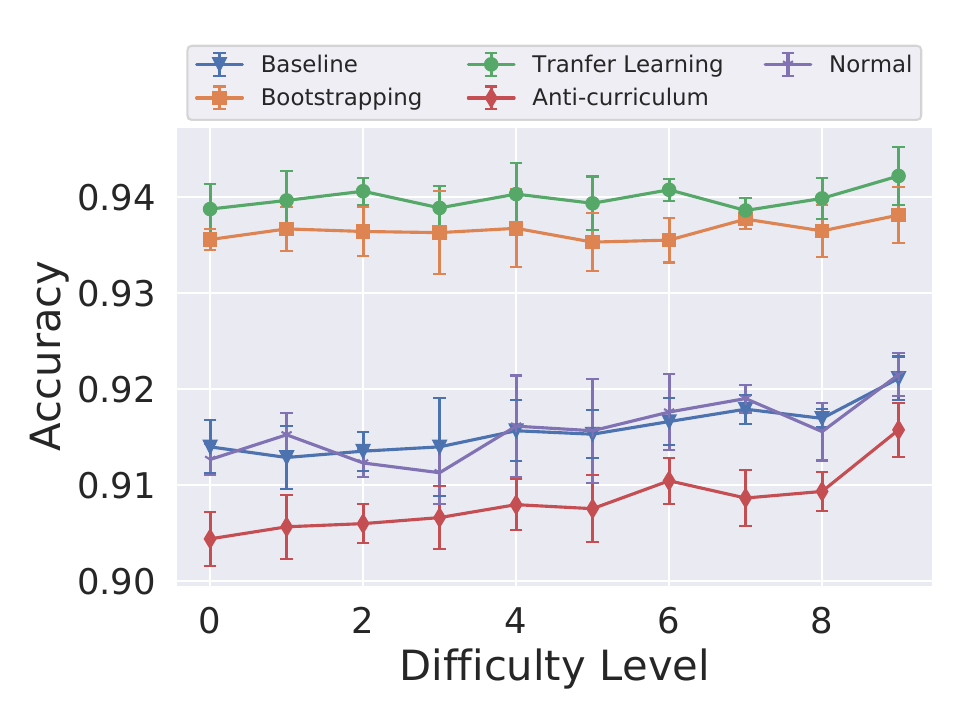}
    \caption{Tiny ImageNet}\label{subfig: ours_mia_tiny}
  \end{subfigure}
    \caption{\attack{}'s accuracy for models trained on CIFAR100 and Tiny ImageNet with ResNet-18.}
    \label{fig: ours_mia_bucket}
\end{figure}

\mypara{Attack Accuracy} ~\autoref{fig: various attack acc} presents the accuracy of \attack{}, which is about 1\% lower compared to NN-based attack on all CL methods. \autoref{fig: ours_mia_bucket} depicts the attack accuracy on CIFAR100 and Tiny ImageNet. Though\attack{} achieves slightly lower (less than 1.44\%) accuracy compared to NN-based attack, with adaptive calibration, we are able to make \textbf{the difficult samples more vulnerable}: e.g., the attack accuracy of difficulty level at 9 and 0 are 86.47\% and 86.32\% for transfer learning under CIFAR100.
The most difficult samples now can be predicted 2.64\% and 2.35\% more accurately for normal and anti-curriculum ML, respectively. Overall, \attack{} is able to overcome the privacy risk discrepancy of different samples through calibration and results in better attack accuracy for difficult samples for normal ML and anti-curriculum ML.

\mypara{Confidence Score} Like the evaluation of basic MIA, we show the confidence scores of samples according to their difficulty level in \autoref{fig: ours_cifar100_confi} and \autoref{fig: ours_tiny_confi} of~\autoref{app:appendix}. Overall, we are able to achieve confidence scores greater than 0.7807 (normal) for CIFAR100 and 0.8678 (normal) for Tiny ImageNet for all member samples, whereas the minimum member confidence score from NN-based is 0.6889 for CIFAR100 and 0.8333 for Tiny ImageNet (\autoref{fig: confidence_score}). In short, we are able to improve the normal training confidence score for all members by 3.29\% for CIFAR100 and 3.45\% for Tiny ImageNet. Similarly, we reduce the confidence score of non-members (note that a lower confidence score means less chance to be misclassified as non-members) by 0.0414 for CIFAR100 and 0.1751 for Tiny ImageNet. Unlike previous NN-based attack, the accuracy of \attack{} does not share a similar trend as the confidence score 
because the final prediction of the membership status of \attack{} is not based on the confidence score solely.

\mypara{TPR at Low FPR}
 In \autoref{fig: tpr_fpr_ours}, we show that \attack{} can achieve much higher TPR at low FPR ($< 10^{-4}$).  We present the ROC curve for the attacks with both linear scaling and log scaling to emphasize the low-FPR regime. \autoref{fig: tpr_tiny_diffcalibased} and ~\autoref{fig: log_tpr_tiny_diffcalibased} demonstrate the ROC curve for \attack{}. The results show that using curriculum increases ROC (\autoref{fig: tpr_tiny_nnbased}, \autoref{fig: tpr_tiny_diffcalibased}).  We observe that our proposed \attack{} performs better at low FPR. More specifically, \autoref{fig: log_tpr_tiny_nnbased} shows that NN-based attack fails to achieve a TPR better than random chance at any FPR below $0.045$ while \attack{} can be better than random guessing at all times. 

\zlbox{Finding 4: \attack{} improves MIA performance in terms of TPR at low FPR, making the difficult samples more vulnerable.}

\subsection{Evaluation of AIA}\label{subsec: evaluation_aia}

\begin{table}[h]
\centering
\footnotesize
\begin{tabular}{l|ccc}
\toprule
{\backslashbox{Method}{Dataset}} & Place100&Place60& UTKFace\\
\midrule
Normal& \textbf{0.107}$\pm$0.003 & \textbf{0.173}$\pm$0.002 & \textbf{0.528}$\pm$0.005\\
Bootstrapping& 0.092$\pm$0.003 & 0.168$\pm$0.004 & 0.515$\pm$0.006\\
Transfer Learning & 0.104$\pm$0.001&  0.150$\pm$0.005& 0.512$\pm$0.006\\
Baseline Curriculum& 0.079$\pm$0.004 &  0.143$\pm$0.001& 0.506 $\pm$0.008\\
Anti-Curriculum&0.033$\pm$0.001& 0.128$\pm$0.005 & 0.517$\pm$0.007\\

\bottomrule
\end{tabular}
\normalsize
\caption{Average accuracy of AIA ($\pm$ standard deviation) on model trained with different methods. ResNet-18 is the target model architecture.}
\label{tb: aia_performance}
\end{table}

We evaluate the 4 CL methods and normal training under the AIA setting described in ~\autoref{subsec:basic_aia} and \autoref{tb: aia_performance} to demonstrate the overall attack accuracy.
Generally, our results indicate that CL does not make the target model more vulnerable.
This somehow contradicts a recent study~\cite{HZ21} showing that a model is more vulnerable under AIA when trained under special settings, i.e., contrastive learning.
Interestingly, the normal training yields the highest average attack accuracy (e.g., 0.107 for Place100), even compared to anti-curriculum. UTKFace has a much higher attack accuracy because the baseline accuracy (random guessing based on majority class labels) of UTKFace is already quite high (42.1\%).
Our further investigation also shows that the attack accuracy is about the same for samples in different groups of difficulty levels (\autoref{fig: aia_app} in \autoref{app:appendix}).
We speculate that this is because the attributes of a sample themselves are already very complex and hard to learn. Besides, the difficulty score (e.g., bootstrapping) is calculated based on the original ML task, which emphasizes the specific attribute the original ML task tries to learn. 
That means the data ranking is effective only for the attribute chosen for the classification task but does not influence the sensitive attribute that one intends to infer.

\zlbox{Finding 5:  The model trained under CL is less vulnerable under AIA compared to MIA.}

\subsection{Evaluation of Defense}\label{subsec: defense}

We evaluate how the defenses including \dpsgd{}, \memguard{}, \mixupMMD{}, and \advreg{} perform under the impact of CL. 
\autoref{tb: defense} shows the attack accuracy on ResNet-18 which is trained with CIFAR100. Because \mixupMMD{} and \advreg{} require reference datasets for defense deployment, we equally divided CIFAR100 into 5 parts for fair comparison among all the defense techniques. More specifically, all target models in \autoref{tb: defense} are trained with only $12,000$ data points, which also explains why the accuracies are lower.
Regarding the setup of the defense methods, bootstrapping and anti-curriculum with \dpsgd{} are trained with the same curriculum as previous experiments. 
\dpsgd{*} uses a \textit{noisy} curriculum for bootstrapping and anti-curriculum, and the difficulty measurer is trained under \dpsgd{}. For transfer learning, it is not impacted as we use a pre-trained model.
$\epsilon$ and $\delta$ in our evaluation are $124,496$ and $1e-5$ for \dpsgd. We have a large $\epsilon$ because we have 200 epochs of training and ResNet-18 contains a large number of parameters. We did not change these settings for a fair comparison with other defense techniques. Previous studies have used large $\epsilon$ for \dpsgd{} in order to achieve good model accuracy~\cite{jayaraman2019evaluating, kurakin2022toward}. Based on a recent work~\cite{bu2022automatic}, we are able to make $\epsilon$ 10 times smaller after proper parameter tuning while achieving similar target accuracy. The $\epsilon$ can be brought down even first with a large batch size. Pulling tricks of \dpsgd{} based on the above recent work can further boost the tradeoff, we do not discuss it here as that is a parallel line of research. Note that in this section, we still use small batch size for \dpsgd{} evaluation though that results in large $\epsilon$. This is because we want to keep parameters across all target models the same for a fair MIA evaluation, and we have limited computing resources for handling large batch numbers.

\begin{table*}[h]
\centering
\scriptsize
\begin{tabular}{lccccccccccccc}
\toprule
\multirow{2}{*}
& \multicolumn{2}{c}{None} & \multicolumn{2}{c}{\dpsgd{*}} & \multicolumn{2}{c}{\dpsgd{}} &\multicolumn{3}{c}{\memguard{}} 
&\multicolumn{2}{c}{\mixupMMD} 
&\multicolumn{2}{c}{\advreg}  \\

 & Target      & MIA    & Target      & MIA  & Target      & MIA & Target      & MIA & Label-only &Target      & MIA& Target      & MIA\\
\midrule
norm  &48.0	& 90.3	&17.4&50.6_{$\pm$0.11}	& 17.4	&50.8_{$\pm$0.07}	&48.0&50.0 &83.0  & 54.1&81.6_{$\pm$0.02}& 51.2&89.2_{$\pm$0.01}\\

bstp  &	51.4 &	91.4_{$\pm$0.03 }&	18.0 &50.6_{$\pm$0.06 }	& 17.2 	&50.7_{$\pm$0.01 }	&51.4 &50.0 &84.5 & 54.4 &83.1_{$\pm$0.02}& 54.2 &91.6_{$\pm$0.02 }\\

tran 	&48.9 &	91.3_{$\pm$0.03 }&	17.2 &50.6_{$\pm$0.01 }	& 17.2 	&50.6_{$\pm$0.01 }	&48.9 &50.0  &84.5 & 55.7 &76.1_{$\pm$0.03 }& 50.4 &92.8_{$\pm$0.04 }\\

base &	50.0 &	91.5_{$\pm$0.02 }&	18.3 &50.4_{$\pm$0.11 }	& 17.6 	&50.4_{$\pm$0.11 }	&50.0 &50.0 &84.0 & 55.0 &84.4_{$\pm$0.02 }& 53.0 &91.6_{$\pm$0.01 }\\

anti &	49.3 &	89.5_{$\pm$0.02 }&	11.2 &50.3_{$\pm$0.11 }	& 17.2 	&50.4_{$\pm$0.10 }	&49.3 &50.0  &81.3 & 52.6 &79.1_{$\pm$0.02 }& 52.1 &87.3 \\

\bottomrule
\end{tabular}
\normalsize
\caption{The average accuracy of MIA ($\pm$ standard deviation (STD)) on target model trained on CIFAR100 with ResNet-18 and different defense methods. All numbers are in percentage, entry without $\pm$ STD means the STD is less than 0.01\%.}
\label{tb: defense}
\end{table*}

\autoref{tb: defense} demonstrates that \dpsgd{} is able to curb the MIA accuracy from 90.8\% to 50.5\% in average, which is close to random guess (i.e., member or non-member), though at the cost of a significant drop in target model's classification accuracy (from 49.52\% to 16.42\% in average).
This observation is consistent with previous works~\cite{li2021membership,kurakin2022toward}. 
We also found \dpsgd{} is effective against \attack{} (e.g., attack accuracy for normal and bootstrapping are dropped to 53.67\% and 53.09\%).
For \dpsgd{*}, due to the introduced noise, the ranking given by its curriculum is less accurate, but \autoref{tb: defense} shows that such change does not impact the MIA accuracy, and the target model accuracy drops by only a small amount (i.e., 0.8\% for bootstrapping and 0.7\% for baseline) except for anti-curriculum. Due to the noise in ranking, the ranking for anti-curriculum is no longer strictly ordered from difficult to easy. Instead, it becomes more random, thus target accuracy of anti-curriculum is even closer to baseline or bootstrapping.
In general, the result suggests using noisy ranking (\dpsgd{*}) as a defense might not be effective.

For \memguard{}, due to its design, NN-based MIA accuracy is fixed to 50\% when the defender knows what MIA method is performed by the attacker. 
In the meantime, the classification task of the target model is not impacted by \memguard{}. 
However, it is not very effective towards label-only attack, as it does not change the label. Our evaluation shows that the overall label-only attack accuracy can still reach up tp 86\% even with \memguard{} deployed.
\mixupMMD{} decreases the MIA accuracy (e.g., 91.4\% to 83.1\% for bootstrapping), and interestingly, it increases the target model accuracy (e.g., from 51.4\% to 54.4\% for bootstrapping), which might be attributed to its new regularizer.
\advreg{} can also increase target accuracy (e.g., 51.4\% to 54.2\% for bootstrapping) but is less effective in mitigating MIA (e.g., MIA accuracy is even increased from 91.4\% to 91.6\% for bootstrapping). This observation concurs with a previous work~\cite{SM21}.

\begin{figure}[!h]
    \centering
    \begin{subfigure}[t]{.49\linewidth}
    \centering\includegraphics[width=\linewidth]{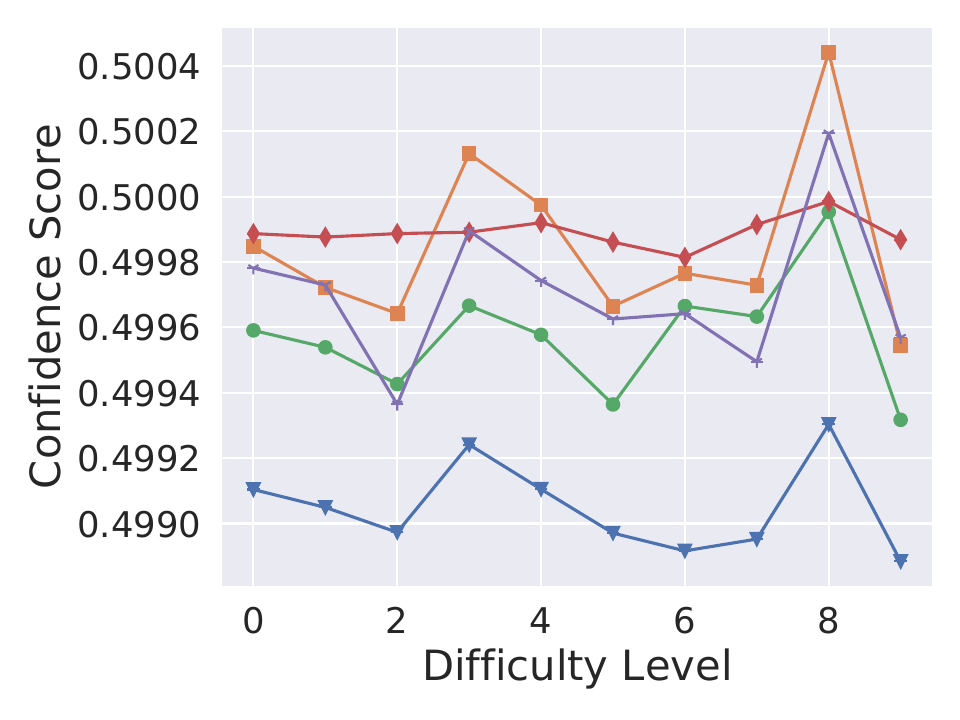}
    \caption{Member of CIFAR100}
  \end{subfigure}
  \begin{subfigure}[t]{.49\linewidth}
    \centering\includegraphics[width=\linewidth]{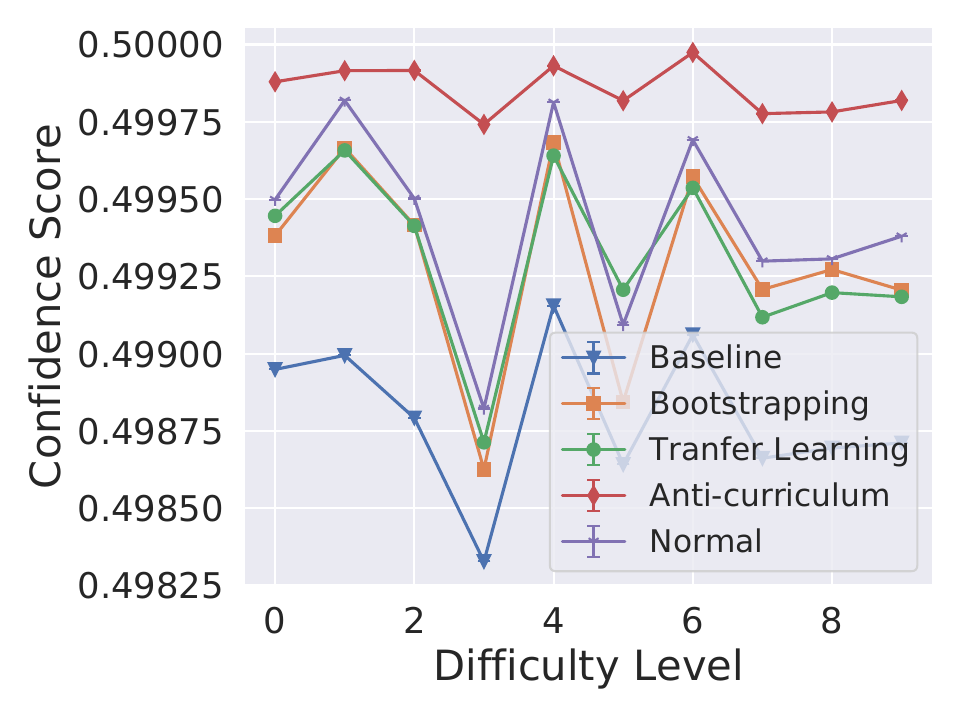}
    \caption{Non-member of CIFAR100}
  \end{subfigure}
    \caption{Attack model's confidence score for member and non-member samples of CIFAR-100 trained on ResNet-18 with \dpsgd. }
    \label{fig: dpsgd}
\end{figure}
\vspace{-3mm}

Given that CL introduces disparate impact on samples under different difficulty groups, we further investigate the relation between difficulty groups and defenses, and we focus on \dpsgd.
\autoref{fig: dpsgd} shows that \dpsgd{} is able to eliminate the disparate impact by CL, essentially making the difficult samples again hard to attack. 
We speculate the reason is that \dpsgd{} adds noise to gradient, which adds randomness to the optimization phase. CL, by introducing a teacher module, reinforces the learning by reducing the randomness. Ultimately, \dpsgd{} and CL are built on two opposite foundations. Thus, \dpsgd{} can eliminate the benefit from CL and achieve significant defense effect.

Overall, there is still room for improvement in defenses. Potential future work is to preserve certain properties brought by an ML technique (e.g., fast convergence and higher final performance by CL) and mitigate privacy risks generically.

\zlbox{Finding 6: None of the studied defenses can significantly drop the MIA accuracy while maintaining the target model accuracy. \dpsgd{} can reverse the impact of CL on MIA.
}
\section{Discussion}
\label{sec:discussion}

\mypara{Limitations} 
1) The research on ML privacy has been growing strong in recent years, and numerous attacks, variations, and defenses have emerged. Admittedly, not all attack methods (e.g., adaptive attack~\cite{SM21} and LiRA~\cite{carlini2021membership}) and defense techniques (e.g., PATE~\cite{PSMRTE18}) have been examined.  Though LiRA is considered state of the art, it requires multiple shadow models while all other attacks on our paper need one. To fairly compare with LiRA,  the current datasets need to be divided into much smaller subsets, which will lead to worse performance of all target models and shadow models. Thus, we didn't investigate LiRA in this work. However, we believe our key conclusions (e.g., the difficult samples become more vulnerable when trained with CL) hold generically, due to the fundamental designs of the curriculum. 2) We mainly evaluated the privacy attack on image and tabular datasets, with widely used models like ResNet and MLP. Admittedly, not all data types (e.g., text~\cite{bengio2009curriculum} and speech~\cite{zheng2019autoencoder}) and models (e.g., VGG) are covered. 
3) Not all ML privacy attacks are tested, such as model inversion attacks~\cite{fredrikson2015model, zhang2020secret}, as we suspect they are less likely to be impacted by CL.
In the end, we want to mention that our efforts are comparable to recent works that study the privacy of special ML settings like contrastive learning~\cite{HZ21}. 

\mypara{Evaluation Metrics} 
For privacy attacks like MIA, whether and how it is effective is determined by the evaluation metrics. Attack accuracy is the one adopted in the beginning and is still widely used today, but recent studies have suggested metrics have to be carefully selected to fully understand the results. Following Carlini et al.~\cite{carlini2021membership}, we adopt TPR at low FPR as another metric. We also view the results under confidence scores to shed light on the divergent impacts of CL into samples, which reveal new insights that are not captured by other metrics. Other metrics like precision/recall~\cite{carlini2021membership} and disparate vulnerability~\cite{YKT19} can be considered and we believe this research direction still needs new input.
\section{Related Work}\label{sec: related_work}

\mypara{Curriculum Learning (CL)}
The idea of CL was first introduced by Bengio et. al~\cite{bengio2009curriculum}. Researchers have then developed many new designs such as predefined CL~\cite{jiang2018mentornet}, self-paced CL learning~\cite{jiang2015self}, CL by transfer learning~\cite{weinshall2018curriculum} and other automated CL~\cite{graves2017automated}. CL is  proved to be effective in the domain of reinforcement learning~\cite{matiisen2019teacher,narvekar2018learning,fournier2019clic,foglino2019gray}, computer vision~\cite{bengio2009curriculum,sakaridis2019guided,duan2020curriculum,soviany2020image}, natural language processing~\cite{bengio2009curriculum,spitkovsky2009baby,zhou2020uncertainty,guo2020fine,liu2020task}, speech~\cite{zheng2019autoencoder,caubriere2019curriculum,lotfian2019curriculum}, etc. Note that a concept called self-paced\cite{kumar2010self} learning can often be confused with CL bootstrapping. They share a similar idea of using an iterative procedure to assign higher weights to training examples that have lower costs with respect to their chosen hypothesis. 
Bootstrapping differs in that the difficulty score is generated based on model accuracy rather than a hypothesis~\cite{hacohen2019power}. 

\mypara{Membership Inference Attack (MIA)}
\autoref{subsec:basic_mia} has surveyed some representative works about MIA. Here we describe other notable works.
On top of the original MIA~\cite{SSSS17}, Salem et al.~\cite{SZHBFB19} proposed three more powerful attacks by relaxing the assumptions made by Shokri et al.~\cite{SSSS17}.
Nasr et al.~\cite{NSH19} investigated privacy risks in centralized as well as federated learning scenarios under both black-box and white-box settings. 
Recent works show that MIA can be further enhanced by adopting flexible thresholds~\cite{jayaraman2020revisiting}, calibrated difficulty level~\cite{watson2021importance}, and loss trajectory~\cite{LZBZ22}.
Besides the general ML settings, recent works examined special settings like contrastive learning~\cite{HZ21,LJQG21}, Generative Adversarial Networks (GAN)~\cite{HHB19,CYZF20,CWGS21}, and Graph Neural Networks (GNN)~\cite{HJBGZ21,HWWBSZ21,WLZL22}. 
However, none of them investigated curriculum learning, and we aim to fill this knowledge gap.
To mitigate MIA, researchers have proposed a few defensive mechanisms, like \dpsgd~\cite{abadi2016deep},  \memguard~\cite{JSBZG19}, \mixupMMD~\cite{li2021membership}, and \advreg
~\cite{NSH18}, as described in \autoref{subsec: defense_methodology}. PATE~\cite{PSMRTE18} uses teacher models to supervise the training of the student model and adds Laplacian noise to the teacher models' output. Salem et al.~\cite{SZHBFB19} leverage model stacking and dropout to reduce overfitting.

\mypara{Attribute Inference Attack (AIA)}
AIA presents another notable threat to ML privacy. \autoref{subsec:basic_aia} surveyed the key works under AIA. In addition, He et al.~\cite{HZ21} show that AIA is more vulnerable to models trained by contrastive learning.
Recently, Song et al.~\cite{SR20} show that AIA is also effective against language models. Jayaraman et al. propose a new white-box AIA method that achieves better accuracy ~\cite{jayaraman2022attribute}. We focus on the black-box setting.

\mypara{Other Attacks Against ML Models}
MIA and AIA can be considered as attacks on the data privacy of ML. Model privacy, integrity, and availability have also been investigated, resulting in numerous studies.
Model stealing aims to learn the parameters~\cite{TZJRR16,OSF19,KTPPI20,KPQ21,SHHZ22} or hyperparameters~\cite{WG18,OASF18} of a target model, and model inversion, whose goal is to recover the training dataset~\cite{fredrikson2015model, zhang2020secret}. 
There also exists some works focus on protecting a model's ownership~\cite{LHZG19,UNSS17,ABCPK18,RCK18,JCCP21,CWPSCJMLS22,CHZ22,LJLK22} to defend against model stealing attacks and other attacks like network pruning and fine-tuning.

\section{Conclusion}
\label{sec:conclusion}
In this work, we perform the first quantitative study to understand how curriculum learning, a widely used technique that accelerates model training, affects the privacy of the trained model. Specifically, we trained target models under 6 image datasets and 3 tabular datasets and performed membership inference attacks (MIA) and attribute inference attacks (AIA) against them to assess the privacy risk in curriculum learning. Our results show that the target model becomes slightly more vulnerable to MIA but not so under AIA. We also found MIA has a significantly larger impact on samples with high difficulty levels. 
By exploiting the leakage from difficulty levels, we design a new MIA, termed \attack{}, which achieves similar overall accuracy with much better TPR at low FPR and can infer difficulty samples from normal ML more accurately.
Finally, we evaluate the existing defenses \dpsgd{}, \memguard{}, \mixupMMD{}, and \advreg{} in the setting of curriculum learning, and our results show that they are still effective against the basic MIA. With this study, we hope to draw attention to the unintended effects of the emerging machine-learning techniques, and more theoretical analysis into the trade-off between privacy, accuracy, and fairness.  

\section*{Acknowledgement}
This work is partially funded by NSF CNS-2220434 and the European Health and Digital Executive Agency (HADEA) within the project ``Understanding the individual host response against Hepatitis D Virus to develop a personalized approach for the management of hepatitis D'' (D-Solve) (grant agreement number 101057917).

\bibliographystyle{plain}
\bibliography{ref,ref_new,normal_generated_py3}

\begin{thebibliography}{10}

\bibitem{CL1}
\url{https://github.com/GuyHacohen/curriculum_learning}.

\bibitem{CL2}
\url{https://github.com/rsundar96/curriculum-learning-acceleration}.

\bibitem{ART}
\url{https://github.com/Trusted-AI/adversarial-robustness-toolbox}.

\bibitem{abadi2016deep}
Martin Abadi, Andy Chu, Ian Goodfellow, H~Brendan McMahan, Ilya Mironov, Kunal Talwar, and Li~Zhang.
\newblock Deep learning with differential privacy.
\newblock In {\em Proceedings of the 2016 ACM SIGSAC conference on computer and communications security}, pages 308--318, 2016.

\bibitem{ABCPK18}
Yossi Adi, Carsten Baum, Moustapha Cisse, Benny Pinkas, and Joseph Keshet.
\newblock {Turning Your Weakness Into a Strength: Watermarking Deep Neural Networks by Backdooring}.
\newblock In {\em {USENIX Security Symposium (USENIX Security)}}, pages 1615--1631. USENIX, 2018.

\bibitem{bengio2009curriculum}
Yoshua Bengio, J{\'e}r{\^o}me Louradour, Ronan Collobert, and Jason Weston.
\newblock Curriculum learning.
\newblock In {\em Proceedings of the 26th annual international conference on machine learning}, pages 41--48, 2009.

\bibitem{bu2022automatic}
Zhiqi Bu, Yu-Xiang Wang, Sheng Zha, and George Karypis.
\newblock Automatic clipping: Differentially private deep learning made easier and stronger.
\newblock {\em arXiv preprint arXiv:2206.07136}, 2022.

\bibitem{carlini2021membership}
Nicholas Carlini, Steve Chien, Milad Nasr, Shuang Song, Andreas Terzis, and Florian Tramer.
\newblock Membership inference attacks from first principles.
\newblock {\em arXiv preprint arXiv:2112.03570}, 2021.

\bibitem{caubriere2019curriculum}
Antoine Caubri{\`e}re, Natalia Tomashenko, Antoine Laurent, Emmanuel Morin, Nathalie Camelin, and Yannick Est{\`e}ve.
\newblock Curriculum-based transfer learning for an effective end-to-end spoken language understanding and domain portability.
\newblock {\em arXiv preprint arXiv:1906.07601}, 2019.

\bibitem{chang2021privacy}
Hongyan Chang and Reza Shokri.
\newblock On the privacy risks of algorithmic fairness.
\newblock In {\em 2021 IEEE European Symposium on Security and Privacy (EuroS\&P)}, pages 292--303. IEEE, 2021.

\bibitem{CYZF20}
Dingfan Chen, Ning Yu, Yang Zhang, and Mario Fritz.
\newblock {GAN-Leaks: A Taxonomy of Membership Inference Attacks against Generative Models}.
\newblock In {\em {ACM SIGSAC Conference on Computer and Communications Security (CCS)}}, pages 343--362. ACM, 2020.

\bibitem{CWPSCJMLS22}
Jialuo Chen, Jingyi Wang, Tinglan Peng, Youcheng Sun, Peng Cheng, Shouling Ji, Xingjun Ma, Bo~Li, and Dawn Song.
\newblock {Copy, Right? A Testing Framework for Copyright Protection of Deep Learning Models}.
\newblock In {\em {IEEE Symposium on Security and Privacy (S\&P)}}. IEEE, 2022.

\bibitem{CWGS21}
Junjie Chen, Wendy~Hui Wang, Hongchang Gao, and Xinghua Shi.
\newblock {{PAR-GAN:} Improving the Generalization of Generative Adversarial Networks Against Membership Inference Attacks}.
\newblock In {\em {ACM Conference on Knowledge Discovery and Data Mining (KDD)}}, pages 127--137. ACM, 2021.

\bibitem{CTCP21}
Christopher A.~Choquette Choo, Florian Tram{\`e}r, Nicholas Carlini, and Nicolas Papernot.
\newblock {Label-Only Membership Inference Attacks}.
\newblock In {\em {International Conference on Machine Learning (ICML)}}, pages 1964--1974. PMLR, 2021.

\bibitem{CHZ22}
Tianshuo Cong, Xinlei He, and Yang Zhang.
\newblock {SSLGuard: A Watermarking Scheme for Self-supervised Learning Pre-trained Encoders}.
\newblock In {\em {ACM SIGSAC Conference on Computer and Communications Security (CCS)}}, pages 579--593. ACM, 2022.

\bibitem{deng2009imagenet}
Jia Deng, Wei Dong, Richard Socher, Li-Jia Li, Kai Li, and Li~Fei-Fei.
\newblock Imagenet: A large-scale hierarchical image database.
\newblock In {\em 2009 IEEE conference on computer vision and pattern recognition}, pages 248--255. Ieee, 2009.

\bibitem{duan2020curriculum}
Yueqi Duan, Haidong Zhu, He~Wang, Li~Yi, Ram Nevatia, and Leonidas~J Guibas.
\newblock Curriculum deepsdf.
\newblock In {\em European Conference on Computer Vision}, pages 51--67. Springer, 2020.

\bibitem{feldman2020does}
Vitaly Feldman.
\newblock Does learning require memorization? a short tale about a long tail.
\newblock In {\em Proceedings of the 52nd Annual ACM SIGACT Symposium on Theory of Computing}, pages 954--959, 2020.

\bibitem{feldman2020neural}
Vitaly Feldman and Chiyuan Zhang.
\newblock What neural networks memorize and why: Discovering the long tail via influence estimation.
\newblock {\em Advances in Neural Information Processing Systems}, 33:2881--2891, 2020.

\bibitem{foglino2019gray}
Francesco Foglino, Matteo Leonetti, Simone Sagratella, and Ruggiero Seccia.
\newblock A gray-box approach for curriculum learning.
\newblock In {\em World Congress on Global Optimization}, pages 720--729. Springer, 2019.

\bibitem{fournier2019clic}
Pierre Fournier, C{\'e}dric Colas, Mohamed Chetouani, and Olivier Sigaud.
\newblock Clic: Curriculum learning and imitation for object control in non-rewarding environments.
\newblock {\em IEEE Transactions on Cognitive and Developmental Systems}, 2019.

\bibitem{fredrikson2015model}
Matt Fredrikson, Somesh Jha, and Thomas Ristenpart.
\newblock Model inversion attacks that exploit confidence information and basic countermeasures.
\newblock In {\em Proceedings of the 22nd ACM SIGSAC conference on computer and communications security}, pages 1322--1333, 2015.

\bibitem{GWYGB18}
Karan Ganju, Qi~Wang, Wei Yang, Carl~A. Gunter, and Nikita Borisov.
\newblock {Property Inference Attacks on Fully Connected Neural Networks using Permutation Invariant Representations}.
\newblock In {\em {ACM SIGSAC Conference on Computer and Communications Security (CCS)}}, pages 619--633. ACM, 2018.

\bibitem{ghorbani2019data}
Amirata Ghorbani and James Zou.
\newblock Data shapley: Equitable valuation of data for machine learning.
\newblock In {\em International Conference on Machine Learning}, pages 2242--2251. PMLR, 2019.

\bibitem{graves2017automated}
Alex Graves, Marc~G Bellemare, Jacob Menick, Remi Munos, and Koray Kavukcuoglu.
\newblock Automated curriculum learning for neural networks.
\newblock In {\em international conference on machine learning}, pages 1311--1320. PMLR, 2017.

\bibitem{guo2020fine}
Junliang Guo, Xu~Tan, Linli Xu, Tao Qin, Enhong Chen, and Tie-Yan Liu.
\newblock Fine-tuning by curriculum learning for non-autoregressive neural machine translation.
\newblock In {\em Proceedings of the AAAI Conference on Artificial Intelligence}, volume~34, pages 7839--7846, 2020.

\bibitem{hacohen2019power}
Guy Hacohen and Daphna Weinshall.
\newblock On the power of curriculum learning in training deep networks.
\newblock In {\em International Conference on Machine Learning}, pages 2535--2544. PMLR, 2019.

\bibitem{hammoudeh2022training}
Zayd Hammoudeh and Daniel Lowd.
\newblock Training data influence analysis and estimation: A survey.
\newblock {\em arXiv preprint arXiv:2212.04612}, 2022.

\bibitem{HZRS16}
Kaiming He, Xiangyu Zhang, Shaoqing Ren, and Jian Sun.
\newblock {Deep Residual Learning for Image Recognition}.
\newblock In {\em {IEEE Conference on Computer Vision and Pattern Recognition (CVPR)}}, pages 770--778. IEEE, 2016.

\bibitem{HJBGZ21}
Xinlei He, Jinyuan Jia, Michael Backes, Neil~Zhenqiang Gong, and Yang Zhang.
\newblock {Stealing Links from Graph Neural Networks}.
\newblock In {\em {USENIX Security Symposium (USENIX Security)}}, pages 2669--2686. USENIX, 2021.

\bibitem{HWWBSZ21}
Xinlei He, Rui Wen, Yixin Wu, Michael Backes, Yun Shen, and Yang Zhang.
\newblock {Node-Level Membership Inference Attacks Against Graph Neural Networks}.
\newblock {\em {CoRR abs/2102.05429}}, 2021.

\bibitem{HZ21}
Xinlei He and Yang Zhang.
\newblock {Quantifying and Mitigating Privacy Risks of Contrastive Learning}.
\newblock In {\em {ACM SIGSAC Conference on Computer and Communications Security (CCS)}}, pages 845--863. ACM, 2021.

\bibitem{HHB19}
Benjamin Hilprecht, Martin H{\"{a}}rterich, and Daniel Bernau.
\newblock {Monte Carlo and Reconstruction Membership Inference Attacks against Generative Models}.
\newblock {\em {Privacy Enhancing Technologies Symposium}}, 2019.

\bibitem{jayaraman2019evaluating}
Bargav Jayaraman and David Evans.
\newblock Evaluating differentially private machine learning in practice.
\newblock In {\em 28th USENIX Security Symposium (USENIX Security 19)}, pages 1895--1912, 2019.

\bibitem{jayaraman2022attribute}
Bargav Jayaraman and David Evans.
\newblock Are attribute inference attacks just imputation?
\newblock In {\em Proceedings of the 2022 ACM SIGSAC Conference on Computer and Communications Security}, pages 1569--1582, 2022.

\bibitem{jayaraman2020revisiting}
Bargav Jayaraman, Lingxiao Wang, Katherine Knipmeyer, Quanquan Gu, and David Evans.
\newblock Revisiting membership inference under realistic assumptions.
\newblock {\em arXiv preprint arXiv:2005.10881}, 2020.

\bibitem{JCCP21}
Hengrui Jia, Christopher~A. Choquette{-}Choo, Varun Chandrasekaran, and Nicolas Papernot.
\newblock {Entangled Watermarks as a Defense against Model Extraction}.
\newblock In {\em {USENIX Security Symposium (USENIX Security)}}, pages 1937--1954. USENIX, 2021.

\bibitem{JSBZG19}
Jinyuan Jia, Ahmed Salem, Michael Backes, Yang Zhang, and Neil~Zhenqiang Gong.
\newblock {MemGuard: Defending against Black-Box Membership Inference Attacks via Adversarial Examples}.
\newblock In {\em {ACM SIGSAC Conference on Computer and Communications Security (CCS)}}, pages 259--274. ACM, 2019.

\bibitem{jia2021scalability}
Ruoxi Jia, Fan Wu, Xuehui Sun, Jiacen Xu, David Dao, Bhavya Kailkhura, Ce~Zhang, Bo~Li, and Dawn Song.
\newblock Scalability vs. utility: Do we have to sacrifice one for the other in data importance quantification?
\newblock In {\em Proceedings of the IEEE/CVF Conference on Computer Vision and Pattern Recognition}, pages 8239--8247, 2021.

\bibitem{jiang2015self}
Lu~Jiang, Deyu Meng, Qian Zhao, Shiguang Shan, and Alexander~G Hauptmann.
\newblock Self-paced curriculum learning.
\newblock In {\em Twenty-Ninth AAAI Conference on Artificial Intelligence}, 2015.

\bibitem{jiang2018mentornet}
Lu~Jiang, Zhengyuan Zhou, Thomas Leung, Li-Jia Li, and Li~Fei-Fei.
\newblock Mentornet: Learning data-driven curriculum for very deep neural networks on corrupted labels.
\newblock In {\em International Conference on Machine Learning}, pages 2304--2313. PMLR, 2018.

\bibitem{KPQ21}
Sanjay Kariyappa, Atul Prakash, and Moinuddin~K. Qureshi.
\newblock {{MAZE:} Data-Free Model Stealing Attack Using Zeroth-Order Gradient Estimation}.
\newblock In {\em {IEEE Conference on Computer Vision and Pattern Recognition (CVPR)}}, pages 13814--13823. IEEE, 2021.

\bibitem{KTPPI20}
Kalpesh Krishna, Gaurav~Singh Tomar, Ankur~P. Parikh, Nicolas Papernot, and Mohit Iyyer.
\newblock {Thieves on Sesame Street! Model Extraction of BERT-based APIs}.
\newblock In {\em {International Conference on Learning Representations (ICLR)}}, 2020.

\bibitem{krizhevsky2009learning}
Alex Krizhevsky, Geoffrey Hinton, et~al.
\newblock Learning multiple layers of features from tiny images.
\newblock 2009.

\bibitem{kumar2010self}
M~Pawan Kumar, Benjamin Packer, and Daphne Koller.
\newblock Self-paced learning for latent variable models.
\newblock In {\em NIPS}, volume~1, page~2, 2010.

\bibitem{kurakin2022toward}
Alexey Kurakin, Steve Chien, Shuang Song, Roxana Geambasu, Andreas Terzis, and Abhradeep Thakurta.
\newblock Toward training at imagenet scale with differential privacy.
\newblock {\em arXiv preprint arXiv:2201.12328}, 2022.

\bibitem{le2015tiny}
Ya~Le and Xuan Yang.
\newblock Tiny imagenet visual recognition challenge.
\newblock {\em CS 231N}, 7(7):3, 2015.

\bibitem{li2021membership}
Jiacheng Li, Ninghui Li, and Bruno Ribeiro.
\newblock Membership inference attacks and defenses in classification models.
\newblock In {\em Proceedings of the Eleventh ACM Conference on Data and Application Security and Privacy}, pages 5--16, 2021.

\bibitem{LHZG19}
Zheng Li, Chengyu Hu, Yang Zhang, and Shanqing Guo.
\newblock {How to Prove Your Model Belongs to You: A Blind-Watermark based Framework to Protect Intellectual Property of DNN}.
\newblock In {\em {Annual Computer Security Applications Conference (ACSAC)}}, pages 126--137. ACM, 2019.

\bibitem{LZ21}
Zheng Li and Yang Zhang.
\newblock {Membership Leakage in Label-Only Exposures}.
\newblock In {\em {ACM SIGSAC Conference on Computer and Communications Security (CCS)}}, pages 880--895. ACM, 2021.

\bibitem{LJQG21}
Hongbin Liu, Jinyuan Jia, Wenjie Qu, and Neil~Zhenqiang Gong.
\newblock {EncoderMI: Membership Inference against Pre-trained Encoders in Contrastive Learning}.
\newblock In {\em {ACM SIGSAC Conference on Computer and Communications Security (CCS)}}. ACM, 2021.

\bibitem{liu2020task}
Jinglin Liu, Yi~Ren, Xu~Tan, Chen Zhang, Tao Qin, Zhou Zhao, and Tie-Yan Liu.
\newblock Task-level curriculum learning for non-autoregressive neural machine translation.
\newblock {\em arXiv preprint arXiv:2007.08772}, 2020.

\bibitem{liu2021self}
Xiao Liu, Fanjin Zhang, Zhenyu Hou, Li~Mian, Zhaoyu Wang, Jing Zhang, and Jie Tang.
\newblock Self-supervised learning: Generative or contrastive.
\newblock {\em IEEE Transactions on Knowledge and Data Engineering}, 35(1):857--876, 2021.

\bibitem{LZBZ22}
Yiyong Liu, Zhengyu Zhao, Michael Backes, and Yang Zhang.
\newblock {Membership Inference Attacks by Exploiting Loss Trajectory}.
\newblock In {\em {ACM SIGSAC Conference on Computer and Communications Security (CCS)}}, pages 2085--2098. ACM, 2022.

\bibitem{LBWBWTGC18}
Yunhui Long, Vincent Bindschaedler, Lei Wang, Diyue Bu, Xiaofeng Wang, Haixu Tang, Carl~A. Gunter, and Kai Chen.
\newblock {Understanding Membership Inferences on Well-Generalized Learning Models}.
\newblock {\em {CoRR abs/1802.04889}}, 2018.

\bibitem{lotfian2019curriculum}
Reza Lotfian and Carlos Busso.
\newblock Curriculum learning for speech emotion recognition from crowdsourced labels.
\newblock {\em IEEE/ACM Transactions on Audio, Speech, and Language Processing}, 27(4):815--826, 2019.

\bibitem{LJLK22}
Nils Lukas, Edward Jiang, Xinda Li, and Florian Kerschbaum.
\newblock {SoK: How Robust is Image Classification Deep Neural Network Watermarking?}
\newblock In {\em {IEEE Symposium on Security and Privacy (S\&P)}}. IEEE, 2022.

\bibitem{matiisen2019teacher}
Tambet Matiisen, Avital Oliver, Taco Cohen, and John Schulman.
\newblock Teacher--student curriculum learning.
\newblock {\em IEEE transactions on neural networks and learning systems}, 31(9):3732--3740, 2019.

\bibitem{MSCS19}
Luca Melis, Congzheng Song, Emiliano~De Cristofaro, and Vitaly Shmatikov.
\newblock {Exploiting Unintended Feature Leakage in Collaborative Learning}.
\newblock In {\em {IEEE Symposium on Security and Privacy (S\&P)}}, pages 497--512. IEEE, 2019.

\bibitem{mireshghallah2020privacy}
Fatemehsadat Mireshghallah, Mohammadkazem Taram, Praneeth Vepakomma, Abhishek Singh, Ramesh Raskar, and Hadi Esmaeilzadeh.
\newblock Privacy in deep learning: A survey.
\newblock {\em arXiv preprint arXiv:2004.12254}, 2020.

\bibitem{narvekar2018learning}
Sanmit Narvekar and Peter Stone.
\newblock Learning curriculum policies for reinforcement learning.
\newblock {\em arXiv preprint arXiv:1812.00285}, 2018.

\bibitem{NSH18}
Milad Nasr, Reza Shokri, and Amir Houmansadr.
\newblock {Machine Learning with Membership Privacy using Adversarial Regularization}.
\newblock In {\em {ACM SIGSAC Conference on Computer and Communications Security (CCS)}}, pages 634--646. ACM, 2018.

\bibitem{NSH19}
Milad Nasr, Reza Shokri, and Amir Houmansadr.
\newblock {Comprehensive Privacy Analysis of Deep Learning: Passive and Active White-box Inference Attacks against Centralized and Federated Learning}.
\newblock In {\em {IEEE Symposium on Security and Privacy (S\&P)}}, pages 1021--1035. IEEE, 2019.

\bibitem{netzer2011reading}
Yuval Netzer, Tao Wang, Adam Coates, Alessandro Bissacco, Bo~Wu, and Andrew~Y Ng.
\newblock Reading digits in natural images with unsupervised feature learning.
\newblock 2011.

\bibitem{OASF18}
Seong~Joon Oh, Max Augustin, Bernt Schiele, and Mario Fritz.
\newblock {Towards Reverse-Engineering Black-Box Neural Networks}.
\newblock In {\em {International Conference on Learning Representations (ICLR)}}, 2018.

\bibitem{OSF19}
Tribhuvanesh Orekondy, Bernt Schiele, and Mario Fritz.
\newblock {Knockoff Nets: Stealing Functionality of Black-Box Models}.
\newblock In {\em {IEEE Conference on Computer Vision and Pattern Recognition (CVPR)}}, pages 4954--4963. IEEE, 2019.

\bibitem{PSMRTE18}
Nicolas Papernot, Shuang Song, Ilya Mironov, Ananth Raghunathan, Kunal Talwar, and {\'{U}}lfar Erlingsson.
\newblock {Scalable Private Learning with {PATE}}.
\newblock In {\em {International Conference on Learning Representations (ICLR)}}, 2018.

\bibitem{RCK18}
Bita~Darvish Rouhani, Huili Chen, and Farinaz Koushanfar.
\newblock {DeepSigns: A Generic Watermarking Framework for IP Protection of Deep Learning Models}.
\newblock {\em {CoRR abs/1804.00750}}, 2018.

\bibitem{rumelhart1986learning}
David~E Rumelhart, Geoffrey~E Hinton, and Ronald~J Williams.
\newblock Learning representations by back-propagating errors.
\newblock {\em nature}, 323(6088):533--536, 1986.

\bibitem{sakaridis2019guided}
Christos Sakaridis, Dengxin Dai, and Luc~Van Gool.
\newblock Guided curriculum model adaptation and uncertainty-aware evaluation for semantic nighttime image segmentation.
\newblock In {\em Proceedings of the IEEE/CVF International Conference on Computer Vision}, pages 7374--7383, 2019.

\bibitem{SBBFZ20}
Ahmed Salem, Apratim Bhattacharya, Michael Backes, Mario Fritz, and Yang Zhang.
\newblock {Updates-Leak: Data Set Inference and Reconstruction Attacks in Online Learning}.
\newblock In {\em {USENIX Security Symposium (USENIX Security)}}, pages 1291--1308. USENIX, 2020.

\bibitem{SZHBFB19}
Ahmed Salem, Yang Zhang, Mathias Humbert, Pascal Berrang, Mario Fritz, and Michael Backes.
\newblock {ML-Leaks: Model and Data Independent Membership Inference Attacks and Defenses on Machine Learning Models}.
\newblock In {\em {Network and Distributed System Security Symposium (NDSS)}}. Internet Society, 2019.

\bibitem{SHZZC18}
Mark Sandler, Andrew~G. Howard, Menglong Zhu, Andrey Zhmoginov, and Liang{-}Chieh Chen.
\newblock {MobileNetV2: Inverted Residuals and Linear Bottlenecks}.
\newblock In {\em {IEEE Conference on Computer Vision and Pattern Recognition (CVPR)}}, pages 4510--4520. IEEE, 2018.

\bibitem{SHHZ22}
Yun Shen, Xinlei He, Yufei Han, and Yang Zhang.
\newblock {Model Stealing Attacks Against Inductive Graph Neural Networks}.
\newblock In {\em {IEEE Symposium on Security and Privacy (S\&P)}}, pages 1175--1192. IEEE, 2022.

\bibitem{SSSS17}
Reza Shokri, Marco Stronati, Congzheng Song, and Vitaly Shmatikov.
\newblock {Membership Inference Attacks Against Machine Learning Models}.
\newblock In {\em {IEEE Symposium on Security and Privacy (S\&P)}}, pages 3--18. IEEE, 2017.

\bibitem{shumailov2021manipulating}
Ilia Shumailov, Zakhar Shumaylov, Dmitry Kazhdan, Yiren Zhao, Nicolas Papernot, Murat~A Erdogdu, and Ross~J Anderson.
\newblock Manipulating sgd with data ordering attacks.
\newblock {\em Advances in Neural Information Processing Systems}, 34:18021--18032, 2021.

\bibitem{SR20}
Congzheng Song and Ananth Raghunathan.
\newblock {Information Leakage in Embedding Models}.
\newblock In {\em {ACM SIGSAC Conference on Computer and Communications Security (CCS)}}, pages 377--390. ACM, 2020.

\bibitem{SS20}
Congzheng Song and Vitaly Shmatikov.
\newblock {Overlearning Reveals Sensitive Attributes}.
\newblock In {\em {International Conference on Learning Representations (ICLR)}}, 2020.

\bibitem{SM21}
Liwei Song and Prateek Mittal.
\newblock {Systematic Evaluation of Privacy Risks of Machine Learning Models}.
\newblock In {\em {USENIX Security Symposium (USENIX Security)}}. USENIX, 2021.

\bibitem{soviany2020image}
Petru Soviany, Claudiu Ardei, Radu~Tudor Ionescu, and Marius Leordeanu.
\newblock Image difficulty curriculum for generative adversarial networks (cugan).
\newblock In {\em Proceedings of the IEEE/CVF Winter Conference on Applications of Computer Vision}, pages 3463--3472, 2020.

\bibitem{soviany2021curriculum}
Petru Soviany, Radu~Tudor Ionescu, Paolo Rota, and Nicu Sebe.
\newblock Curriculum learning: A survey.
\newblock {\em arXiv preprint arXiv:2101.10382}, 2021.

\bibitem{spitkovsky2009baby}
Valentin~I Spitkovsky, Hiyan Alshawi, and Daniel Jurafsky.
\newblock Baby steps: How “less is more” in unsupervised dependency parsing.
\newblock 2009.

\bibitem{szegedy2016rethinking}
Christian Szegedy, Vincent Vanhoucke, Sergey Ioffe, Jon Shlens, and Zbigniew Wojna.
\newblock Rethinking the inception architecture for computer vision.
\newblock In {\em Proceedings of the IEEE conference on computer vision and pattern recognition}, pages 2818--2826, 2016.

\bibitem{TZJRR16}
Florian Tram{\`e}r, Fan Zhang, Ari Juels, Michael~K. Reiter, and Thomas Ristenpart.
\newblock {Stealing Machine Learning Models via Prediction APIs}.
\newblock In {\em {USENIX Security Symposium (USENIX Security)}}, pages 601--618. USENIX, 2016.

\bibitem{UNSS17}
Yusuke Uchida, Yuki Nagai, Shigeyuki Sakazawa, and Shin'ichi Satoh.
\newblock {Embedding Watermarks into Deep Neural Networks}.
\newblock In {\em {International Conference on Multimedia Retrieval (ICMR)}}, pages 269--277. ACM, 2017.

\bibitem{WG18}
Binghui Wang and Neil~Zhenqiang Gong.
\newblock {Stealing Hyperparameters in Machine Learning}.
\newblock In {\em {IEEE Symposium on Security and Privacy (S\&P)}}, pages 36--52. IEEE, 2018.

\bibitem{wang2021survey}
Xin Wang, Yudong Chen, and Wenwu Zhu.
\newblock A survey on curriculum learning.
\newblock {\em IEEE Transactions on Pattern Analysis and Machine Intelligence}, 2021.

\bibitem{watson2021importance}
Lauren Watson, Chuan Guo, Graham Cormode, and Alexandre Sablayrolles.
\newblock On the importance of difficulty calibration in membership inference attacks.
\newblock In {\em International Conference on Learning Representations (ICLR)}, 2022.

\bibitem{weinshall2018curriculum}
Daphna Weinshall, Gad Cohen, and Dan Amir.
\newblock Curriculum learning by transfer learning: Theory and experiments with deep networks.
\newblock In {\em International Conference on Machine Learning}, pages 5238--5246. PMLR, 2018.

\bibitem{WLZL22}
Fan Wu, Yunhui Long, Ce~Zhang, and Bo~Li.
\newblock {LinkTeller: Recovering Private Edges from Graph Neural Networks via Influence Analysis}.
\newblock In {\em {IEEE Symposium on Security and Privacy (S\&P)}}, pages 2005--2024. IEEE, 2022.

\bibitem{wu2021when}
Xiaoxia Wu, Ethan Dyer, and Behnam Neyshabur.
\newblock When do curricula work?
\newblock In {\em International Conference on Learning Representations}, 2021.

\bibitem{YKT19}
Mohammad Yaghini, Bogdan Kulynych, and Carmela Troncoso.
\newblock {Disparate Vulnerability: on the Unfairness of Privacy Attacks Against Machine Learning}.
\newblock {\em {CoRR abs/1906.00389}}, 2019.

\bibitem{yang2016participatory}
Dingqi Yang, Daqing Zhang, and Bingqing Qu.
\newblock Participatory cultural mapping based on collective behavior data in location-based social networks.
\newblock {\em ACM Transactions on Intelligent Systems and Technology (TIST)}, 7(3):1--23, 2016.

\bibitem{zhang2020secret}
Yuheng Zhang, Ruoxi Jia, Hengzhi Pei, Wenxiao Wang, Bo~Li, and Dawn Song.
\newblock The secret revealer: Generative model-inversion attacks against deep neural networks.
\newblock In {\em Proceedings of the IEEE/CVF Conference on Computer Vision and Pattern Recognition}, pages 253--261, 2020.

\bibitem{zhang2017age}
Zhifei Zhang, Yang Song, and Hairong Qi.
\newblock Age progression/regression by conditional adversarial autoencoder.
\newblock In {\em Proceedings of the IEEE conference on computer vision and pattern recognition}, pages 5810--5818, 2017.

\bibitem{zheng2019autoencoder}
Siqi Zheng, Gang Liu, Hongbin Suo, and Yun Lei.
\newblock Autoencoder-based semi-supervised curriculum learning for out-of-domain speaker verification.
\newblock {\em System}, 3:98, 2019.

\bibitem{zhou2017places}
Bolei Zhou, Agata Lapedriza, Aditya Khosla, Aude Oliva, and Antonio Torralba.
\newblock Places: A 10 million image database for scene recognition.
\newblock {\em IEEE transactions on pattern analysis and machine intelligence}, 40(6):1452--1464, 2017.

\bibitem{zhou2020uncertainty}
Yikai Zhou, Baosong Yang, Derek~F Wong, Yu~Wan, and Lidia~S Chao.
\newblock Uncertainty-aware curriculum learning for neural machine translation.
\newblock In {\em Proceedings of the 58th Annual Meeting of the Association for Computational Linguistics}, pages 6934--6944, 2020.

\end{thebibliography}

\appendix

\begin{figure*}[]
    \centering
    \begin{subfigure}[h]{.32\linewidth}
    \centering\includegraphics[width=\linewidth]{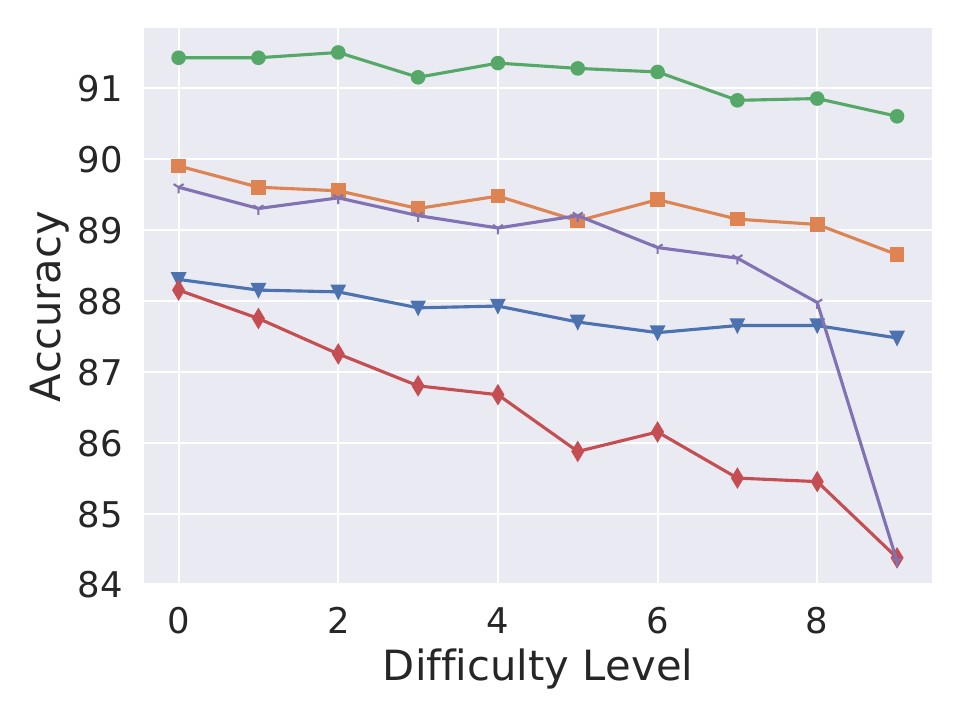}
    \caption{MIA accuracy on Place60}
  \end{subfigure}
  \begin{subfigure}[h]{.32\linewidth}
    \centering\includegraphics[width=\linewidth]{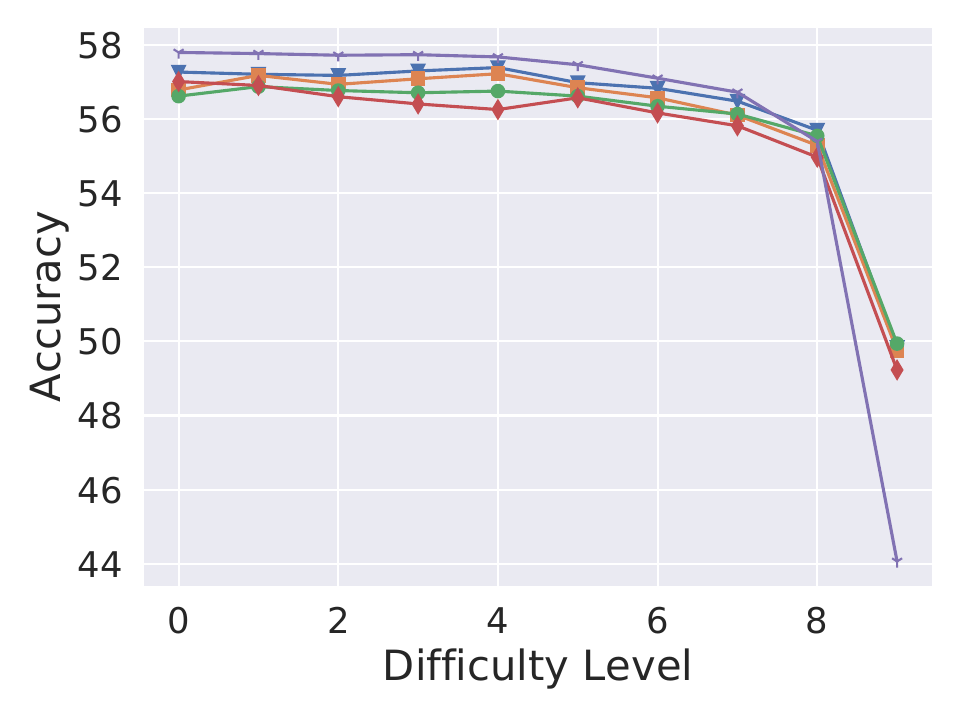}
    \caption{MIA accuracy on SVHN}
  \end{subfigure}
  \begin{subfigure}[h]{.32\linewidth}
    \centering\includegraphics[width=\linewidth]{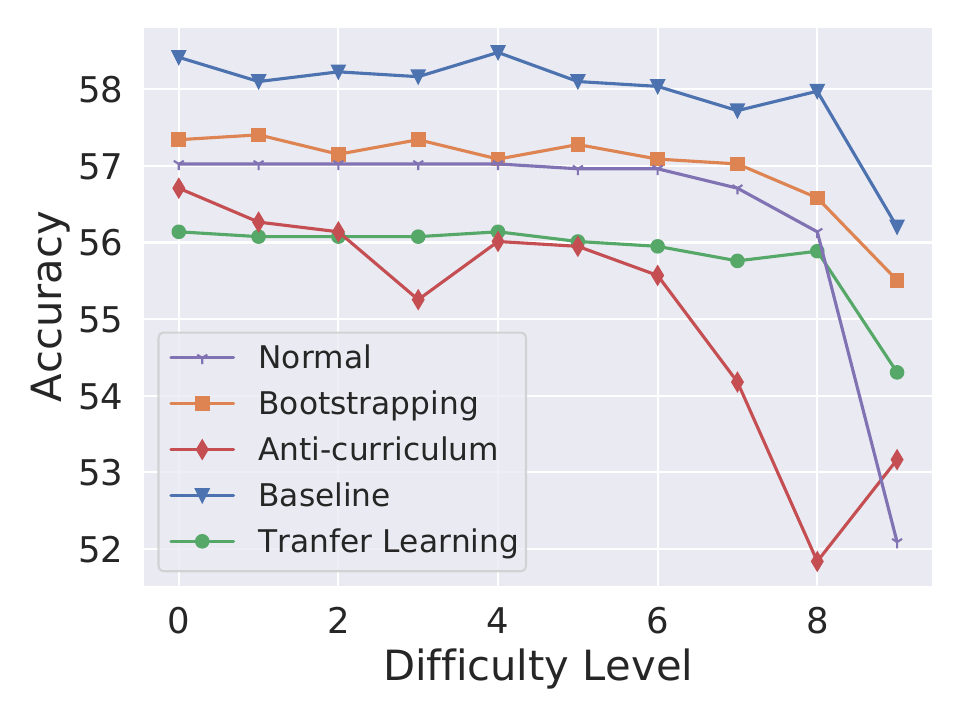}
    \caption{MIA accuracy on UTKFace}
  \end{subfigure}
    \caption{MIA accuracy on Place60, SVHN and UTKFace. ResNet-18 is used for target model training and bootstrapping is used for CL. The x-axis represents the difficulty level of the data, and the y-axis represents the attack accuracy.}
    \label{fig: bucket_attack_acc_app}
\end{figure*}
\section{More Results}
\label{app:appendix}
\subsection{t-SNE Study} 
\label{app:t-SNE}

\begin{figure}[H]
    \centering
    \begin{subfigure}[t]{.46\linewidth}
    \includegraphics[width=\linewidth]{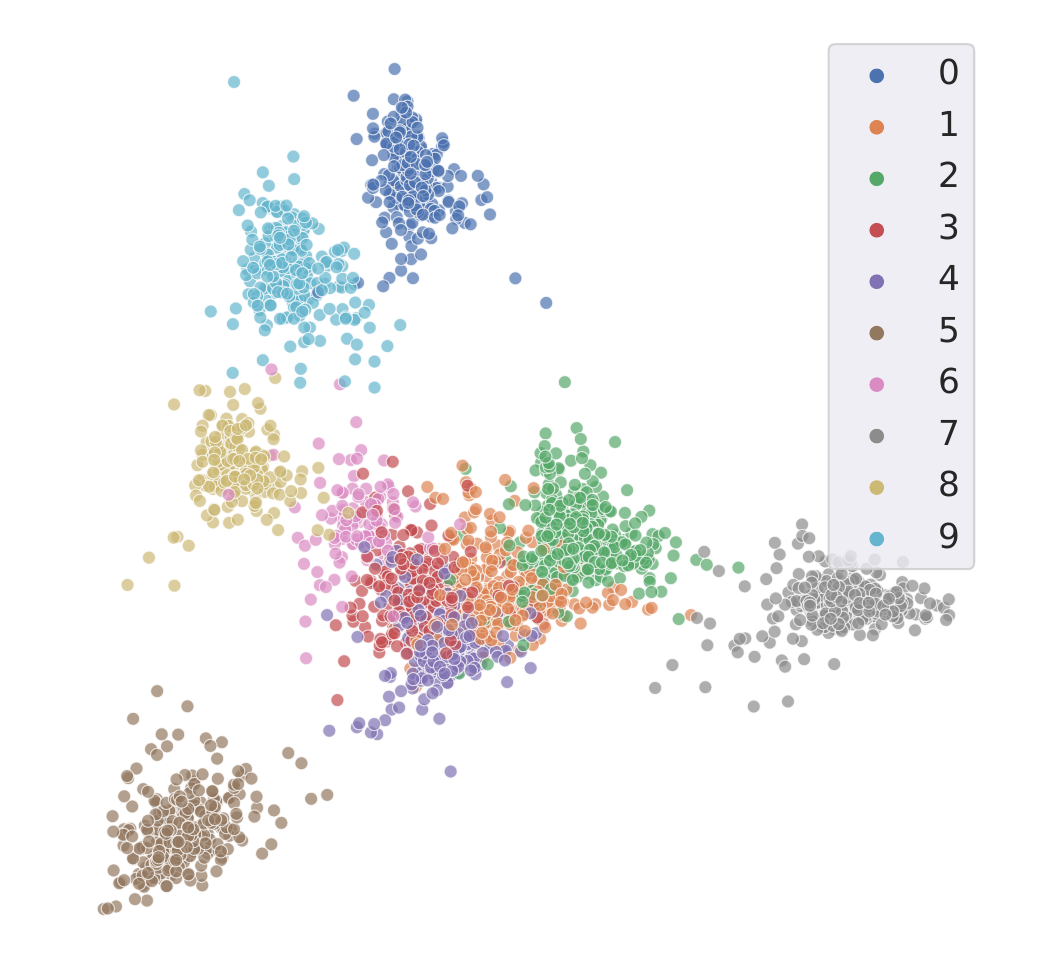}
    \caption{Normal training}\label{subfig: tsne_SVHN_normal}
  \end{subfigure}
  \begin{subfigure}[t]{.46\linewidth}
    \includegraphics[width=\linewidth]{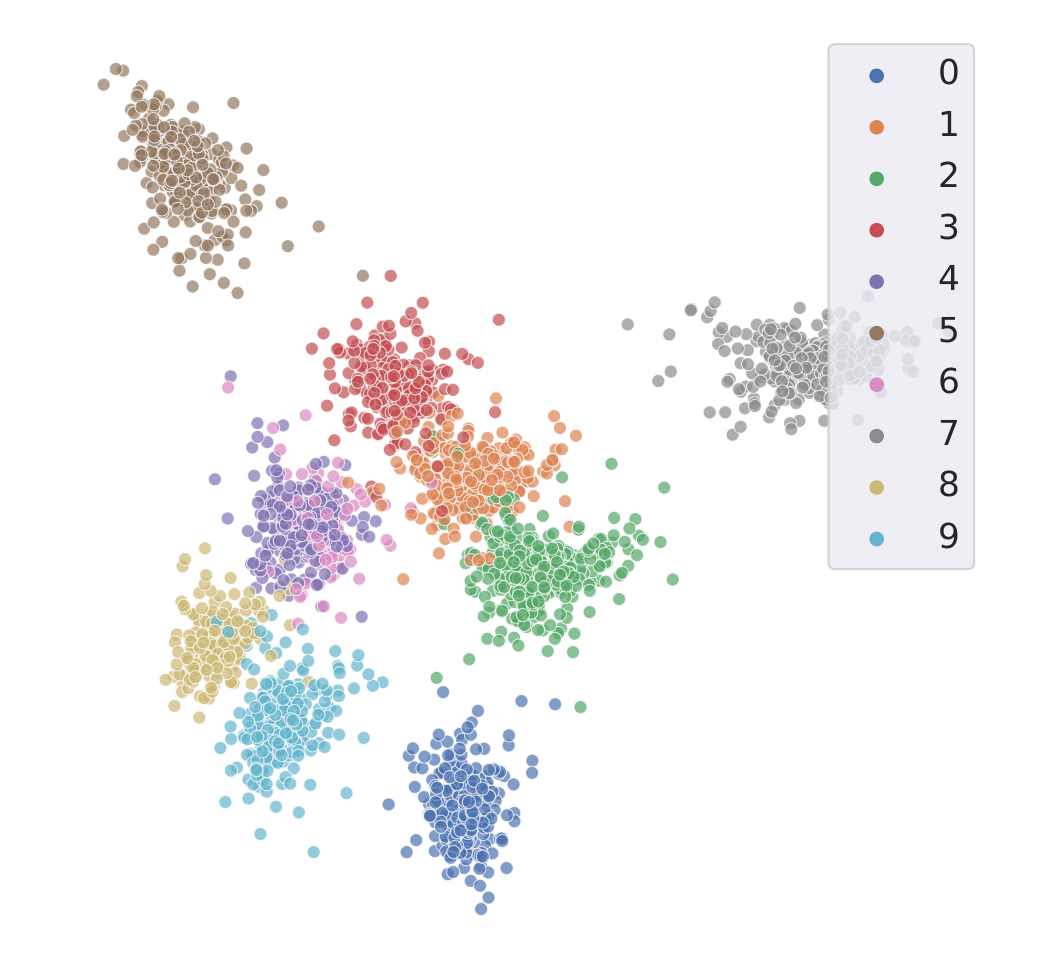}
    \caption{Bootstrapping}\label{subfig: tsne_SVHN_bootstrapping}
  \end{subfigure}
    \caption{t-SNE of the classification results on the difficult batch of SVHN.
    }
    \label{fig: tsne_SVHN}
\end{figure}

To investigate the disparate impact CL has on the classification accuracy across samples. we use t-distributed stochastic neighbor embedding (t-SNE) to visualize the classification tasks carried out by bootstrapping and normal ML on the most difficult batch of data of SVHN. \autoref{fig: tsne_SVHN} shows all samples within the difficult batch, and it turns out bootstrapping can separate  samples from group ``1'', ``2'' and ``3'' better than normal training. 

\subsection{More Results of Confidence Score} 
\label{app:appendix}

\begin{table*}[t]
\centering
\begin{tabular}{l|cccccccccccccc}
\toprule
{\backslashbox[26mm]{Method}{Dataset}}    & {Tiny ImageNet} & {CIFAR100} & {Place100} &{Place60} 
&{SVHN} 
&{Purchase} &{Texas}&{Location} \\
\midrule
Normal &100.0&100.0&100.0&100.0&100.0&100.0&96.770&100.0 \\

Bootstrapping &100.0&100.0&100.0&99.996&100.0&100.0&94.030&100.0 \\

Transfer &100.0&99.997&100.0&99.972&100.0&/&/&/\\

Baseline &100.0&99.993&100.0&100.0&100.0&99.990&95.600&100.0\\

Anti-curriculum &99.963&100.0&100.0&99.918&100.0&100.0&97.410&100.0\\
\bottomrule
\end{tabular}
\normalsize
\caption{The average training accuracy of datasets in \autoref{fig: target_acc}. Image datasets are trained on ResNet-18 while non-image datasets are trained on MLP. Numbers are all in percentage. We observe that all training accuracies are nearly 100\%. Note that for non-image datasets, we skip the transfer method as there is no commonly used pre-trained model for the tabular dataset.
}
\label{tb: trainning_acc}
\end{table*}

\begin{figure}[h]
    \centering
    \begin{subfigure}[h]{.49\linewidth}
    \centering\includegraphics[width=\linewidth]{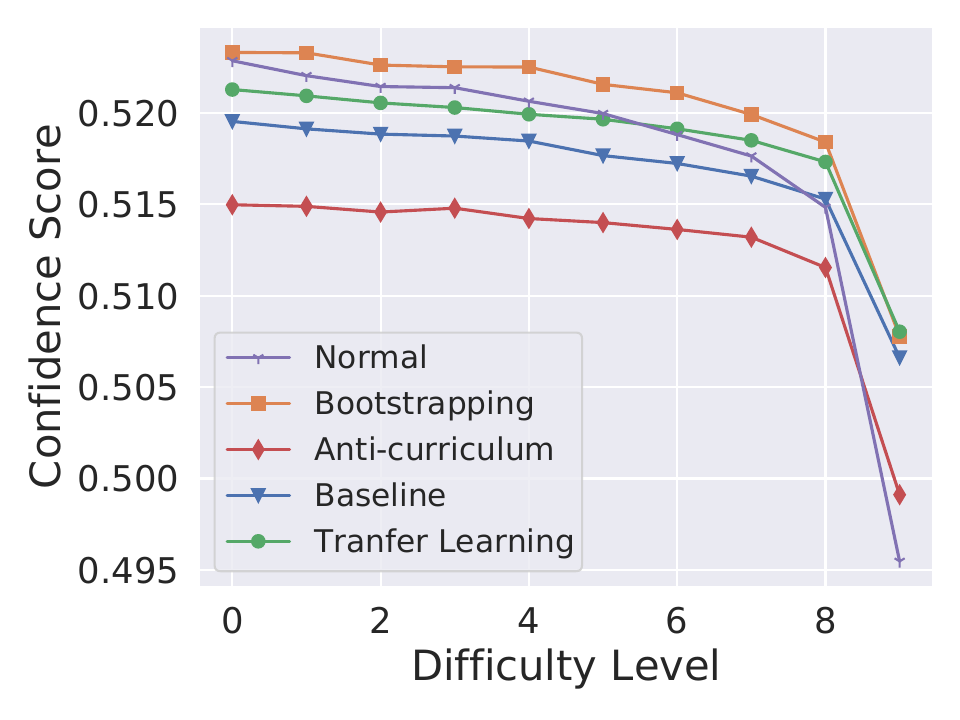}
    \caption{SVHN member}
  \end{subfigure}
  \begin{subfigure}[h]{.49\linewidth}
\centering\includegraphics[width=\linewidth]{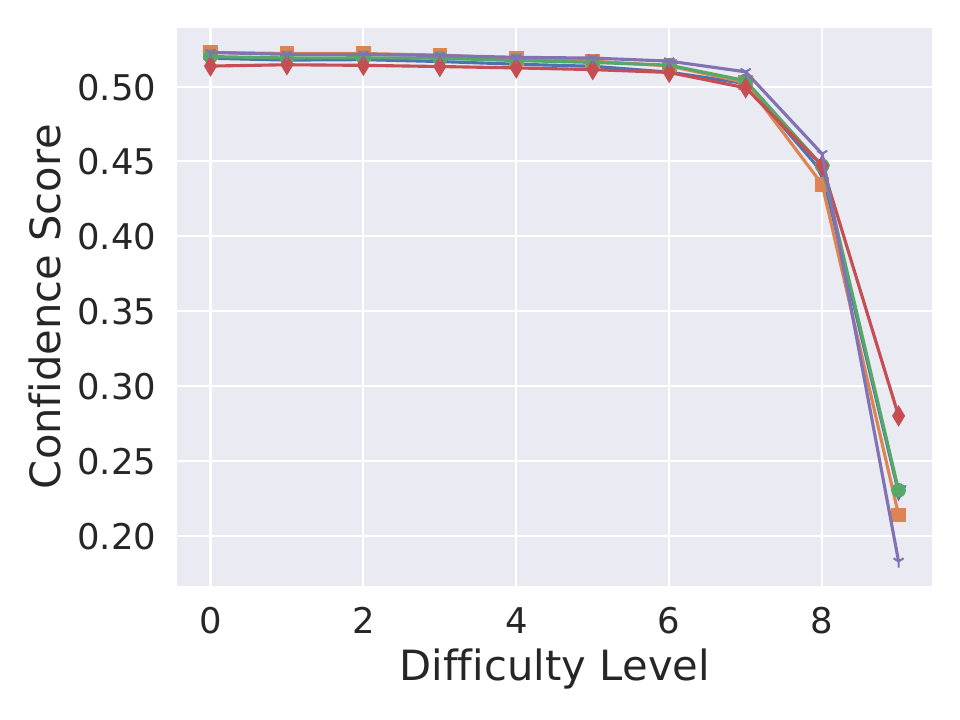}
    \caption{SVHN non-member}
  \end{subfigure}
    \caption{Attack model's confidence score for both member and non-member samples on SVHN. ResNet-18 is used for target model training, and data samples are arranged according to their difficulty scores from bootstrapping.}
\end{figure}

\begin{figure}[h]
    \centering
    \begin{subfigure}[h]{.49\linewidth}
    \centering\includegraphics[width=\linewidth]{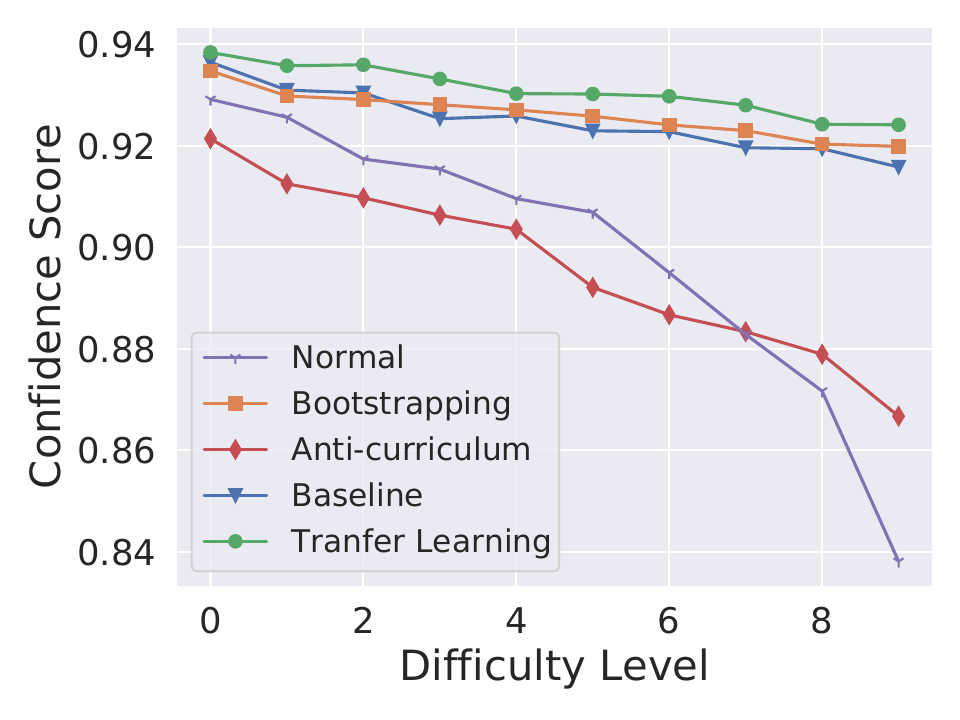}
    \caption{Place100 member}
  \end{subfigure}
  \begin{subfigure}[h]{.49\linewidth}
    \centering\includegraphics[width=\linewidth]{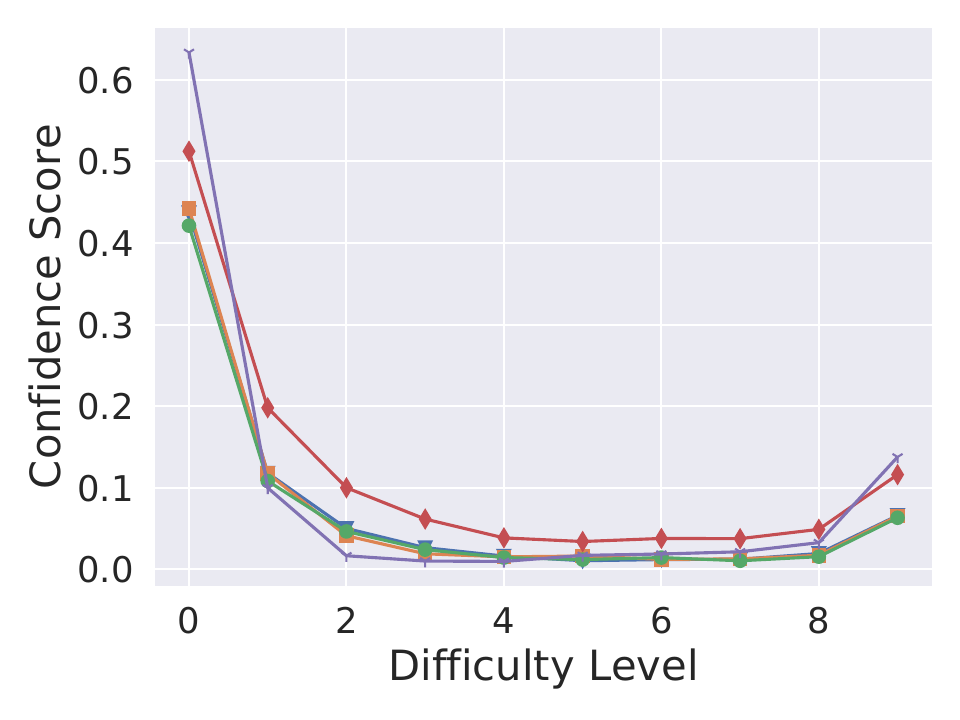}
    \caption{Place100 non-member}
  \end{subfigure}
    \caption{Attack model's confidence score for both member and non-member samples on Place100. ResNet-18 is used for target model training, and data samples are arranged according to their difficulty scores from bootstrapping.}
\end{figure}

\begin{figure}[H]
    \centering
    \begin{subfigure}[h]{.49\linewidth}
    \centering\includegraphics[width=\linewidth]{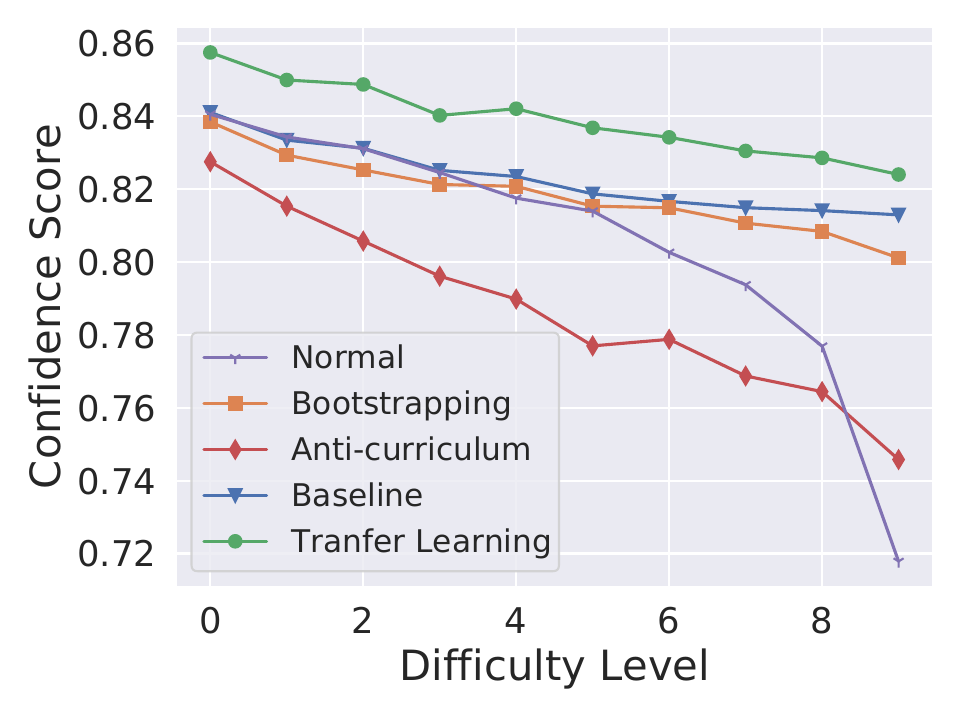}
    \caption{Place60 member}
  \end{subfigure}
  \begin{subfigure}[h]{.49\linewidth}
    \centering\includegraphics[width=\linewidth]{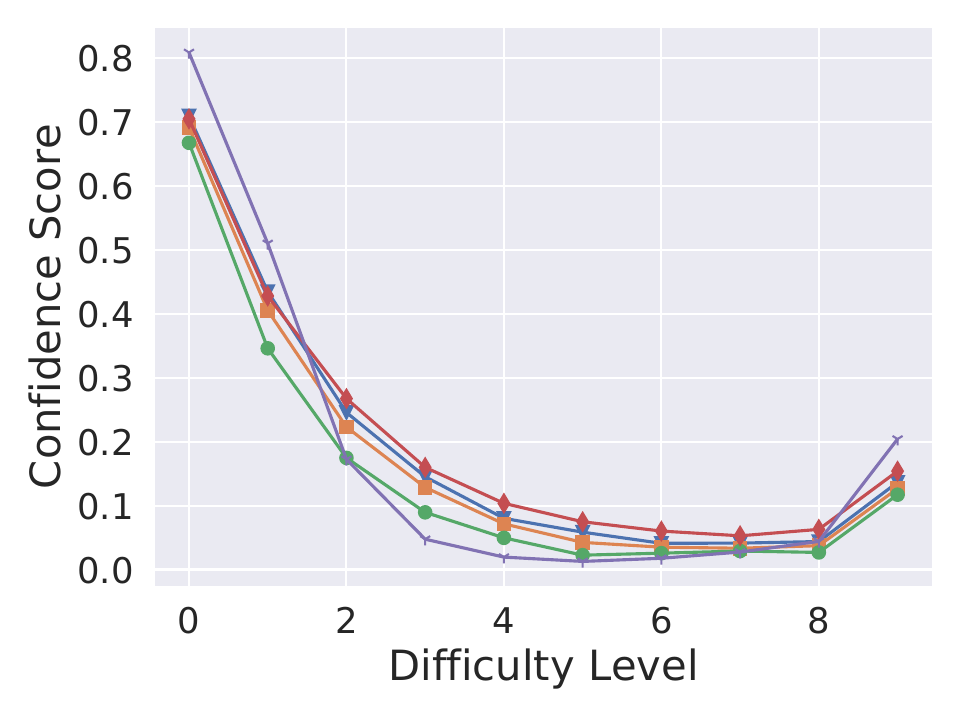}
    \caption{Place60 non-member}
  \end{subfigure}
    \caption{Attack model's confidence score for both member and non-member samples on Place60. ResNet-18 is used for target model training, and data samples are arranged according to their difficulty scores from bootstrapping.}
\end{figure}
\begin{figure}[H]
    \centering
    \begin{subfigure}[h]{.49\linewidth}
    \centering\includegraphics[width=\linewidth]{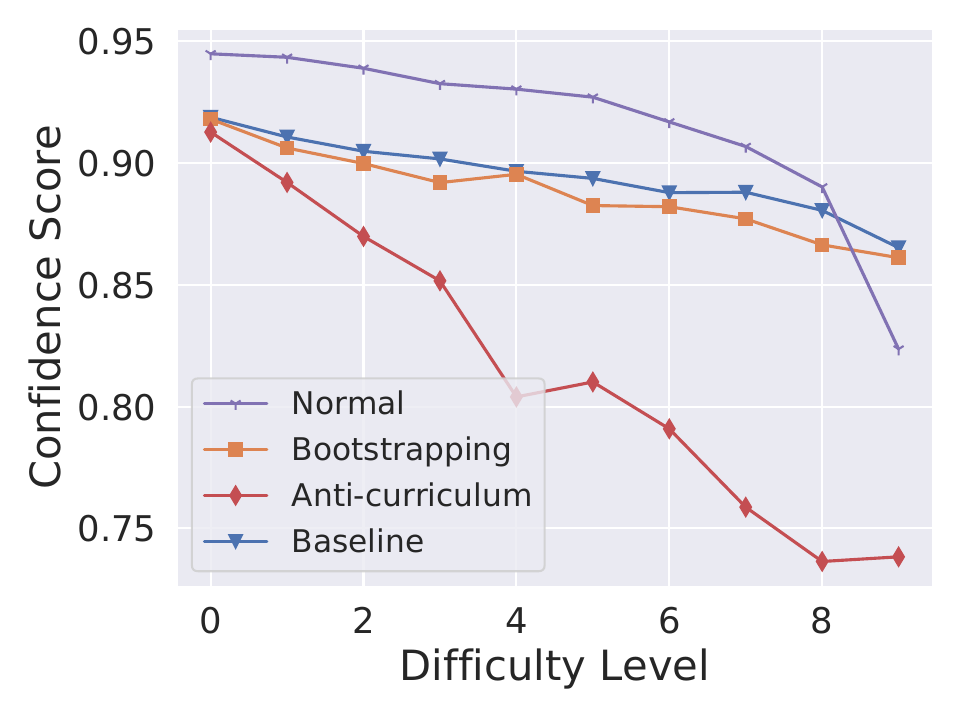}
    \caption{Purchase member}
  \end{subfigure}
  \begin{subfigure}[h]{.49\linewidth}
    \centering\includegraphics[width=\linewidth]{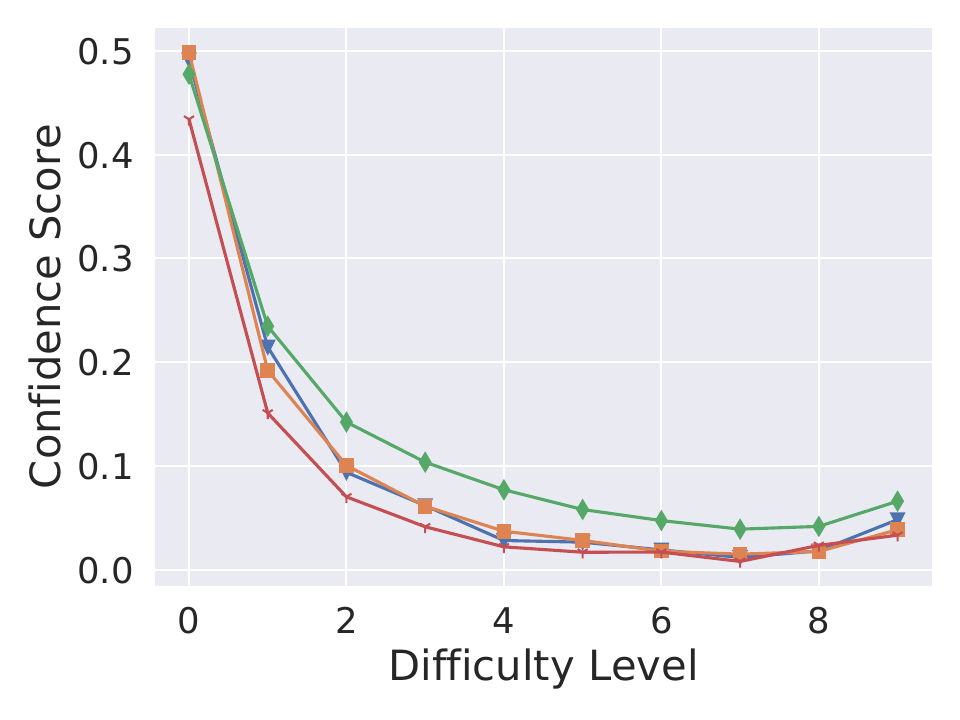}
    \caption{Purchase non-member}
  \end{subfigure}
    \caption{Attack model's confidence score for both member and non-member samples on Purchase. MLP is used for target model training, and data samples are arranged according to their difficulty scores from bootstrapping.}
    \label{app: purchase_fig}
\end{figure}

\begin{figure}[H]
    \centering
    \begin{subfigure}[t]{.49\linewidth}
    \centering\includegraphics[width=\linewidth]{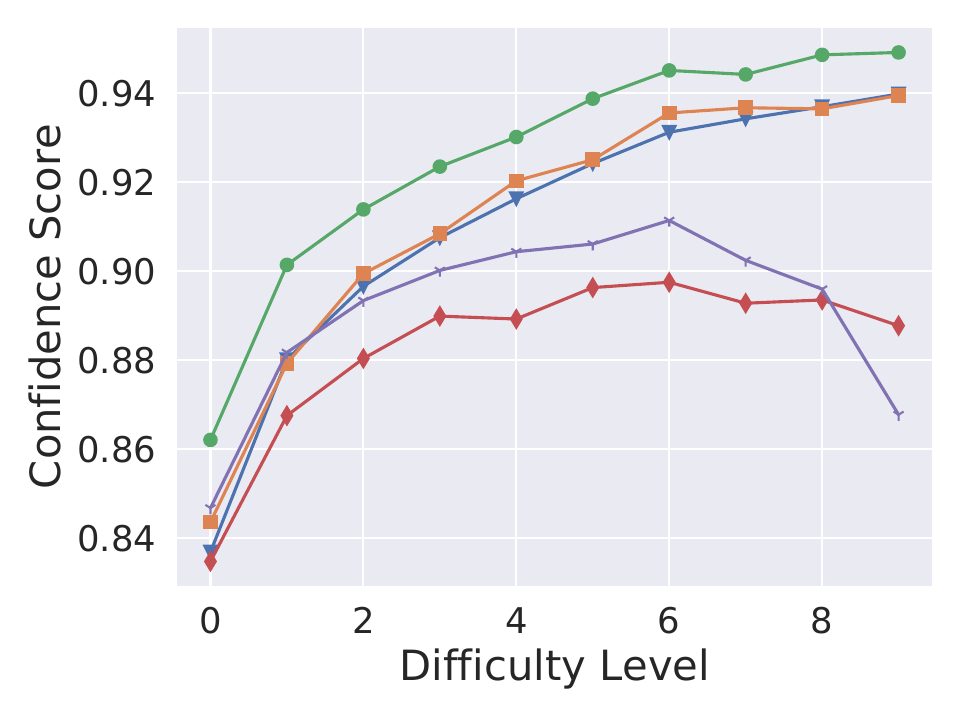}
    \caption{ Member}\label{subfig: ours_mia_cifar100}
  \end{subfigure}
  \begin{subfigure}[t]{.49\linewidth}
    \centering\includegraphics[width=\linewidth]{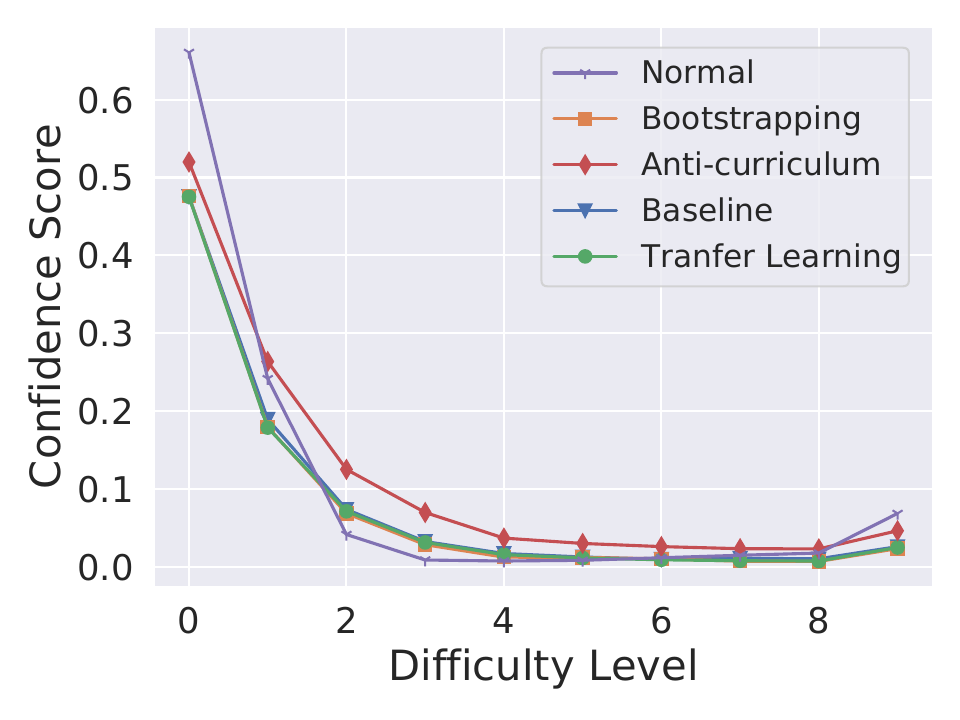}
    \caption{Non-member}\label{subfig: ours_mia_tiny}
  \end{subfigure}
    \caption{\attack{}'s member and non-member confidence score for models trained on Tiny ImageNet with ResNet-18.}
    \label{fig: ours_tiny_confi}
\end{figure}

\begin{figure}[H]
    \centering
    \begin{subfigure}[t]{.49\linewidth}
    \centering\includegraphics[width=\linewidth]{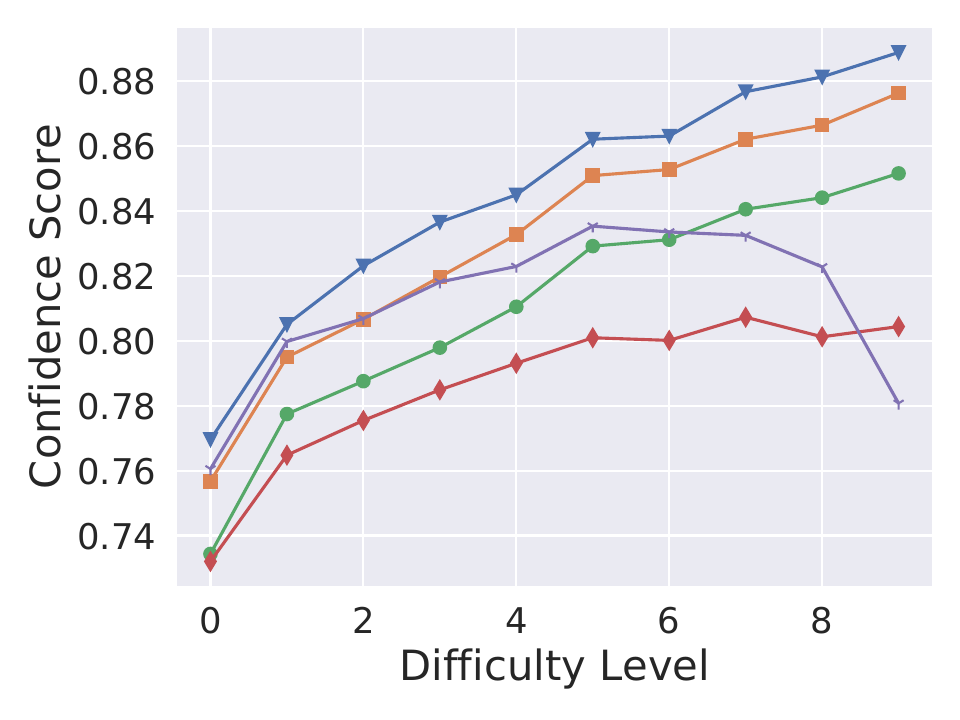}
    \caption{ Member}\label{subfig: ours_mia_cifar100}
  \end{subfigure}
  \begin{subfigure}[t]{.49\linewidth}
    \centering\includegraphics[width=\linewidth]{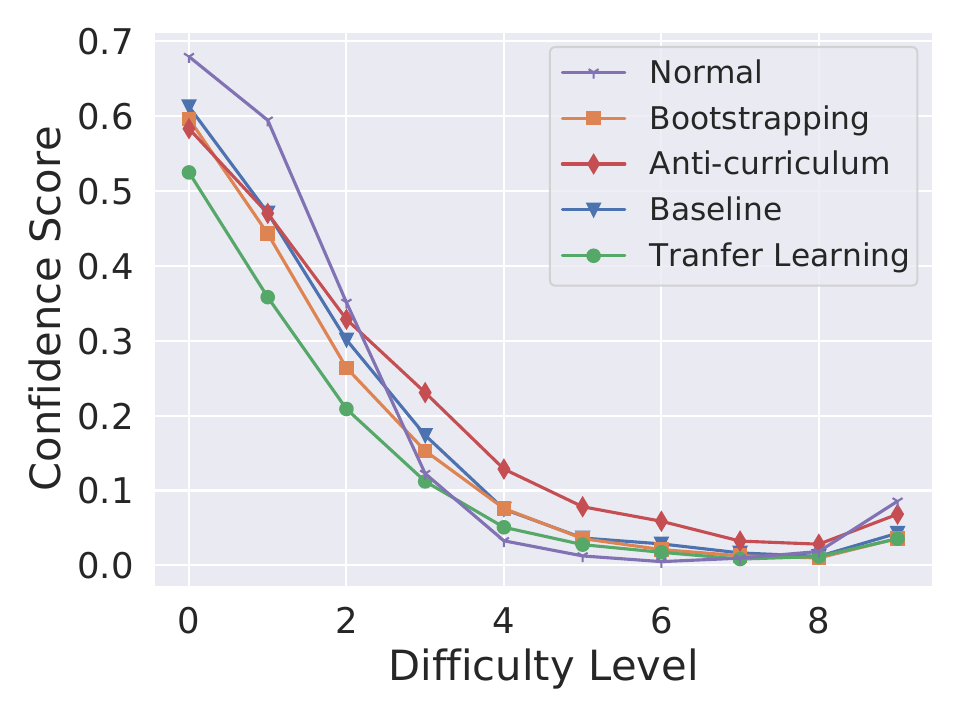}
    \caption{Non-member}\label{subfig: ours_mia_tiny}
  \end{subfigure}
    \caption{\attack{}'s member and non-member confidence score for models trained on CIFAR100 with ResNet-18.
    }
    \label{fig: ours_cifar100_confi}
\end{figure}

\begin{figure}[H]
    \centering
    \includegraphics[width=\linewidth]{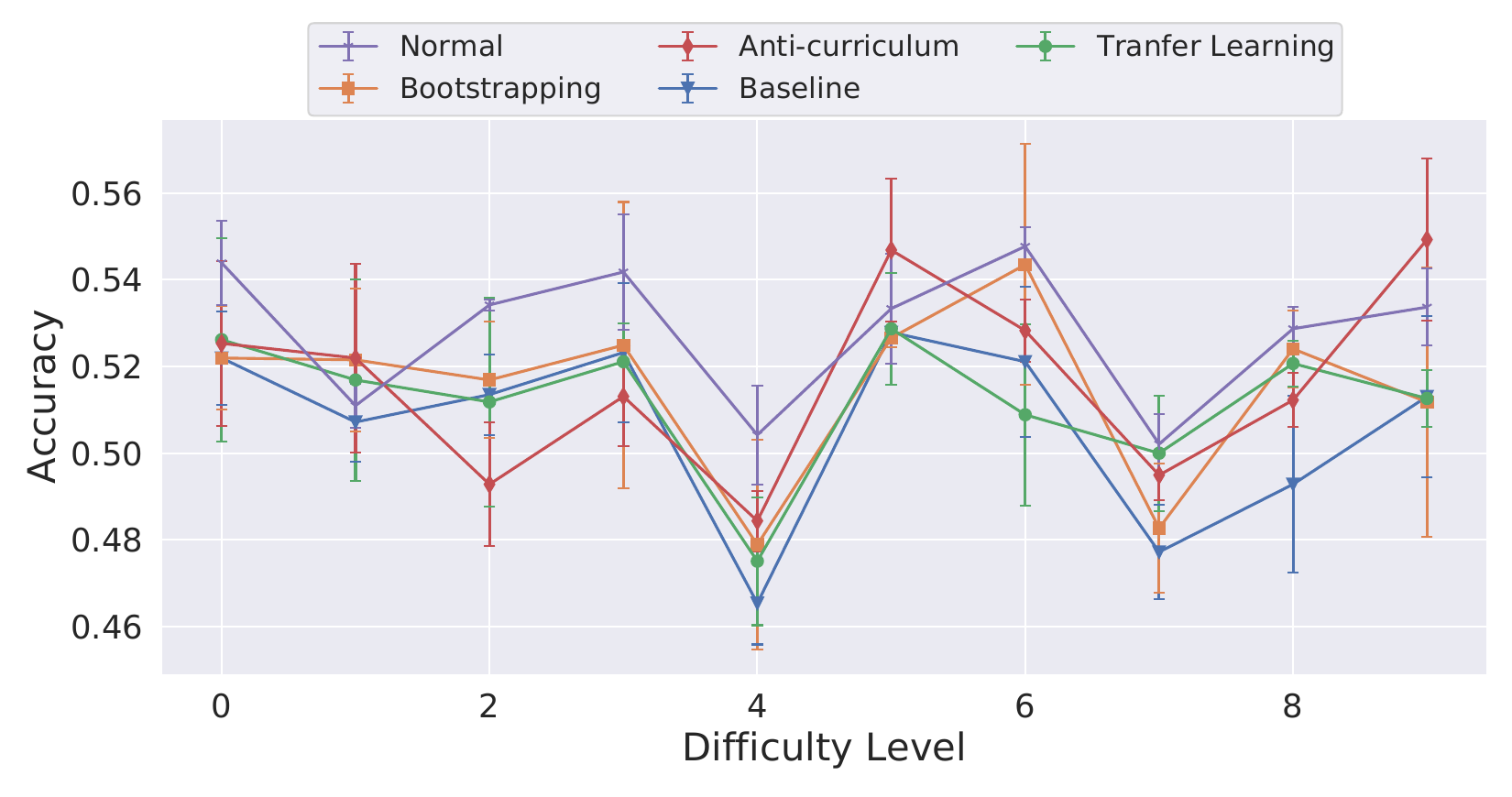}
    \caption{Attribute inference attack accuracy on UTKFace}\label{fig: aia_app}
\end{figure}
\subsection{AIA Accuracy}\label{app:aia}
\autoref{fig: aia_app} shows that CL does not necessarily make the target model more vulnerable to AIA.

\subsection{Training Accuracy}\label{app:training_acc}
Training accuracy corresponding to datasets in ~\autoref{fig: target_acc} are listed in ~\autoref{tb: trainning_acc}. All numbers are in percentage.

\end{document}